\documentclass[fleqn,usenatbib]{mnras}

\usepackage[T1]{fontenc}

\DeclareRobustCommand{\VAN}[3]{#2}
\let\VANthebibliography\thebibliography
\def\thebibliography{\DeclareRobustCommand{\VAN}[3]{##3}\VANthebibliography}

\usepackage{graphicx}	
\usepackage{amsmath}	
\usepackage{amssymb}	
\usepackage{xspace}
\usepackage[colorinlistoftodos]{todonotes}

\newcommand{\V}{$V$\xspace}	
\newcommand{\sig}{$\sigma$\xspace}	
\newcommand{\h}[1]{$h_{#1}$\xspace}
\newcommand{\kms}{km\ s$^{-1}$\xspace}	
\newcommand{\Fe}{[M/H]\xspace}	
\newcommand{\al}{[Mg/Fe]\xspace}	
\newcommand{\msun}{M$_\odot$\xspace}	
	
\usepackage{newtxtext,newtxmath}

\title[NGC 5746]{NGC 5746: formation history of a massive disc-dominated galaxy}

\author[M. Martig et al.]{Marie Martig,$^{1}$\thanks{E-mail: m.martig@ljmu.ac.uk}, Francesca Pinna$^{2}$, Jes{\'u}s Falc{\'o}n-Barroso$^{3,4}$, Dimitri A. Gadotti$^{5}$, Bernd Husemann$^{2}$, \newauthor
Ivan Minchev$^{6}$, Justus Neumann$^{7}$, Tom\'as Ruiz-Lara$^{8}$, Glenn van de Ven$^{9}$
\\
$^{1}$Astrophysics Research Institute, Liverpool John Moores University, 146 Brownlow Hill, Liverpool L3 5RF, UK\\
$^{2}$Max-Planck-Institut f\"{u}r Astronomie, K\"{o}nigstuhl 17, D-69117 Heidelberg, Germany\\
$^{3}$Instituto de Astrof{\'i}sica de Canarias, Calle Via L{\'a}ctea s/n, 38200 La Laguna, Tenerife, Spain\\
$^{4}$Depto. Astrof{\'i}sica, Universidad de La Laguna, Calle Astrof{\'i}sico Francisco S{\'a}nchez s/n, 38206 La Laguna, Tenerife, Spain\\
$^{5}$ European Southern Observatory, Karl-Schwarzschild-Str. 2, D-85748 Garching bei M\"{u}nchen, Germany\\
$^{6}$Leibniz-Institut f\"{u}r Astrophysik Potsdam (AIP), An der Sternwarte 16, D-14482 Potsdam, Germany\\
$^{7}$ Institute of Cosmology and Gravitation, University of Portsmouth, Burnaby Road, Portsmouth, PO1 3FX, UK\\
$^{8}$Kapteyn Astronomical Institute, University of Groningen, Landleven 12, 9747 AD Groningen, The Netherlands \\
$^{9}$Department of Astrophysics, University of Vienna, T\"{u}rkenschanzstrasse 17, 1180 Wien, Austria\\
}
\date{Accepted XXX. Received YYY; in original form ZZZ}

\pubyear{2021}

\begin{document}
\label{firstpage}
\pagerange{\pageref{firstpage}--\pageref{lastpage}}
\maketitle

\begin{abstract}
The existence of massive galaxies lacking a classical bulge has often been proposed as a challenge to $\Lambda$CDM. However, recent simulations propose that a fraction of massive disc galaxies might have had very quiescent merger histories, and also that mergers do not necessarily build classical bulges.  We test these ideas with deep MUSE observations of NGC 5746, a massive ($\sim 10^{11}$ \msun) edge-on disc galaxy with no classical bulge. We analyse its stellar kinematics and stellar populations, and infer that a massive and extended disc formed very early: 80\% of the galaxy’s stellar mass formed more than 10 Gyr ago. Most of the thick disc and the bar formed during that early phase. The bar drove gas towards the center and triggered the formation of the nuclear disc followed by the growth of a boxy/peanut-shaped bulge. Around $\sim$ 8 Gyr ago,  a $\sim$1:10 merger happened, possibly on a low-inclination orbit. The satellite did not cause significant vertical heating, did not contribute to the growth of a classical bulge, and did not destroy the bar and the nuclear disc.  It was however an important event for the galaxy: by depositing its stars throughout the whole galaxy it contributed $\sim 30$\% of accreted stars to the thick disc. NGC 5746 thus did not completely escape mergers, but the only relatively recent significant merger did not damage the galaxy and did not create a classical bulge. Future observations will reveal if this is representative of the formation histories of massive disc galaxies.
\end{abstract}




\section{Introduction}

Disc galaxies host a variety of components, and the relation between their detailed  morphology and their formation history is complex, and still not fully understood. The general picture of disc galaxy formation starts with an early violent phase, with mergers or violent internal instabilities creating thick discs and bulges. Then, at lower redshift, as the merger rate decreases, thinner discs can settle, form bars and spiral arms and are dominated by secular evolution (slow changes driven by internal processes, see \citealp{Kormendy2004,Sellwood2014}). Those mechanisms all act to shape galaxies, and create the diversity of morphologies observed along the Hubble Sequence.

Most spiral galaxies, including  the Milky Way, contain both a thin and thick disc \citep{Gilmore1983,Dalcanton2002,Yoachim2006,Comeron2018}, although it is not always clear if they are clearly meaningful distinct components (see discussions in \citealp{Bovy2012a,Mackereth2017} for the Milky Way). One way to form thick discs is through mergers: mergers can thicken pre-existing discs \citep{Quinn1993,Read2008,Villalobos2008} and/or add accreted stars in a thick configuration \citep{Abadi2003}. Alternatively, discs can form already thick from turbulent gas, either following gas-rich mergers \citep{Brook2004} or violent disc instabilities in gas-rich discs \citep{Bournaud2009}. Thick discs typically have older, more metal-poor populations than thin discs, although there exists some diversity \citep{Mould2005, Rejkuba2009,Yoachim2008b,Comeron2015,Comeron2016, Kasparova2016,Kasparova2020, Pinna2019a,Pinna2019b,Scott2021}. In particular, some thick discs are complex components, with their outer parts being formed from flared younger populations \citep{Minchev2015, Garcia2021}, creating radial age gradients like the one observed in the Milky Way \citep{Martig2016}. Whether this picture applies to thick discs in general is unknown, but it is likely that thick discs could arise from several different formation channels.

Bulges are also complex, and are usually classified in two main categories: classical and pseudobulges (see reviews by \citealp{Kormendy2004,Kormendy2013, Fisher2016}). Classical bulges have a spherical shape, a high S{\'e}rsic index, and are dominated by velocity dispersion. They are formed during violent processes, either galaxy mergers \citep{Walker1996,Aguerri2001}, or after the coalescence of giant clumps in high redshift galaxies \citep{Bournaud2007,Ceverino2015}. By contrast, pseudobulges have a low S{\'e}rsic index, are dominated by rotation, and arise from bar-driven secular processes within discs. Discy pseudobulges (sometimes also called nuclear discs) form following the inflow of gas along bars, with gas settling and forming stars in a ring or disc configuration \citep{Kormendy2004,Athanassoula2005}. The other type of pseudobulges are boxy/peanut shaped bulges (B/P bulges, see e.g., \citealp{Bureau1999,Lutticke2000}), and they form through vertical thickening of a bar, either via the buckling instability or vertical resonances \citep{Combes1990,Debattista2004,Martinez2006,Sellwood2020}.

Galaxies can host different types of bulges at the same time \citep{Kormendy2010a,Mendez2014, Erwin2015b}, but most nearby disc galaxies are either bulgeless or only have a pseudobulge \citep{Fisher2011}. The Milky Way itself does not have a classical bulge (or maybe only a very low mass one, see \citealp{Shen2010}). The frequency of classical bulges increases with stellar mass, and classical bulges are the most frequent bulge type in galaxies with a stellar mass above $10^{10.5}$ \msun \citep{Gadotti2009,Fisher2011}. There are however massive disc galaxies with no signs of a classical bulge \citep{Barentine2012}.

The absence of classical bulges in massive disc galaxies has often been described as a challenge to $\Lambda$CDM, and to the idea of a hierarchical build up of galaxies \citep{Kautsch2006,Shen2010,Kormendy2010b,Peebles2020}. 
The existence of very massive, disc-dominated galaxies is indeed interesting, given that the fraction of accreted stars typically increases with stellar mass \citep{Lackner2012,Lee2013,Rodriguez2016,Davison2020,Remus2021}. It has been proposed that those very massive discs either have very unusually quiescent merger histories, or reform discs after gas-rich mergers \citep{Springel2005,Font2017,Martin2018,Jackson2020}.

In parallel with improving predictions from simulations on the formation histories of disc galaxies, it is crucial to establish the connection between galaxy morphology, stellar populations, and accretion history from an observational point of view.
For nearby galaxies, past merger events might be identified with deep photometry, resolving individual stars in stellar halos \citep{McConnachie2009,Bailin2011, Monachesi2016}. In particular, the total mass, the mean metallicity and the metallicity gradient of stellar halos can be compared to simulations to constrain the fraction of accreted stars, and the typical mass of the accreted satellites \citep{Bell2017,Harmsen2017,DSouza2018,Elias2018,Monachesi2019}. This is possible because of the steep relation between the stellar mass of galaxies and their mean metallicity \citep{Gallazzi2005, Kirby2013, Delgado2014}.  Additionally, mergers might also be identified and characterized thanks to the globular cluster systems that satellite galaxies bring with them \citep{Blom2012,Beasley2018,Hughes2019,Mackey2019,Alabi2020}. For distant galaxies, stellar halos cannot be resolved into individual stars, but past accretion events might still be identified with deep observations. Indeed, recent mergers leave signatures in the form of stellar streams, plumes, shells or tidal tails, depending on the merger ratio and orbit \citep{Martinez2010,Duc2015,Merritt2016,Hood2018,Mancillas2019,Fensch2020}. Those structures do not however trace the whole merger history of a galaxy, and do not necessarily allow us to identify the main contributors to the accreted stellar halo \citep{Vera2021}.  

Ultimately, to study the connection between morphology and accretion events, it is also important to identify accreted stars within the discs and bulges of galaxies. This can be done through stellar kinematics, enabling us to detect counter-rotating components in galaxies. A famous example is NGC 4550, a lenticular galaxy hosting two counter-rotating discs \citep{Rubin1992,Rix1992}, which could be the result of a past merger \citep{Crocker2009}. Such clear cases are rare (see also \citealp{Yoachim2008a} for a galaxy with a counter-rotating thick disc, for instance),  and the imprints of mergers on discs are often subtle: models of the velocity and velocity dispersion profiles of nearby galaxies have found less than 5\% of counter-rotating stars in thick discs \citep{Comeron2019}. Even if and when counter-rotating components are identified (or not), this does not provide the final answer on the contribution of mergers to the build-up of discs since accreted stars may be on prograde orbits \citep{Gomez2017}, and counter-rotating discs can be formed via gas accretion \citep{Coccato2015}. Instead of simply identifying counter-rotating stars, careful dynamical modelling coupled with comparisons with numerical simulations is a promising avenue to uncover the merger history of galaxies (\citealp{Poci2021}, Zhu et al in prep.). 

Finally, mergers are also expected to leave traces in the stellar populations of galaxies because of the relation between mass and metallicity: the accretion of a low mass galaxy will result in a population of metal-poor stars, distinct from in-situ stars of a similar age. These younger, metal-poor components can be identified through spectroscopy and full spectrum fitting. From the decomposition of spectra into populations of different ages and metallicities, \cite{Boecker2020} proposed a method to recover the accretion history of a galaxy. A simpler version of this idea was applied by \cite{Pinna2019a, Pinna2019b}, identifying in edge-on S0 galaxies some metal-poor components that they argue correspond to  ancient merger events (see also \citealp{Davison2021} for a similar idea).

Integral field spectrographs like MUSE (the Multi-Unit Spectroscopic Explorer, \citealp{Bacon2010}), installed on the ESO Very Large Telescope (VLT) thus allow us in one go to determine the kinematics and stellar populations in different galactic components, and to potentially  uncover past accretion events. This is a key step towards a better understanding of the build-up of disc galaxies. In this paper, we study a very massive edge-on disc galaxy, NGC 5746. With a stellar mass of $\sim 10^{11}$\msun \citep{Jiang2019}, it is definitely in the regime where most galaxies are bulge-dominated. However, it does not contain a classical bulge, but only a nuclear disc and a B/P bulge, as well as a bar \citep{Barentine2012}. Its disc also has a complex structure, with both a thin and thick components, while the stellar halo does not show any signs of recent accretion events \citep{Morales2018}. We have thus obtained MUSE data covering the central regions and part of the disc of NGC 5746, to better characterize the build-up of its diverse components, and to try and uncover how they relate to each other, and how they formed.

In Section 2, we start by presenting what is already known about NGC 5746, before describing in Section 3 our data and the analysis we performed. In Section 4, we then present maps of the kinematics and stellar populations throughout the galaxy. In Section 5, we study in more detail the stellar populations in each of the components (thin and thick disc, B/P bulge and nuclear disc). We finish in Section 6 by gathering all the available information to put together a picture of the formation history of NGC 5746: its mass growth and merger history, as well as the formation of its thick disc and its bar.

\section{NGC 5746}\label{sec:ngc5746}

NGC 5746 is a nearly edge-on, massive disc galaxy that forms a very wide galaxy pair with NGC 5740. It is classified as an SAB(rs)b galaxy in the Third Reference Catalog of Bright Galaxies \citep{RC3}, and as (R')SBx(r,nd)0/a by \cite{Buta2015}.
The circular velocity from the flat section of the HI rotation curve is $\sim 310$ \kms \citep{Rand2008}, and the stellar mass is $\sim 10^{11}$ \msun \citep{Jiang2019}.  We adopt a distance of 26.5 Mpc from \cite{Springob2007} and \cite{Tully2016}, so that 1 arcsec on the sky corresponds to 128 pc. We show in Figure \ref{fig:pointings} an SDSS image of the galaxy, together with Spitzer 3.6 and 8 $\mu$m images.

The few inner arcseconds of the galaxy correspond to a clear minimum in velocity dispersion
\citep{Falcon2003}. In addition to this, a double-humped velocity profile and a central anticorrelation between \V and \h3 confirm that NGC 5746 hosts a cold, rapidly rotating, and axisymmetric nuclear disc \citep{Chung2004,Molaeinezhad2016}. This component is also seen in photometry: using Hubble Space Telescope NICMOS images, \cite{Balcells2007} and \cite{Barentine2012} have identified in the inner 2 arcsec of NGC 5746 the presence of a discy pseudo-bulge, with a  S\'ersic index close to 1. For simplicity, we call this component a nuclear disc in the rest of this paper. 

There are no signs of a classical bulge in the galaxy \citep{Barentine2012}, but the nuclear disc is embedded within a boxy bulge with a peanut shape extending out to $\sim 35$ arcsec along the major axis of the galaxy \citep{Kuijken1995, Bureau1999}. This structure actually corresponds to a bar seen at an intermediate angle, close to end-on \citep{Peters2017}. Signatures of the bar are also seen in the kinematics of gas  \citep{Kuijken1995} and stars, with in particular a correlation between \V and \h3 which is expected  from  a  barred disc seen in projection \citep{Chung2004,Molaeinezhad2016}. \cite{Peletier1999} used HST colours to show that stellar populations in the central regions of NGC 5746 are old and metal-rich. This was confirmed by \cite{Kormendy2019} using line indices, with the additional findings of a metallicity gradient (with the center being the most metal-rich region) and of a general high \al  abundance ratio in the whole region.

Spitzer 3.6 $\mu$m images reveal a bright inner ring with a radius of $\sim$1 arcmin 
\citep{Barentine2012,Comeron2014}. The ring is brighter in the 8$\mu$m images tracing PAH emission (see also Figure \ref{fig:pointings}), indicating that it hosts star forming regions \citep{Barentine2012}, which are also apparent in an analysis of the ionized gas \citep{DeVis2019}. A similar structure is found in HI images: the HI is distributed in a large disc with a central depression with a radius of 1 arcmin \citep{Rand2008}. These observations suggest that face-on we would see a bright ring of star formation around the bar (as for instance in NGC 2523, as explained in \citealp{Kormendy2010a}), and a deficit of star formation within the ring, which corresponds to the "star formation desert" regions described by \cite{James2018} and \cite{Donohoe2019}.
The rest of the disc harbours low levels of star formation, for a total star formation rate slightly below 1 \msun yr$^{-1}$ \citep{Jiang2019}. The ionized gas in the disc has a high metallicity ($\mathrm{12+ \log(O/H)=8.6}$), and a very small metallicity gradient \citep{DeVis2019}.

In addition to the thin disc, there is evidence for a complex vertical structure: \cite{Barentine2012} identified a thick disc with a scale height of $\sim 8.6$ arcsec (1.1 kpc),  and \cite{Rich2019} found a very large oval-shaped envelope around the disc out to a radius of 5.5 arcmin, or 42 kpc (at a surface brightness level of 28 mag arcsec$^{-2}$). No other faint structures are found around the galaxy, and in particular there are no traces of tidal features that would have traced recent accretion events \citep{Morales2018}.

\begin{table}
\caption{Main properties of NGC 5746 (references: $^1$\citealp{Springob2007}, $^2$\citealp{Bianchi2007}, $^3$\citealp{Jiang2019}, $^4$\citealp{RC3})}
\label{tab:properties}
\begin{tabular}{lc}
\hline
Distance$^1$  & 26.5 Mpc\\
Inclination$^2$ & 86 deg \\
Stellar mass$^3$ & 1--1.2$\times 10^{11}$ \msun\\
SFR$^3$ & 0.8 -- 0.9 \msun yr$^{-1}$\\
$D_{25} ^4$ & 7.4' (59 kpc)\\
$R_{50}$ (W1,W2)$^3$ & 28--30" (3.6--3.8 kpc)\\
$R_{90}$ (W1,W2)$^3$ & 128" (16 kpc)\\
\hline
\end{tabular}
\end{table}

\section{Data and analysis}

\subsection{Observations and data reduction}
\begin{figure}
\includegraphics[width=\columnwidth]{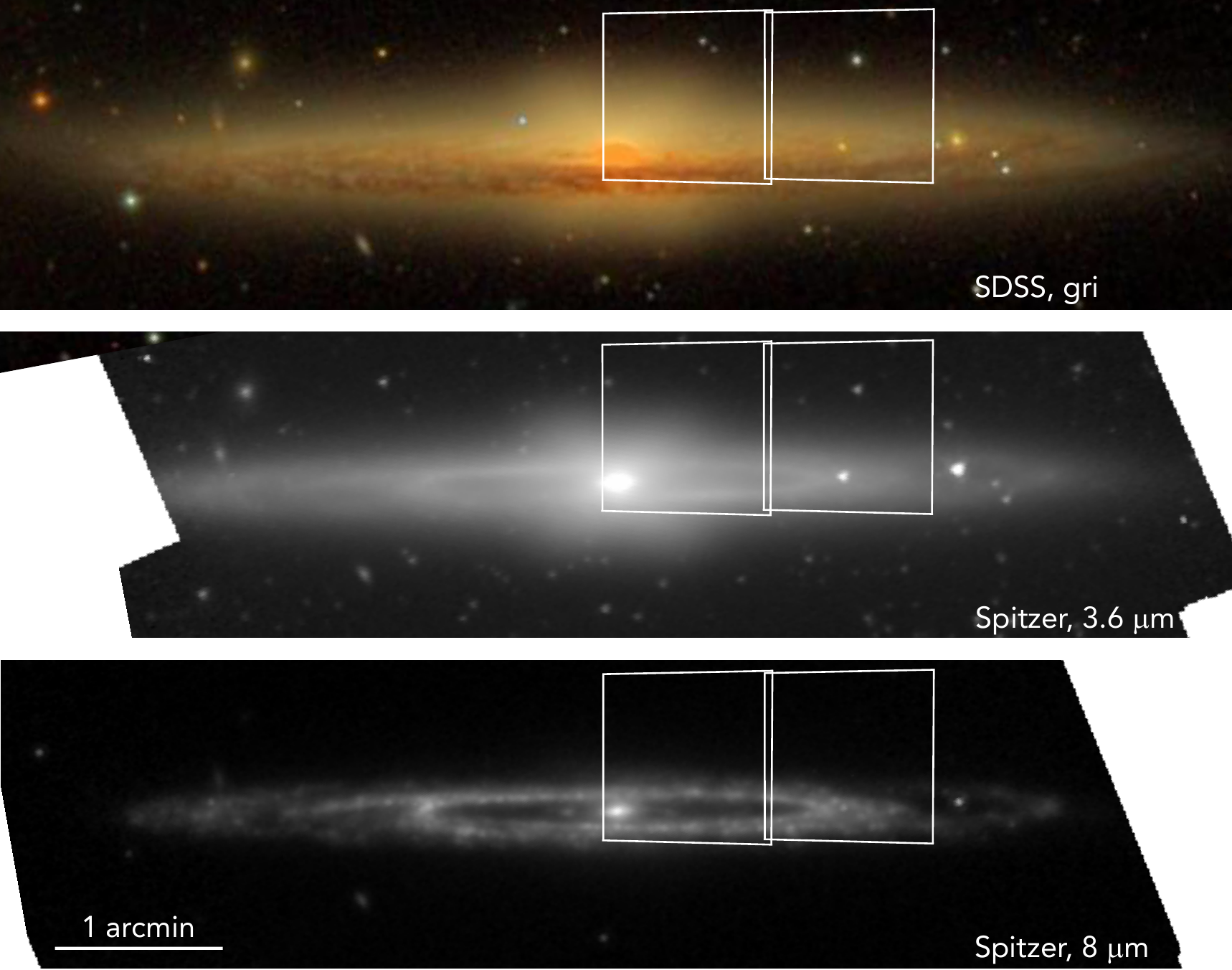}
\caption[]{Our two MUSE pointings superimposed on a $gri$ SDSS image (top panel), and two Spitzer images (middle and bottom panel), at 3.6 and 8 $\mu$m (the Spitzer images were downloaded from DustPedia\protect\footnotemark, \citealp{Clark2018}).} 
\label{fig:pointings}
\end{figure}
\footnotetext{\url{http://dustpedia.astro.noa.gr/Data}}

The MUSE instrument has a wavelength range extending from 4750~\AA\ to 9350~\AA\ with a spectral sampling of 1.25~\AA. In wide field mode, the field of view corresponds to 1 arcmin$^2$ with a spatial sampling of 0.2 arcsec.
We present MUSE data obtained in wide field mode in period P95 during the nights of June 11$^\mathrm{th}$, June 17$^\mathrm{th}$, July 11$^\mathrm{th}$, and July 17$^\mathrm{th}$ 2015. The observations were carried out in Service Mode, with a seeing requested to be below 1.2 arcsec.

We obtained two different pointings on NGC 5746: a central pointing and a disc pointing. These pointings (shown in Figure \ref{fig:pointings}) only cover one side of the galaxy, and do not cover its whole radial extent: they only reach out to about half of the optical radius (which is equal to 3.7 arcmin, see Table \ref{tab:properties}). However, they include a  significant fraction of the total stellar mass, since the radius enclosing 90\% of the luminosity in the WISE W1 and W2 bands (tracing stellar mass) is around 130 arcsec (Table \ref{tab:properties} and \citealp{Jiang2019}).

\begin{figure}
\includegraphics[width=\columnwidth]{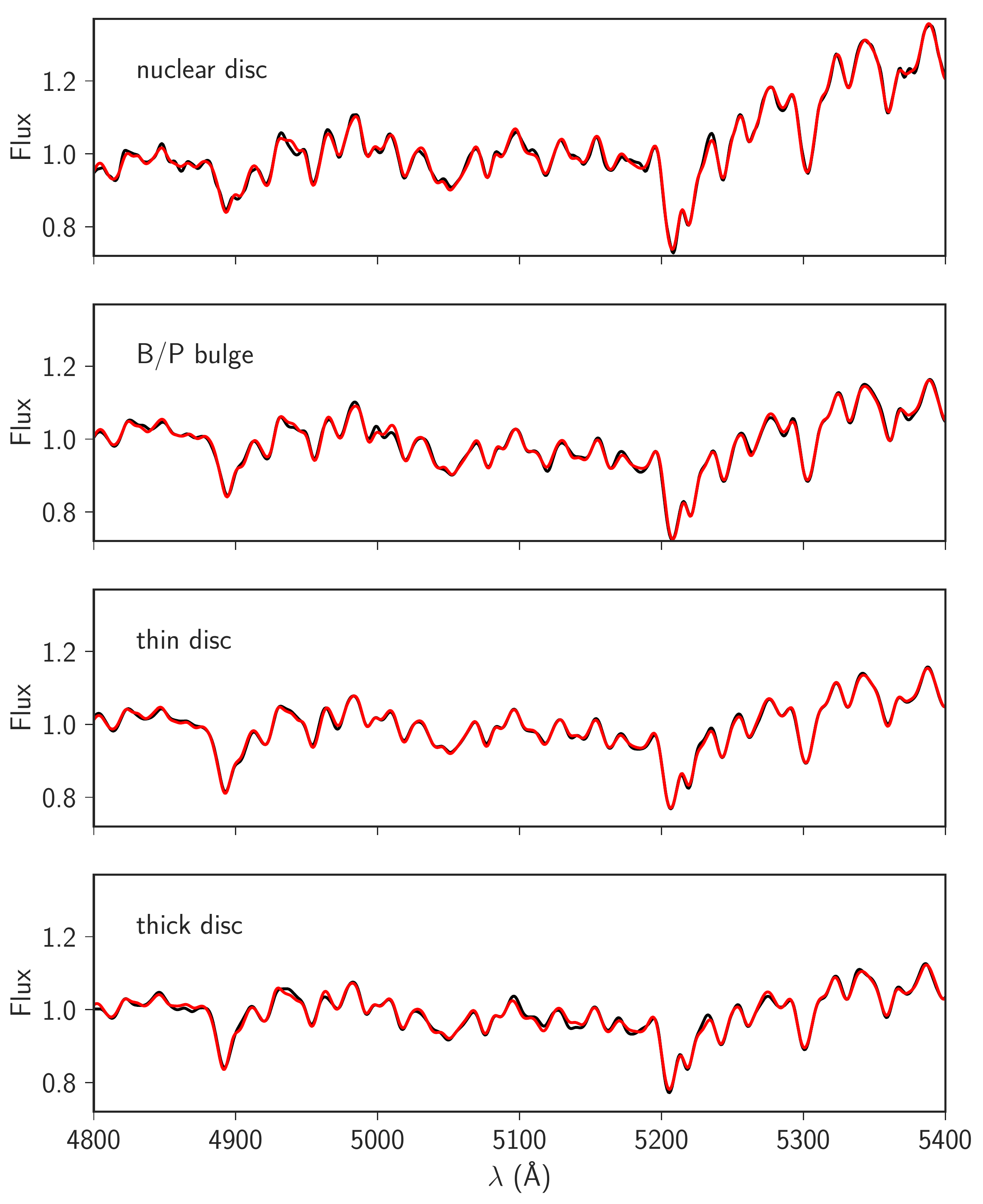}
\caption{Examples of spectra from individual Voronoi bins in the four different morphological components of the galaxy (black lines), with the best fit from pPXF in red. The spectra have all been cleaned from emission lines and convolved to the same 8.4 \AA\ spectral  resolution.}
\label{fig:spectra}
\end{figure}

The total observing time for the program was 6 h, divided in 5 observation blocks of 1 h and 12 min, including one block for the central pointing and four blocks for the disc pointing.
Within each observation block, we used a standard strategy to alternate object and sky pointings (in an OSOOSO pattern), and dither/rotate between each object pointing. In total, the science exposure time is 45 minutes for the central pointing and 180 minutes for the disc pointing.

The data was then reduced with the standard MUSE pipeline, version 1.6 \citep{Weilbacher2014,Weilbacher2020}. This includes flux and wavelength calibration as well as bias, flat-fielding, and illumination corrections. The data is finally registered astrometrically and the two pointings are stitched together.

\subsection{Voronoi binning and signal-to-noise ratio}

We use the Voronoi tessellation software from \cite{Cappellari2003} to spatially re-bin  the combined MUSE data cube: this is a way to perform an adaptive spatial binning of the spectra using a target signal-to-noise ratio (SNR) per bin. In our case, we require a minimum target SNR of 100, which results in 1415 new Voronoi bins. We only present in this paper results for bins with a SNR above 80, and discard bins around a bright foreground star that is superimposed on the disc of NGC 5746. After this, we are left with 1170 bins, whose SNR  goes from 80 to 135, with a median at 105. 

These SNR values obtained from the formal error cube are probably overestimated since they only reflect Poisson uncertainties and uncorrelated noise. As in \cite{Garcia2015} and  \cite{Sarzi2018}, we re-estimate the SNR in each spaxel after the stellar kinematics fit (see next Section for  a description of the fits) by computing the difference between each spectrum and the best fitting model: those more realistic SNR values range from 40 to 105 with a median at 73. This is enough to reliably extract kinematics and stellar populations using full spectrum fitting (e.g., \citealp{Sanchez2014,Ruiz2015,Gadotti2019}).

Our Voronoi bins can be seen in maps of the kinematics and stellar populations that we present later in this paper, and we also show example spectra from bins in different galactic components in Figure \ref{fig:spectra}.

\subsection{Full spectrum fitting}

We  determine the stellar kinematics and stellar populations using full-spectrum  fitting, with the  Penalized  Pixel-Fitting code (pPXF) initially described in \cite{Cappellari2004}, then upgraded  in  \cite{Cappellari2017}. We select a wavelength range from 4750 to 5500 \AA\ to perform our analysis since the blue region of the MUSE spectra contains most of the features with good sensitivity to the parameters we want to measure, and also avoids regions of the spectra where there could be residuals from the sky subtraction (this is similar to what was done in \citealp{Pinna2019a,Pinna2019b}, and also for instance in the TIMER project, see \citealp{Gadotti2019}). For our fits, we use the MILES models, with a mean spectral resolution of 2.51\,\AA\ \citep{Falcon2011}. They are broadened to the spectral resolution of the MUSE data in our wavelength range (2.65\,\AA) before the fitting process.  We show in Figure \ref{fig:spectra} some examples of fits to spectra in each of the main components of NGC 5746.

To obtain those fits, we first measure the stellar kinematics by fitting each spectrum with stellar templates convolved with a line-of-sight velocity  distribution (LOSVD) described by a Gauss-Hermite expansion. We use a 4$^\mathrm{th}$ order expansion, so that the LOSVD of each spectrum is described by the mean line-of-sight velocity \V, the velocity dispersion \sig and the 3$^\mathrm{rd}$ and 4$^\mathrm{th}$ order Gauss-Hermite moments, \h3 and \h4. This first fit employs no regularization, and includes a 6$^\mathrm{th}$ order additive Legendre polynomial to correct for uncertainties in the spectral calibration.

We clean the spectra of emission lines using the Gas and Absorption Line Fitting code (GandALF \citep{Sarzi2006}, assuming the same kinematics for all emission lines within our wavelength range (we present and discuss the maps of ionized gas in Appendix \ref{appendix:emission}). We then convolve all spectra to the same 8.4 \AA\ spectral resolution (similar to the Lick/IDS system). The main purpose of this is to simplify the analysis of stellar populations using line-strength indices, which we use to test for biases in our full spectrum fitting analysis (see next Section and Appendix \ref{appendix:indices}.)

We finally use pPXF to fit the spectra cleaned from emission lines and derive the star formation history and chemical composition in each Voronoi bin, using 8$^\mathrm{th}$ order multiplicative polynomials to correct for uncertainties in the spectral calibration.
The stellar templates we choose to fit the spectra are the MILES single stellar population (SSP) models in their \al-variable version \citep{Falcon2011,Vazdekis2015}. We use the  BaSTI stellar isochrones \citep{Pietrinferni2004} and assume a  Kroupa Universal Initial Mass Function,  with a slope of 1.3 \citep{Kroupa2001}.  The 1272  templates include 12 values of total metallicity \Fe from -2.27 to 0.40 dex,  53  values  of  age  between  0.03  and  14 Gyr, and two values of \al, of 0 and 0.4 dex.  While the templates with [M/H] below -1.5 do not give safe predictions (see Table 2 in \citealt{Vazdekis2015}), we have found that including them or not in the fits does not bias our results, and, when they are included, they only contribute a negligible fraction of the total mass. Hence, for simplicity, we present here results obtained with the full library, but do not show or discuss the contribution of SSPs with [M/H] below -1.5.

Each SSP model is normalized to its total stellar mass, so that mass-weighted results are  obtained from pPXF: the weight in each SSP directly represents the fraction of the mass in that combination of age, \Fe and \al. While the models only include two values of \al, this procedure allows to interpolate between the two extreme values, and to find a range of \al values for the spectra. We verify that the \al values we obtain this way are consistent with values derived from an analysis based on line indices (see next Section and Appendix \ref{appendix:indices}).

For the stellar populations fit with pPXF, we use a regularization parameter $R=10$. This value was chosen following the procedure described in \cite{Pinna2019a} to find a compromise between smoothing the solution and not loosing any information. The first step is to determine the maximum regularisation parameter allowed, giving the smoothest solution consistent with the data (as described in \citealp{Cappellari2017}). As in \cite{Pinna2019a}, we then tested different levels of regularization between 0 and this maximum value, and noticed that some features in the star formation histories disappeared above $R\simeq10$. We thus choose to use a regularization parameter $R=10$ for all the spectra, which gives us smooth solutions while still showing interesting features in the SFH.

A final quantity we determine is the stellar mass in each spaxel. This is done by using the mass-to-light ratio of the best fitting combination of SSPs in each spaxel and a $g$-band SDSS image giving us the absolute surface brightness of those spaxels (see also \citealp{Pinna2019a}). If we sum the mass for all the good spaxels in our data cubes, we find a total stellar mass of  5.4$\times 10^{10}$ \msun for the region covered by our pointings, which is roughly consistent with a total stellar mass of $\sim 10^{11}$ \msun for NGC 5746. 

\subsection{Estimating uncertainties and biases} \label{sec:uncertainties}
The statistical uncertainties on stellar kinematics can be estimated via Monte Carlo simulations, as suggested by \cite{Cappellari2004}. \cite{Pinna2019a} showed that for MUSE data  with a lower SNR compared to ours and analysed with the same method, those uncertainties are of the order of $\sim 5$ \kms for \V, $\sim 10$ \kms for \sig, and 0.05 for \h3 and \h4 (see their Appendix A). Given that the results presented in this paper do not require a very fine understanding of kinematics uncertainties, we do not perform any further analysis on this topic.

By contrast, we pay a lot more attention to uncertainties in the stellar populations recovered, particularly since some of our results rely on fine details seen in the distribution of age,  metallicity, and \al.
We estimate the statistical uncertainties on the stellar populations using a bootstrap analysis similar to the one described in \cite{Kacharov2018}. For each spectrum, we first perform an unregularized fit and compute the difference between the best fit solution and the initial spectrum. We then resample the initial spectrum 100 times by adding those residuals on top of the best fit spectrum, with the sign of the residuals at each wavelength randomly flipped. Each resampled spectrum is fitted with pPXF using a very small regularization parameter $R=0.1$: this ensures that we explore the uncertainties that are due to variance in the spectrum itself, without being biased toward smooth solutions \citep{Cappellari2004,Kacharov2018}. 

We use those 100 bootstrap realisations in different ways: to compare the distribution of weights between the regularized solution and the mean of the bootstrap samples, to estimate uncertainties on the corresponding distributions of ages and metallicities, and to create maps of the uncertainties on the mean stellar populations parameters. All of those will be further discussed in Sections 4 and 5.
The overall mean age uncertainty is of the order of 1 Gyr, and the mean uncertainties on [M/H] and \al are 0.04 and 0.02 dex, respectively.
Those are of course only statistical uncertainties for a given setup of pPXF. \cite{Pinna2019a} discuss further tests varying the degree of the multiplicative polynomial used and varying the regularization parameter: they conclude that those do not significantly affect the order of magnitude of the uncertainties on stellar populations.

Systematic uncertainties due to our choice of method for the analysis are harder to constrain. We have chosen to perform two series of tests to increase our confidence in our results. First, we re-analysed all the spectra using line-strength indices to determine an SSP-equivalent age, metallicity and [Mg/Fe] for each spaxel. The results are discussed in Appendix \ref{appendix:indices}: we do not find a good correlation between the mean mass-weighted age from pPXF and the SSP-equivalent age (as expected, see the discussion in Appendix \ref{appendix:indices}), but the SSP-equivalent [M/H] and [Mg/Fe] are close to the mass-weighted averages from pPXF. There is a slight offset in [Mg/Fe], with SSP-equivalent values larger by $\sim 0.07$ compared to the mass-weighted average from pPXF. This also confirms that the pPXF analysis interpolating between only two values of \al recovers reasonable values for the mass-weighted \al of a stellar population.

The other test that we perform is that we combine all spectra in the thick disc region and analyse the combined spectrum with an entirely different code, STECKMAP (``STEllar Content and Kinematics via Maximum A Posteriori likelihood'', \citealp{Ocvirk2006}). We show in Appendix \ref{appendix:steckmap} that one of the main results in our paper (the presence of younger metal-poor stars in the thick disc) still holds when we perform our analysis with STECKMAP, using a different version of the BaSTI MILES models (the \textit{baseFe} version), and a different set of stellar isochrones (the Padova00 isochrones). 

We also note that full spectral fitting techniques like the ones used in this paper have been recently tested against the, in principle more reliable, Colour-Magnitude Diagram (CMD) fitting approach \citep[][]{Ruiz2015, Ruiz2018}. Such techniques, in particular STECKMAP, provide compatible results to those found from CMD fitting in systems with complex SFHs (LMC and Leo~A, both dwarf galaxies in the Local Group for which deep CMDs and high quality spectra can be observed), for SNR values similar to, or even slightly lower than, the ones we use in this paper. The reliability of these techniques, together with the positive comparison between codes performed in this work increases our confidence in our results.

\subsection{Morphological decomposition} \label{sec:masks}

To perform a morphological decomposition of NGC 5746, we use a Spitzer 3.6 micron image from the S$^4$G survey \citep{Sheth2010}, and fit multiple components with IMFIT \footnote{\url{https://www.mpe.mpg.de/~erwin/code/imfit/}}  \citep{Erwin2015}. While the galaxy is slightly inclined, if we let the inclination free, the preferred value given by IMFIT is 90 deg. After trying various combinations, we found that the best fit model corresponds to a sum of 5 components: a nuclear disc, a box/peanut bulge, two thin disc components, and a thick disc. The B/P bulge component represents both the main bar itself and its vertical extension into a B/P shape (as discussed in Section \ref{sec:ngc5746}, the bar is seen close to end-on, so it is impossible to discriminate between the B/P bulge and parts of the bar that might not have been thickened vertically). All discs are modelled as exponentials in the radial direction and hyperbolic secants in the vertical direction. The two thin disc components are needed to properly account for the deficit of light within the inner ring. The second thin disc component is purely artificial and has a negative contribution to the flux: this only reflects the fact that the main thin disc deviates from an exponential in the central regions. 

The best fit parameters show that the thin disc has a scale-length of 47.8 arcsec (6.1 kpc), and a scale-height of 3.9 arcsec (0.5 kpc). It contributes 50\% of the total 3.6 micron luminosity of NGC 5746. The thick disc has a scale-length of 64.2 arcsec (8.2 kpc), a scale-height of 11.2 arcsec (1.4 kpc), and contributes 21\% of the total luminosity. We do not find signs of flaring either in the thin or thick disc with our simple analysis.

\begin{figure}
\includegraphics[width=\columnwidth]{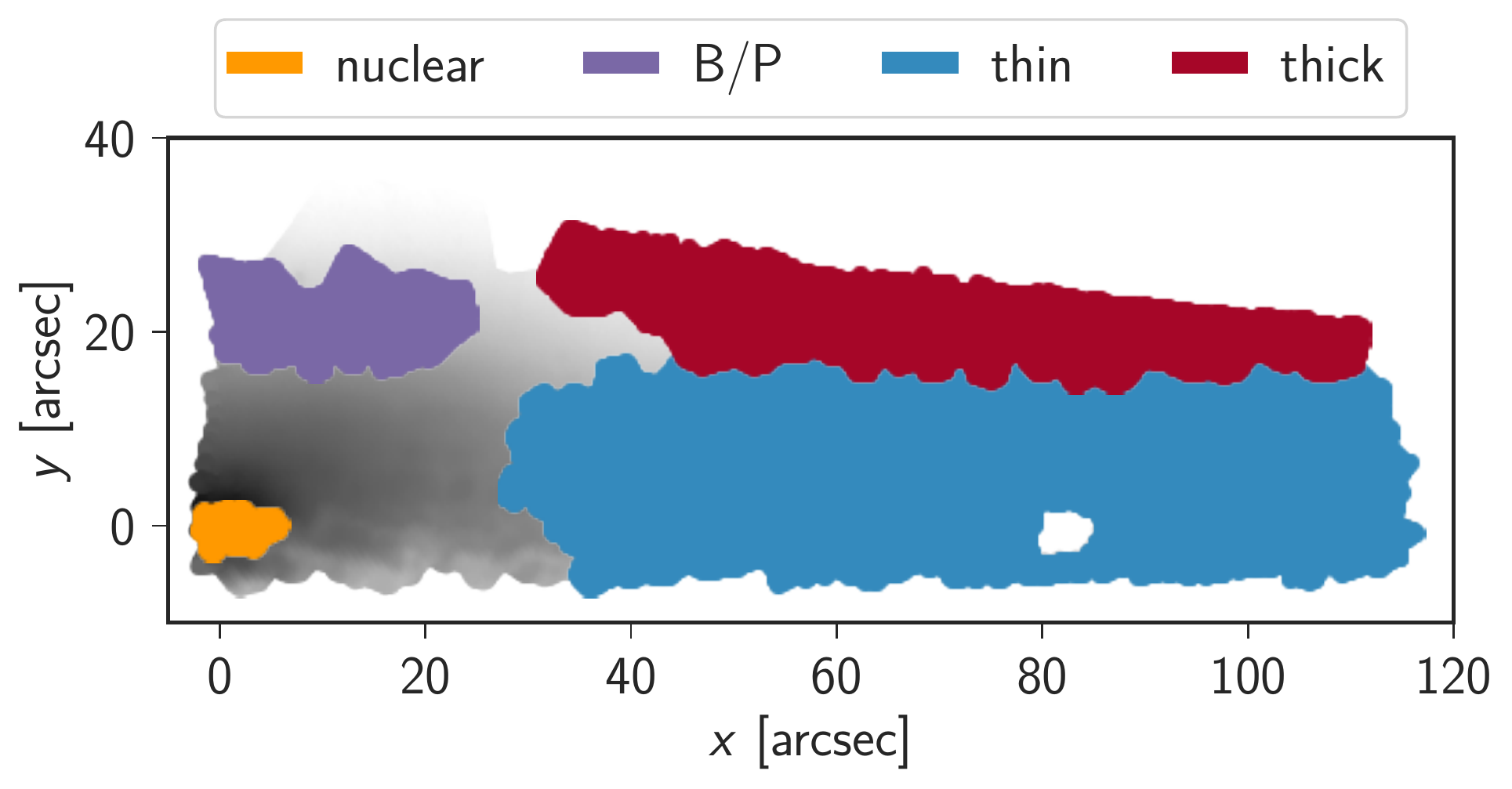}
\caption{On top of a grey scale image of NGC 5746 (computed from the flux extracted from our MUSE data cube), we show which Voronoi bins we have attributed to each galactic component (nuclear disc, B/P bulge, thin and thick disc)  after following the procedure described in Section \ref{sec:masks}.}
\label{fig:masks}
\end{figure}
We then attribute Voronoi bins to each galactic component, by assigning a bin to a component if that component contributes more than 50\% of the light in that bin. For the thin disc, we add an additional criterion, including only bins with $|V|>\sigma$ (this removes the very inner regions of the disc), and for the nuclear disc we only include bins with $\sigma<200$ \kms (the photometric decomposition produced quite a thick nuclear disc that did not match what we noticed in the kinematics and stellar population maps shown later in this paper).
We show in Figure \ref{fig:masks} the result of that procedure: some bins have not been attributed to any particular components because they correspond to the superposition of several components. The nuclear disc dominates the light in the center, while only the very top part of the B/P bulge is cleanly attributed to that component (the B/P bulge and the bar are of course further extended towards the midplane but those regions are much more contaminated by the disc). Most of the other bins belong to the thin disc, and a stripe corresponding to the thick disc is found at high altitude, thus confirming that our MUSE data is deep enough to reach the thick disc. 

\section{Global structure of NGC 5746}
\begin{figure*}
\includegraphics[width=1.8\columnwidth]{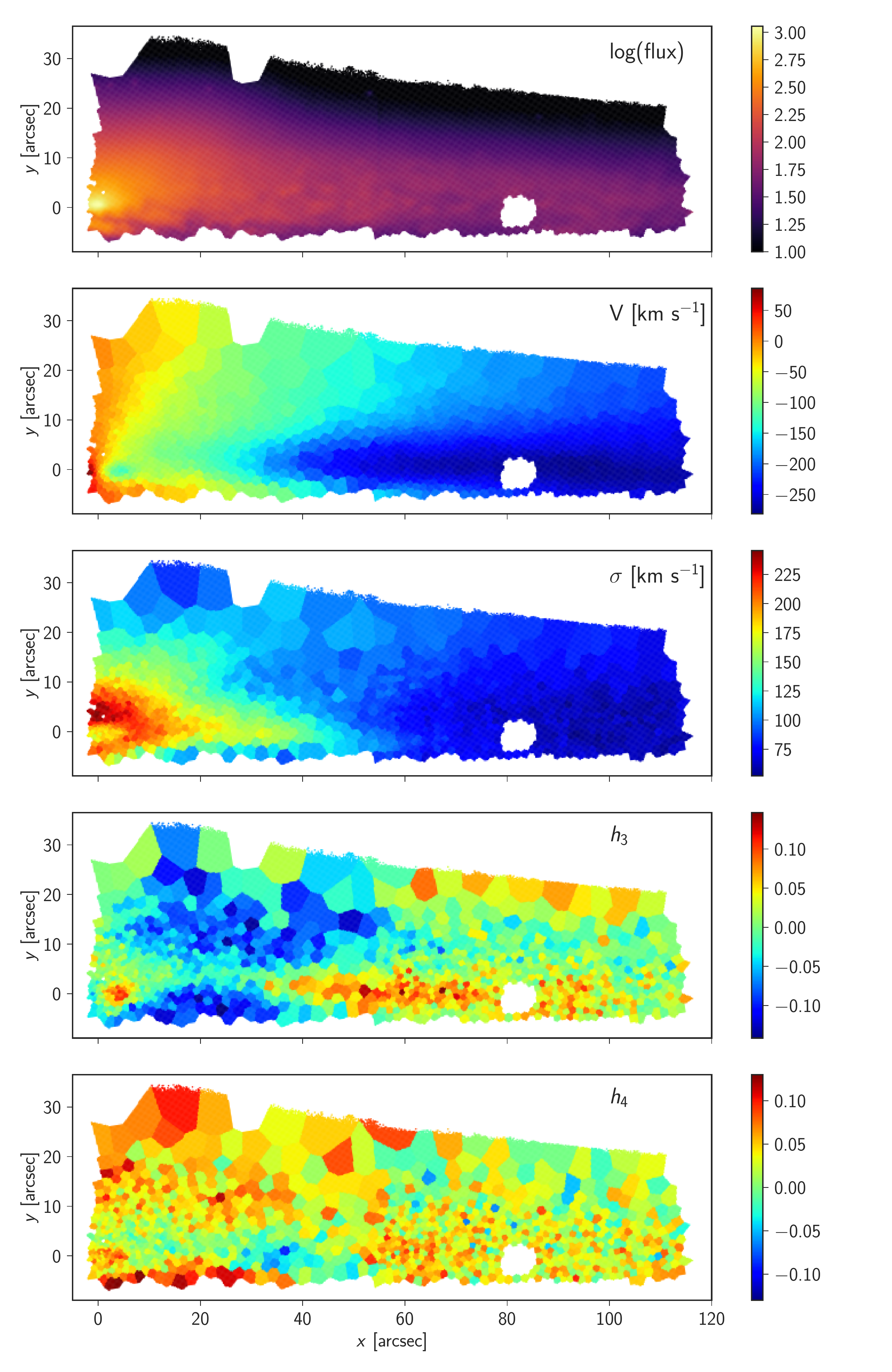}
\caption{Stellar kinematic maps. To guide the eye, the top panel represents the flux extracted from the MUSE datacube. The following panels show the mean line-of-sight velocity, velocity dispersion, and the third and fourth order Gauss-Hermite moments.  }
\label{fig:kin_maps}
\end{figure*}
\subsection{Stellar kinematics}
We present in Figure \ref{fig:kin_maps} the stellar kinematics maps showing the spatial distribution of \V, \sig, \h3, and \h4. To guide the eye, the top panel in that Figure also shows the flux extracted from the MUSE cube. An important feature to note is that because NGC 5746 is not exactly edge-on, any vertical trends that are seen in the disc are not only due to vertical gradients, but also to the inclination angle (at least for most of the thin disc-dominated region).

\begin{figure}
\includegraphics[width=0.95\columnwidth]{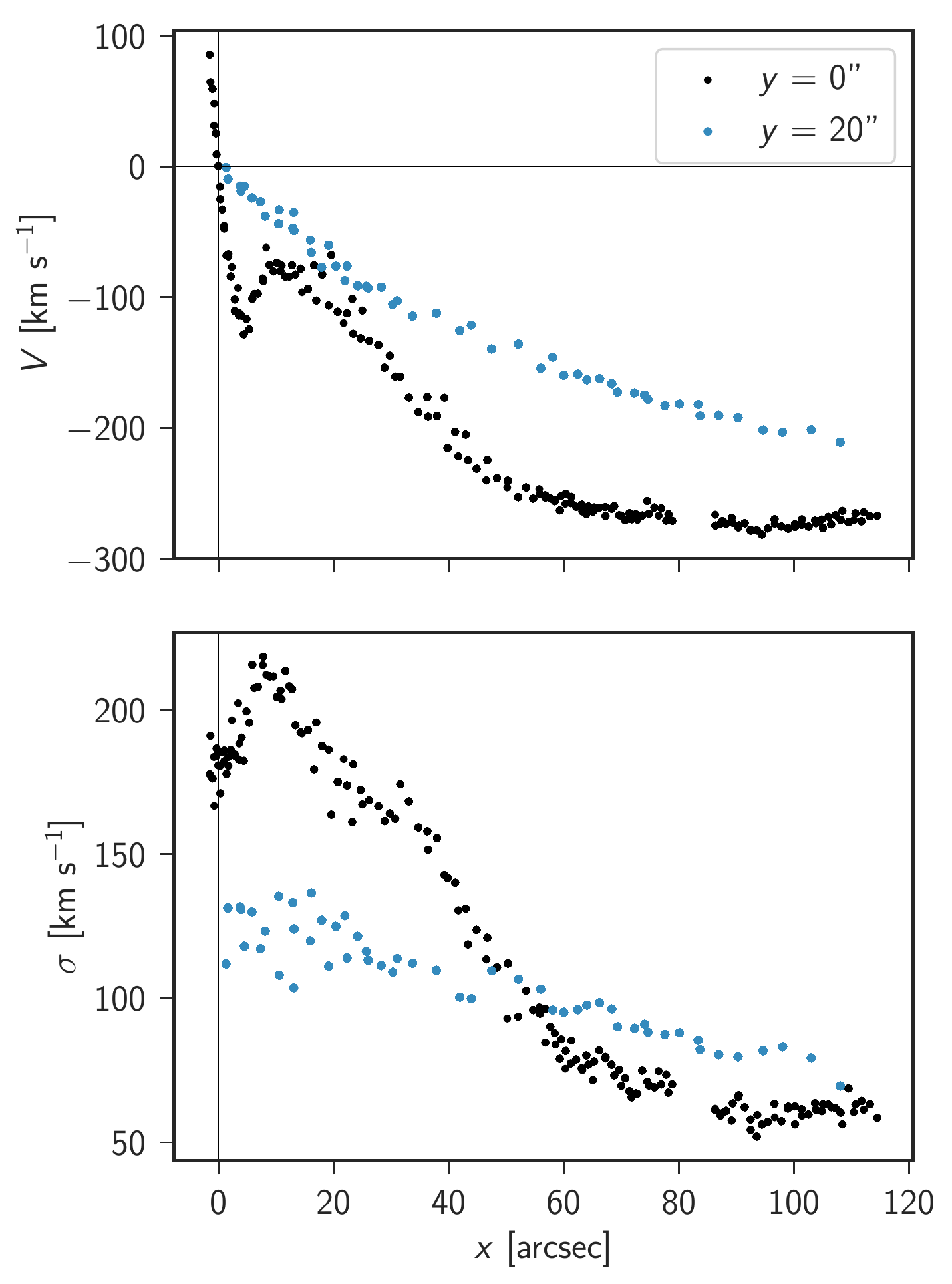}
\caption{Radial profiles of the stellar mean radial velocity (top panel) and velocity dispersion (bottom panel), in the midplane (black symbols), and 20 arcsec above the midplane, a region typical of the thick disc (blue symbols). The midplane curves show the rotation of the nuclear disc as well as its lower $\sigma$ compared to the rest of the disc. Outside of the nuclear disc, the rotation curve dips slightly, creating a ``double-hump" shape typical of barred galaxies. }
\label{fig:rot_curve}
\end{figure}

The velocity map shows a global rotation pattern, with the strongest rotation along the midplane of the galaxy, and a velocity  decreasing with height above the midplane. This effect is clearer in the top panel of Figure \ref{fig:rot_curve}, where we show the line-of-sight velocity as a function of radius for two different heights (in the midplane, and 20 arcsec above the midplane, which is typical of the thick disc-dominated region). The rotation curve in the midplane is flat in the outer disc, reaching values around -273 \kms. By contrast, the thick disc rotation curve decreases smoothly with radius, to values around \mbox{-200 \kms} in the outer disc. The rotation velocity in the thick disc is lower than in the midplane because it has a higher velocity dispersion: stars in the thick disc are on more eccentric orbits, so that their rotation velocity represents a smaller fraction of their total velocity (this effect is called asymmetric drift, see e.g. \citealp{Stroemberg1946}). The asymmetric drift seen in the thick disc decreases with radius because its velocity dispersion decreases with radius.
The velocity map additionally shows in the very inner galaxy a region that rotates faster than its surroundings: this corresponds to the nuclear disc. The rotation of the nuclear disc can also be compared to the rotation of the main disc in Figure \ref{fig:rot_curve}: the nuclear disc has a very steep rotation curve with a maximum velocity of 130 \kms.
The kinematic radius of the nuclear disc, defined as the radius within the nuclear disc where \V/\sig is maximum, is around 3 arcsec, or 380 pc. This is within the range of values typically observed for nuclear discs in nearby galaxies \citep{Gadotti2020}. Outside of the nuclear disc, the rotation curve dips slightly, creating a ``double-hump" shape typical of barred galaxies \citep{Bureau2005}, as also measured by \cite{Williams2011}.

In the third panel in Figure \ref{fig:kin_maps} we show the map of the stellar velocity dispersion: \sig is generally greater towards the inner galaxy, but shows several interesting features. First of all, while \sig is generally high in the inner 10 arcsec of NGC 5746, the very central regions correspond to a clear drop in \sig: this is the nuclear disc already seen in the velocity map (as also previously discussed by \citealp{Falcon2003}, \citealp{Chung2004}, and \citealp{Molaeinezhad2016}). The radial \sig profile in the bottom panel of Figure \ref{fig:rot_curve} also displays lower values in the nuclear disc region, with typical values of \sig around 180 {\kms}. Outside of the nuclear disc, the bar and B/P region has an increased \sig following a boxy shape corresponding to the boxy shape of the overall B/P structure. This increased \sig also extends in the midplane of the galaxy, and seems to correspond to the region inside the inner ring that has a diameter of $~1$ arcmin (see also Figure \ref{fig:pointings}). In the rest of disc, \sig smoothly decreases towards the outer disc, reaching values of 60 \kms in the midplane and 80 \kms in the thick disc region.

\begin{figure}
\includegraphics[width=0.95\columnwidth]{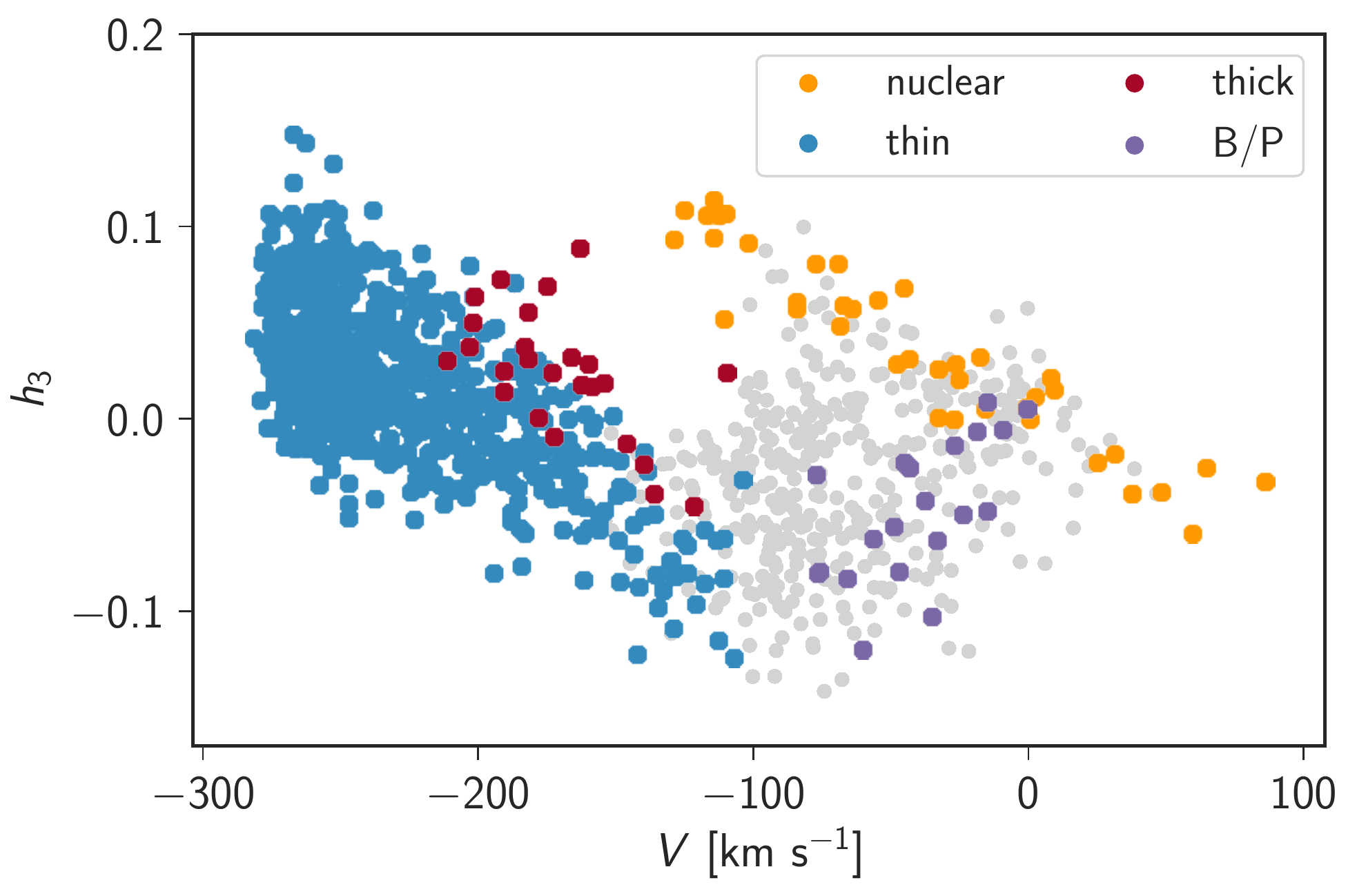}
\caption{\h3 as a function of \V for bins in the thin disc (blue), the thick disc (red), the nuclear disc (orange) and the B/P bulge (purple). The grey points are bins that cannot be unambiguously attributed to a single component, most of them are in the inner region, particularly in the bar. As expected, both the main disc (thin and thick) and the nuclear disc show a clear anti-correlation between \V and \h3, while the bar (most of the grey points)  and the B/P bulge show a distinct behaviour with a correlation between \V and \h3.}
\label{fig:V_h3}
\end{figure}

The fourth panel in Figure \ref{fig:kin_maps} represents the spatial distribution of \h3, the third order Gauss-Hermite moment. It indicates the skewness of the line-of-sight velocity distribution. The most apparent feature in this map is the anti-correlation between \h3 and \V in the main disc, and in the nuclear disc (as already shown by \citealp{Chung2004} and \citealp{Molaeinezhad2016}). We plot \h3 as a function of \V for all the bins in Figure \ref{fig:V_h3}: those two quantities are indeed clearly anti-correlated in the main disc (thin and thick, blue and red points) and in the nuclear disc (orange points). This anti-correlation is a general feature of rotating discs \citep{Chung2004,Bureau2005,Falcon2006,Gadotti2020}. We note that at a given velocity, \h3 seems slightly higher in the thick disc than in the thin disc, but given the uncertainties in \h3 this result might not be very significant. In addition to this, \V and \h3 are correlated in the bar region (most of the grey points in Figure \ref{fig:V_h3}) and particularly in the B/P bulge (purple points). This is expected for regions where a bar is superimposed over a disc \citep{Bureau2005, Iannuzzi2015, Li2018}.

Finally, the fifth panel in Figure \ref{fig:kin_maps} shows the spatial distribution of \h4, the fourth order Gauss-Hermite moment. It indicates the kurtosis of the line-of-sight velocity distribution. A high value of \h4 is generally interpreted as indicating the superposition of several dynamical components (e.g., \citealp{Bender1994}). We indeed find high values of \h4 in the nuclear disc region, where the cold circular orbits are also superimposed with bulge/bar orbits and global disc orbits from the foreground (as also found for nuclear discs in \citealp{Gadotti2020}).

The kinematic maps thus highlight the different components of NGC 5746: we have identified kinematic signatures of the nuclear disc, bar and B/P bulge that had already been observed by \cite{Falcon2003}, \cite{Chung2004}, and \cite{Molaeinezhad2016}. We also extend previous kinematic analyses into the disc region, where we find a velocity dispersion decreasing with radius and increasing with height above the midplane, and the corresponding asymmetric drift in the rotation curve.

\subsection{Stellar populations}

\begin{figure*}
\includegraphics[width=2\columnwidth]{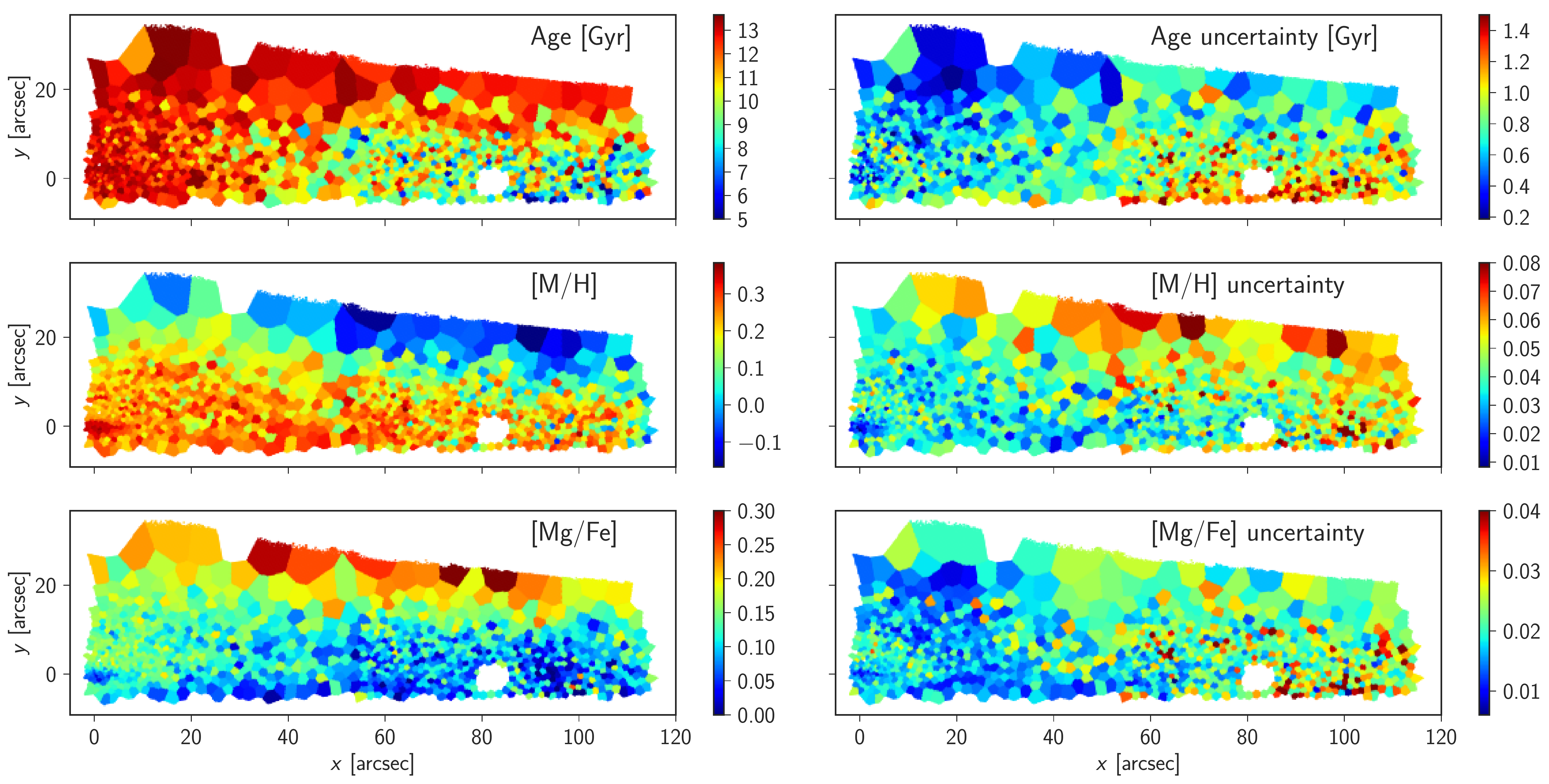}
\caption{Stellar populations maps. The left column shows the spatial distribution of the mean mass-weighted age, [M/H] and  \al obtained from full spectrum fitting. The right column shows the corresponding uncertainties derived from our bootstrap analysis described in Section \ref{sec:uncertainties}. }
\label{fig:pop_maps}
\end{figure*}

For each spatial bin, we compute the mean mass-weighted age, \Fe and \al from the distribution of SSP weights given by pPXF. The spatial distribution of those three quantities is shown in Figure \ref{fig:pop_maps}, together with the spatial distribution of the corresponding uncertainties from our bootstrap analysis (see Section \ref{sec:uncertainties}).

In the mean mass-weighted age map (top panel), a first striking feature is that  most of the stars in NGC 5746 are very old: most of the bins have a mean mass-weighted age above 8 Gyr (as expected for massive galaxies, e.g. \citealp{Trager2000, Gallazzi2005, Kuntschner2010}). It still is possible to distinguish two separate regions: the thin disc is the youngest, with mean ages around 8-9 Gyr, while the rest of the galaxy is older. The thick disc, bar, B/P bulge and nuclear disc all have mean mass-weighted ages above 11-12 Gyr. The nuclear disc does not appear very distinct from the regions around it.

We find an overall mean age uncertainty of 0.8 Gyr. This is a slight underestimate: for the oldest bins, the uncertainties are artificially small because the age grid is limited to values up to 14 Gyr (this is why some of the oldest bins in the central regions have very small age uncertainties). If we compute the mean age uncertainty only including bins with a mean age smaller than 10 Gyr, we find a value of 1.1 Gyr. All of these values are of course lower limits on the real uncertainties since they do not account for systematic uncertainties due to for instance the choice of set of SSP models, isochrones, IMF, or fitting algorithm (as discussed in Section \ref{sec:uncertainties}).

The mean mass-weighted \Fe map shows overall very high values, above -0.2 dex  everywhere, even in the thick disc region. Such a high metallicity is actually not surprising for a massive galaxy like NGC 5746, given the steep relation between stellar mass and metallicity \citep[e.g.,][]{Gallazzi2005, Delgado2014}, and is also consistent with the high gas-phase metallicity \citep{DeVis2019}. The most metal-rich region is the nuclear disc, which appears distinct from the surrounding regions. The thin disc and the bar appear similar to each other, with mean metallicities between 0.1 and 0.3 dex, while the thick disc is more metal-poor, with \Fe $\sim -0.1$ dex. The mean \Fe uncertainty is 0.04 dex. Similarly to what we find for age, the uncertainties are slightly underestimated in the very metal-rich regions (e.g., the nuclear disc) where our metallicities reach the edge of the SSP grid.

Finally, the mean mass-weighted \al map is shown in the bottom of Figure \ref{fig:pop_maps}. The bins have a wide range of \al values. In the thin disc, \al is typically below 0.15 dex, and the values of \al increase towards the thick disc, where \al is above 0.2 dex. The bar and B/P regions have intermediate values of \al, while the nuclear disc is easily seen in the map due to its very low \al. The mean uncertainty on \al is only 0.02 dex, with small spatial variations.

To summarize, the stellar populations maps clearly highlight differences between the different components. The thin disc is younger, more metal-rich, and more $\alpha$-poor than the thick disc. The bar and B/P bulge are old, metal-rich and slightly more $\alpha$-enhanced than the thin disc. Finally, the nuclear disc is old, metal-rich and $\alpha$-poor. There are no very obvious radial gradients in the properties of each component (similarly, the ionized gas only shows a weak radial metallicity gradient, see \citealp{DeVis2019}). These general trends are also seen in the SSP-equivalent age, metallicity and [Mg/Fe] maps from our line-strength analysis shown in Appendix \ref{appendix:indices} (and Figure \ref{fig:comp_indices_maps}). Those results are also consistent with the ones obtained by \cite{Molaeinezhad2017} using a line-strength analysis of SAURON data.

\subsection{Global age and metallicity distributions}
\begin{figure*}
\includegraphics[width=2\columnwidth]{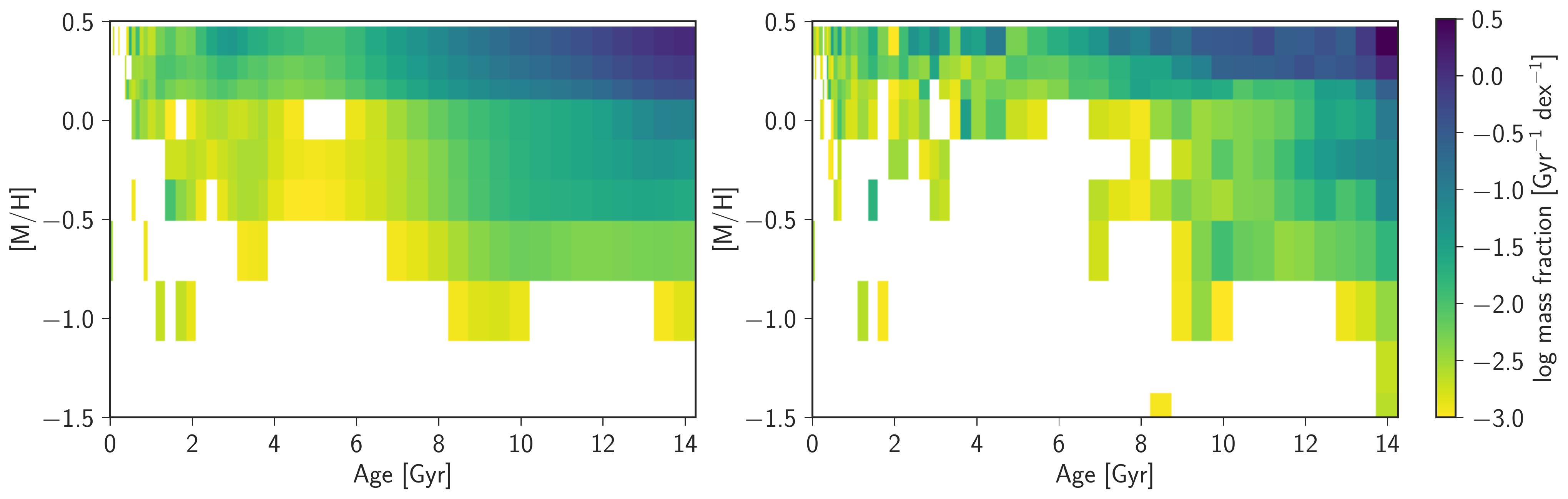}
\caption{Mass distribution in age and \Fe space, taking into account all the spatial bins for NGC 5746 (each bin is weighted by its stellar mass). We only plot the weights corresponding to mass fraction above 0.001 Gyr$^{-1}$ dex$^{-1}$. We compare the regularized solution from pPXF (left panel) to the mean of 100 bootstrap realizations (right panel), and find a good agreement between the two.  Most of the mass is found at old ages and high metallicities, but stellar populations of all ages are present in the galaxy.}
\label{fig:age_fe_global}
\end{figure*}

In addition to computing the mass-weighted averages of age and \Fe, we can also study for each bin the full distribution of weights in age and \Fe space. We combine the weights distributions of all our Voronoi bins to obtain the global age and metallicity distribution for NGC 5746 (for this, we weight each bin by its stellar mass). The result of this process is shown in Figure \ref{fig:age_fe_global}: this figure represents the fraction of the mass present in each combination of age and \Fe. Because the bins in age and \Fe are not spaced regularly, we divide each weight by its area. This is not a choice commonly made by studies using pPXF, but we find that it gives a fairer visual representation of the stellar populations in the galaxy. For instance, if a young bin with a width of 0.05 Gyr and an old bin with a width of 0.5 Gyr had the same weight, simply plotting the weights as a function of age would give the impression of a flat star formation history, while in reality the star formation rate would be 10 times higher for the younger bin. This is corrected by dividing each weight by its width in age (and in metallicity).

\begin{figure}
\includegraphics[width=0.95\columnwidth]{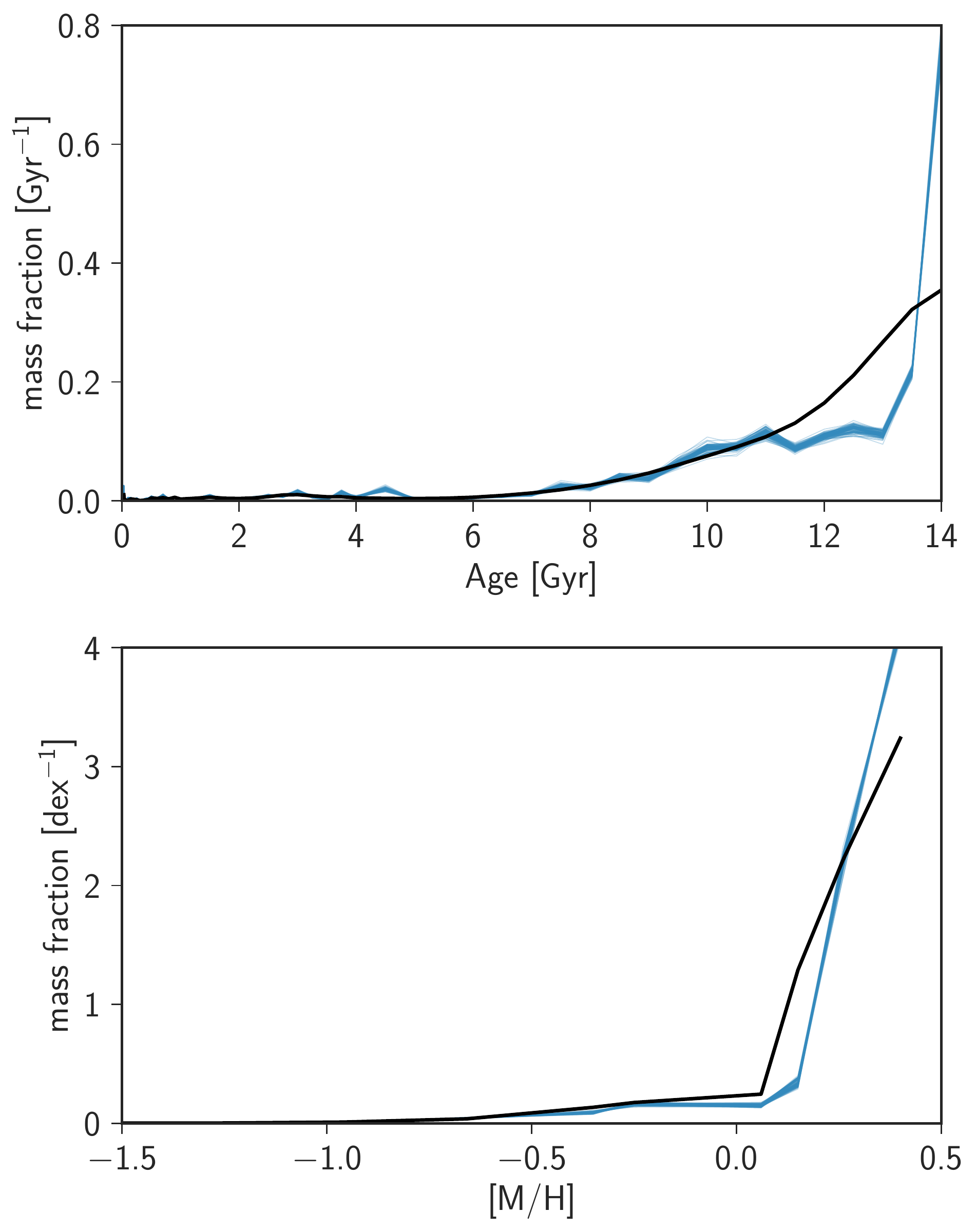}
\caption{Age distribution (top panel) and \Fe distribution (bottom panel) for all the bins in NGC 5746 (each bin is weighted by its stellar mass). In each panel, the black line shows the regularized pPXF best fit, while the 100 thin blue lines correspond to the 100 bootstrap realizations.}
\label{fig:sfh_mdf_all}
\end{figure}

The left panel in Figure \ref{fig:age_fe_global} corresponds to the weights from the standard regularized fit, while the right panel shows the mean of our 100 bootstrap realizations. While the two distributions are different (if only because in practice they correspond to different levels of regularization), they share many common features and their overall shapes are very similar, which is very reassuring. This can also be seen in Figure \ref{fig:sfh_mdf_all}, showing the age and \Fe distributions separately (obtained by integrating the array shown in Figure \ref{fig:age_fe_global} over \Fe and age, respectively). In Figure \ref{fig:sfh_mdf_all}, the black curves correspond to the regularized solution and we plot the 100 bootstrap realizations with thin blue lines: we find that the two approaches give similar results, as we will now describe in more details.

From the mean age and \Fe maps, we had concluded that NGC 5746 was mostly an old and metal-rich galaxy. This is also clear in Figures \ref{fig:age_fe_global} and \ref{fig:sfh_mdf_all}, where most of the mass is found at old ages and high metallicities. However, a more complex picture is also emerging: stellar populations of all ages are present in the galaxy. The phase of intense star formation seems to last until $\sim 8$ Gyr ago, but star formation continues after that, with a small secondary peak between 2 and 4-5 Gyr ago. Those features are present both in the regularized solution and in the bootstrapped distribution.

We also find a broader distribution of \Fe for older stars, with \Fe reaching down to -1 dex for stars between 8 and 14 Gyr old. By contrast, the youngest stars have mostly super-solar metallicities.

In the following section, we further investigate the stellar populations of NGC 5746 by studying each of its components separately.

\section{Stellar populations in each component of NGC 5746}\label{sec:pops_components}

\begin{figure*}
\includegraphics[width=2\columnwidth]{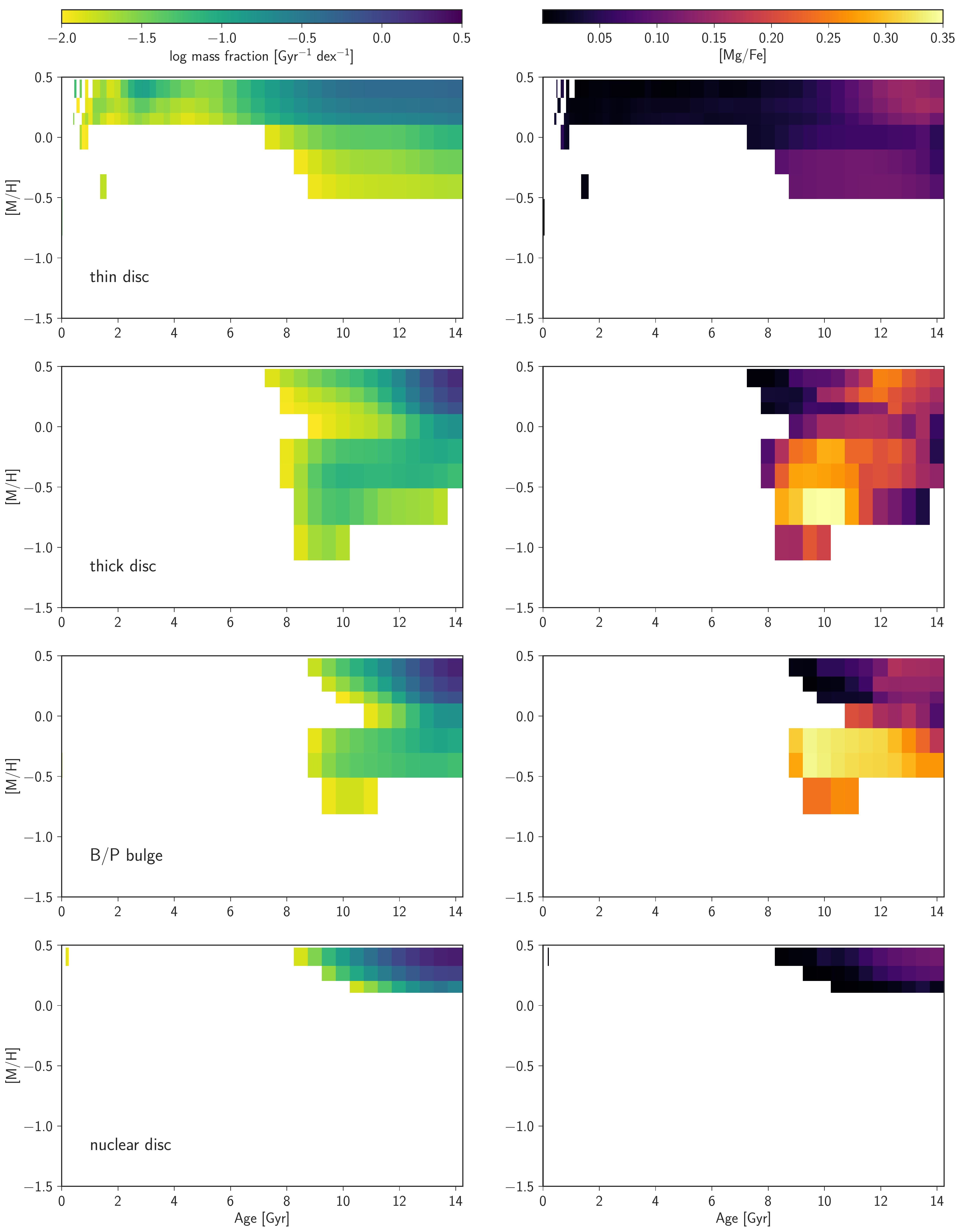}
\caption{Left column: mass distribution in age and \Fe space for each component of NGC 5746 (each bin is weighted by its stellar mass) --- from top to bottom: the thin disc, the thick disc, the B/P bulge and the nuclear disc. Right column: corresponding values of \al as a function of age and \Fe. We only plot the weights and \al values corresponding to mass fraction above 0.01 Gyr$^{-1}$ dex$^{-1}$.}
\label{fig:age_fe_components}
\end{figure*}

\begin{figure*}
\includegraphics[width=2\columnwidth]{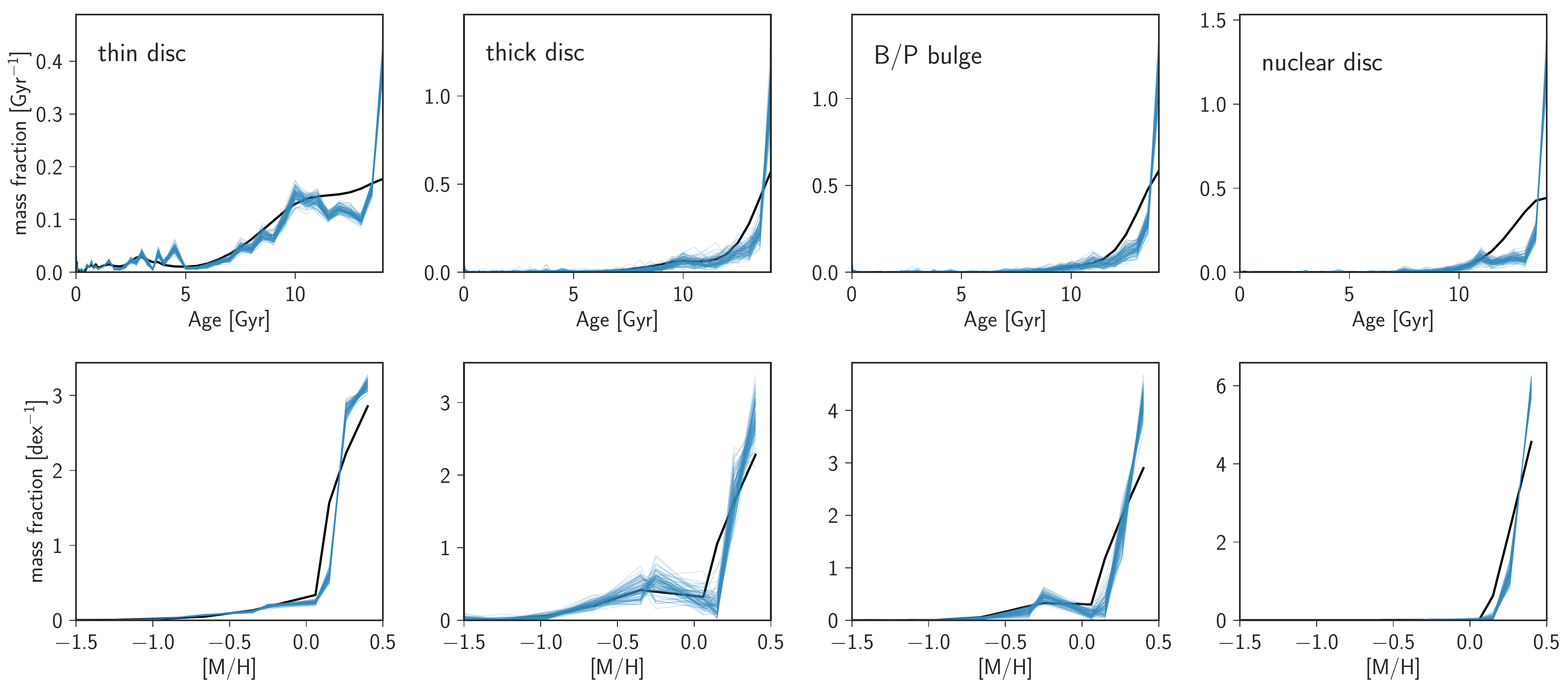}
\caption{Age distribution (top row) and \Fe distribution (bottom row) for each component of NGC 5746 (each bin is weighted by its stellar mass) --- from left to right: the thin disc, the thick disc, the B/P bulge and the nuclear disc. In each panel, the black line shows the regularized pPXF best fit, while the 100 thin blue lines correspond to the 100 bootstrap realizations.}
\label{fig:sfh_mdf_components}
\end{figure*}

In this section, we present the stellar populations found in the thin and thick disc, in the B/P bulge and in the nuclear disc. To that aim, we sum the weight distributions for the bins attributed to each component (we take into account the mass in each bin). In the left column of Figure \ref{fig:age_fe_components}, we show the distribution of the weights in age and \Fe space. Those distributions can be integrated over age and \Fe to produce the plots in Figure \ref{fig:sfh_mdf_components}, where we show the distribution of ages and \Fe for the four different components. Finally, while the distributions shown so far were obtained by summing the weights corresponding to \al= 0 and \al= 0.4, we can also compute a mean \al for each bin in age and \Fe, this is shown in the right column of Figure \ref{fig:age_fe_components}.

\subsection{Thin disc}
As can be seen in Figures \ref{fig:age_fe_components} and \ref{fig:sfh_mdf_components}, the thin disc has an extended star formation history. While most of the stars are older than 8 Gyr (with possibly a peak of star formation around 10 Gyr), the thin disc also shows more recent star formation, from about 4-5 Gyr ago down to the current time. The shapes of the age distributions are consistent between the regularized pPXF run and the bootstrap analysis (see also Appendix \ref{appendix:bootstrap} for the comparison of the regularized and bootstrapped distributions in the \Fe vs age plane). One aspect to note is that the weight distribution is more peaked at very old ages for the bootstrap analysis (for all components, not just the thin disc): this is because it employs a smaller level of regularization, so that the solutions are less smoothed.

Most of the stars in the thin disc are metal-rich, even at old ages (Figure \ref{fig:age_fe_components}), and the metallicity distribution shows that most stars have \Fe$>0$ (Figure \ref{fig:sfh_mdf_components}). The distribution of \al shows a smooth transition from \al $\sim 0.1-0.15$ dex at old ages down to \al $\sim 0$ dex for the youngest stars.

\subsection{Thick disc}\label{sec:thick_disk_pops}
The thick disc is showing complex stellar populations, that appear nearly bimodal in Figure \ref{fig:age_fe_components}. The main component of the thick disc has a super-solar metallicity and is very old (ages typically above 12 Gyr, with a tail down to 8 Gyr). In that component, \al is correlated with age: stars older than $\sim 12$ Gyr are the most $\alpha$-rich, and \al decreases to $\sim0$ for stars 8 Gyr old.

The second component of the thick disc is more metal-poor (\Fe between -1 and 0 dex), and its age distribution is roughly flat from 8 to 14 Gyr old (at the oldest ages, it is impossible to determine for sure if the two components are still distinct or if they overlap). A striking feature of that second component is its very unique distribution of \al: it is significantly more $\alpha$-rich than the rest of the disc, with \al$>0.25$ for the younger stars in that component. We show in Appendix \ref{appendix:bootstrap} that the mean weights distribution from the bootstrap analysis gives results that are consistent with the regularized solution presented here, which shows that the distribution we have discussed is not an artifact of our particular choice of regularization parameter. We have done additional 
tests, where starting from an unregularized solution, we gradually increase the level of regularization: we find that the second component is always present, and at a similar location, until regularization becomes so high that everything gets blurred. We also discuss in Appendix \ref{appendix:steckmap} some tests that we have performed using another code, STECKMAP \citep{Ocvirk2006}. STECKMAP does not compute a full distribution of weights in age and metallicity, but computes an age distribution, and the mean metallicity as a function of age. In Figure \ref{fig:comp_steckmap}, we show that for the thick disc STECKMAP finds that the oldest stars are metal-rich, and that the metallicity decreases for stars younger than 10 Gyr (we test two different sets of isochrones, which gives similar results). This is consistent with our pPXF analysis and reinforces our confidence in our results.

This second component is not consistent with the global chemical evolution of the disc: it is younger, more metal-poor and more $\alpha$-rich than most of the stars in the disc. Our best interpretation is that it corresponds to stars accreted from a smaller satellite galaxy, similarly to what was also found in FCC 170, 153, and 177 by  \cite{Pinna2019a,Pinna2019b}, and in NGC 7135 by \cite{Davison2021}. This interpretation will be further discussed in Section \ref{sec:accretion}.

\subsection{B/P bulge}
The B/P bulge is showing stellar populations that are very similar to the thick disc (Figure \ref{fig:age_fe_components}): a main component that is old and metal-rich, with \al correlated with age, and a second component that is younger, more metal-poor, and more $\alpha$-rich. The similarity with the thick disc is also apparent in Figure \ref{fig:sfh_mdf_components}.

This could mean that the B/P bulge and the thick disc genuinely host similar populations (with for instance some accreted stars also present in the B/P bulge), but part of this similarity could be only due to the physical superposition of thick disc populations in front of the B/P bulge. However, our morphological decomposition using IMFIT suggests that only 23\% of the light in the B/P region can be attributed to the thick disc. The similarity between the populations in the thick disc and the B/P bulge is thus  genuine, and not an effect of superposition. 

\subsection{Nuclear disc}
The nuclear disc is showing some of the simplest stellar populations in the galaxy: nearly all of the stars are older than 10 Gyr, with \Fe$>0$, and a low \al abundance. There are hints of a slight decline of \al from the oldest to the youngest stars.

\section{The formation history of NGC 5746}

\begin{figure}
\includegraphics[width=0.95\columnwidth]{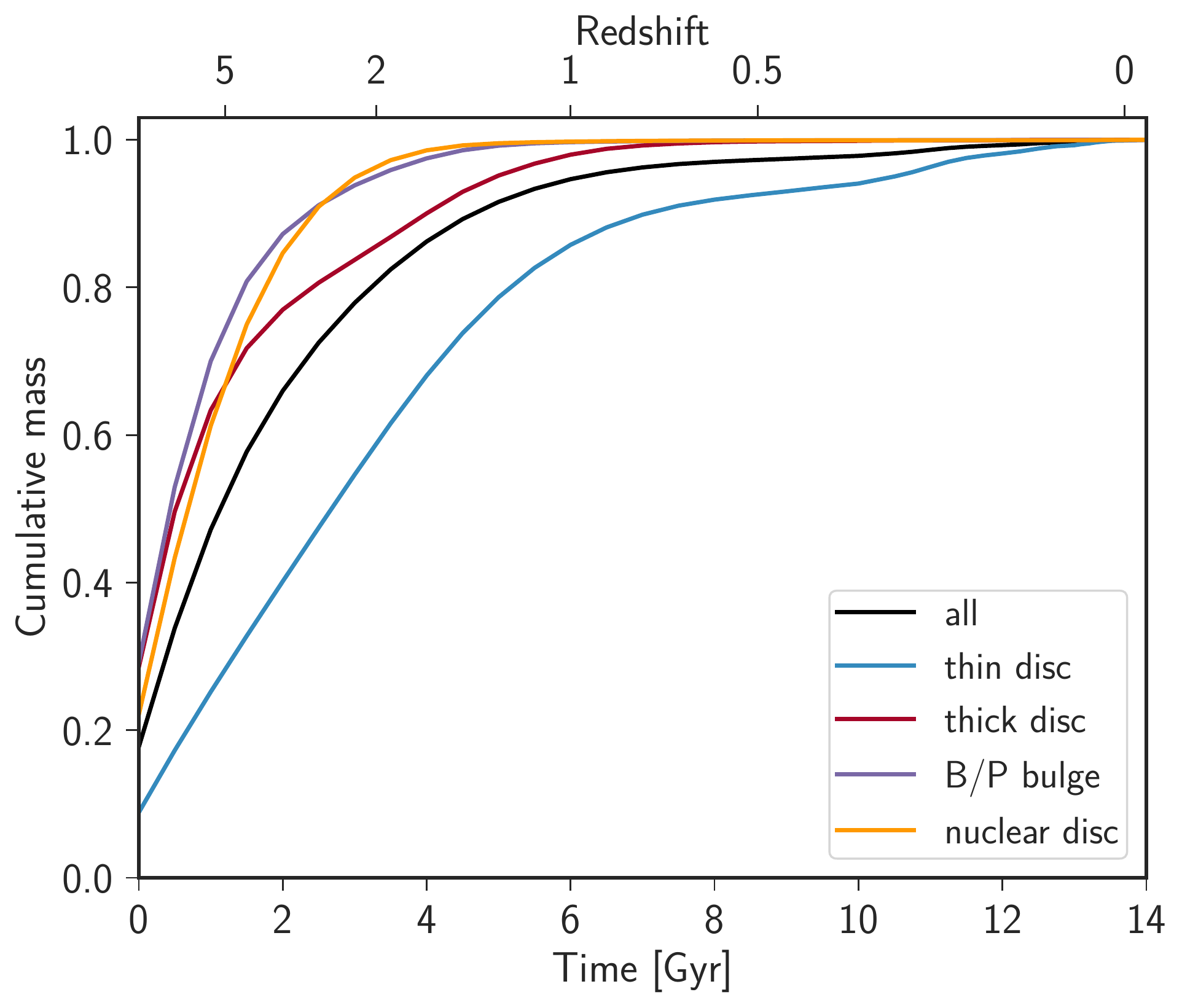}
\caption{Cumulative mass growth of NGC 5746 (black line) and its different components. This shows that 80\% of the stellar mass formed before $z=2$ (10 Gyr ago) and that the nuclear disc, the B/P bulge and the thick disc are all in place by $z=1$ (or 8 Gyr ago). By contrast, star formation continues in the thin disc down to the present time. }
\label{fig:cumul_mass}
\end{figure}

\begin{figure*}
\includegraphics[width=2\columnwidth]{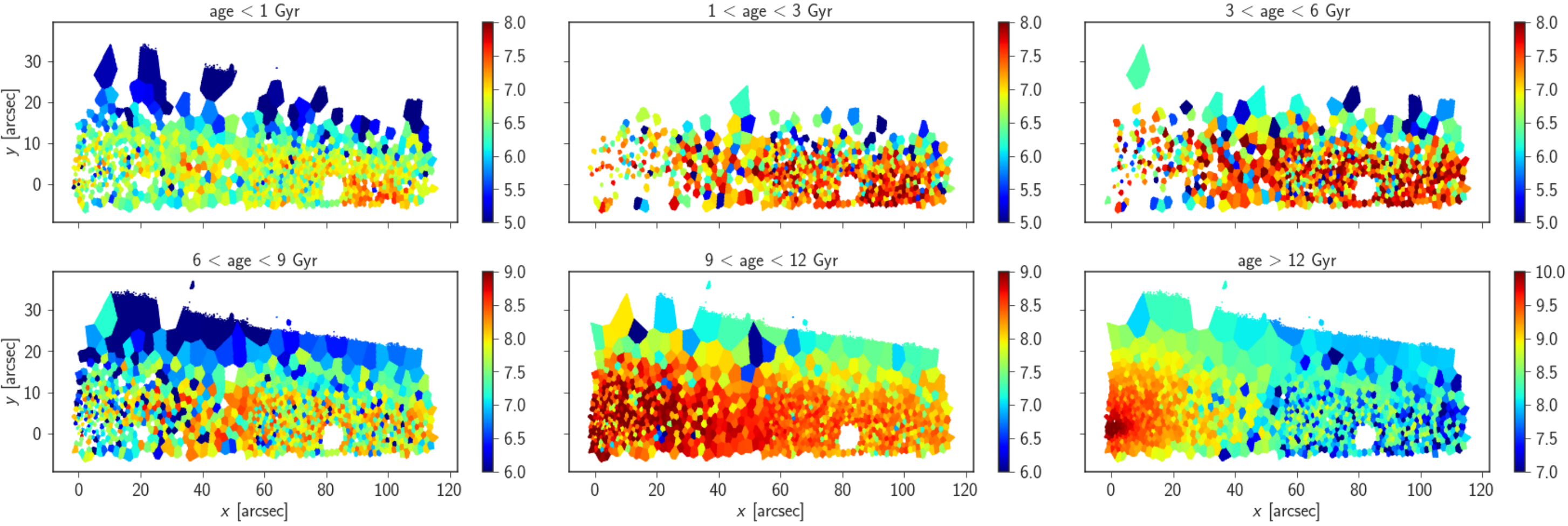}
\caption{Surface density maps of stellar populations binned by age, from the youngest stars in the top left panel, to the oldest in the bottom right panel. The colorbars correspond to the logarithm of the surface density, in units of \msun kpc$^{-2}$. Stars older than 12 Gyr are very centrally concentrated, while stars younger than 9 Gyr are mostly found in the thin disc.}
\label{fig:mass_maps}
\end{figure*}

\begin{figure}
\includegraphics[width=0.9\columnwidth]{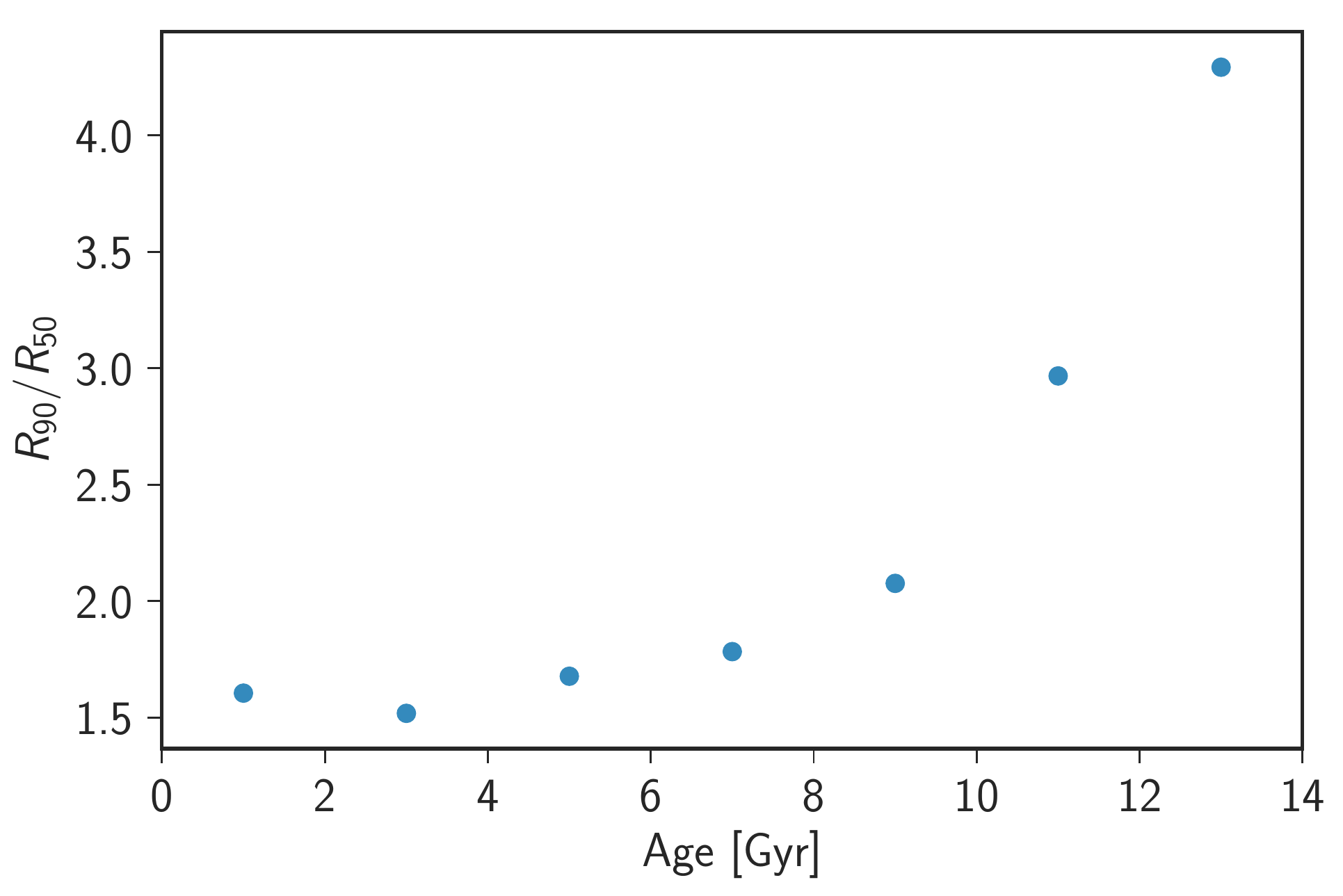}
\caption{Ratio of the radii enclosing 90\% and  50\% of the mass as a function of age, showing that concentration increases steeply with age.}
\label{fig:concentration}
\end{figure}

\subsection{Mass growth}
To summarize our results, we show in Figure \ref{fig:cumul_mass} the cumulative mass growth history of NGC 5746 and its individual components: this is another way of showing that the galaxy is overall very old, with 80\% of its stellar mass formed before $z=2$ (or 10 Gyr ago). The nuclear disc, the B/P bulge and the thick disc are all in place by $z=1$ (or 8 Gyr ago). By contrast, star formation continues in the thin disc down to the present time.
We note that we probably underestimate the contribution of young stars to the thin disc: dust is present throughout the thin disc (see Figure \ref{fig:pointings}), and absorbs a larger fraction of the light emitted by young stars compared to old stars (as shown by \citealp{Nersesian2019} who modelled the SED of NGC 5746). This might affect the exact shape of the star formation histories for spaxels in the thin disc, but none of the main results of the paper depends strongly on this. The effect should be negligible in the other regions of the galaxy, where dust is absent.

We show in Figure \ref{fig:mass_maps} the spatial distribution of stars in different age bins, from the youngest stars in the upper left panel to the oldest in the lower right panel. Stars older than 12 Gyr are very centrally concentrated: even though they are present everywhere in the galaxy, their spatial distribution is heavily biased towards the inner galaxy. A change is already observed for stars between 9 and 12 Gyr old, whose distribution still peaks in the central regions but is significantly more extended towards the disc. Stars younger than 9 Gyr are themselves mostly found in the thin disc, and we find little difference between the populations of different ages. The very youngest stars (below an age of 1 Gyr) possibly show a slight overdensity in a ring-like feature (the orange-red region at radii of 80--100 arcsec, extending towards the inner disc in an arc at a height of $\sim$10 arcsec): this seems to correspond to the outer region of the  ring seen in the 8 $\mu$m Spitzer image (Figure \ref{fig:pointings}), and to the distribution of star-forming regions seen in the ionized gas maps (Appendix \ref{appendix:emission}), but projection effects make it hard to see this clearly. 

We quantify the spatial distribution of stars of different ages by computing the radii enclosing 50\% and 90\% of the mass of each population ($R_{50}$ and $R_{90}$, respectively): their ratio is an indicator of concentration. We note that these values do not directly correspond to the global concentration values we would obtain if our field of view encompassed the whole galaxy: our overall value of $R_{90}$ (including all stellar populations) is only 72'', while WISE observations in the W1 and W2 bands (tracing stellar mass) give values of $R_{90}$ around 130'' \citep{Jiang2019}. This means that we cannot directly compare our concentration values to photometric studies encompassing the whole galaxy, but it still remains meaningful to compare populations of different ages. We show in Figure \ref{fig:concentration} that $R_{90}$/$R_{50}$ increases steeply with age. More precisely, populations younger than 8 Gyr have very similar low concentrations around 1.5--2, while populations older than 8--10 Gyr are very concentrated.

\subsection{Merger history} \label{sec:accretion}

\begin{figure*}
\includegraphics[width=2\columnwidth]{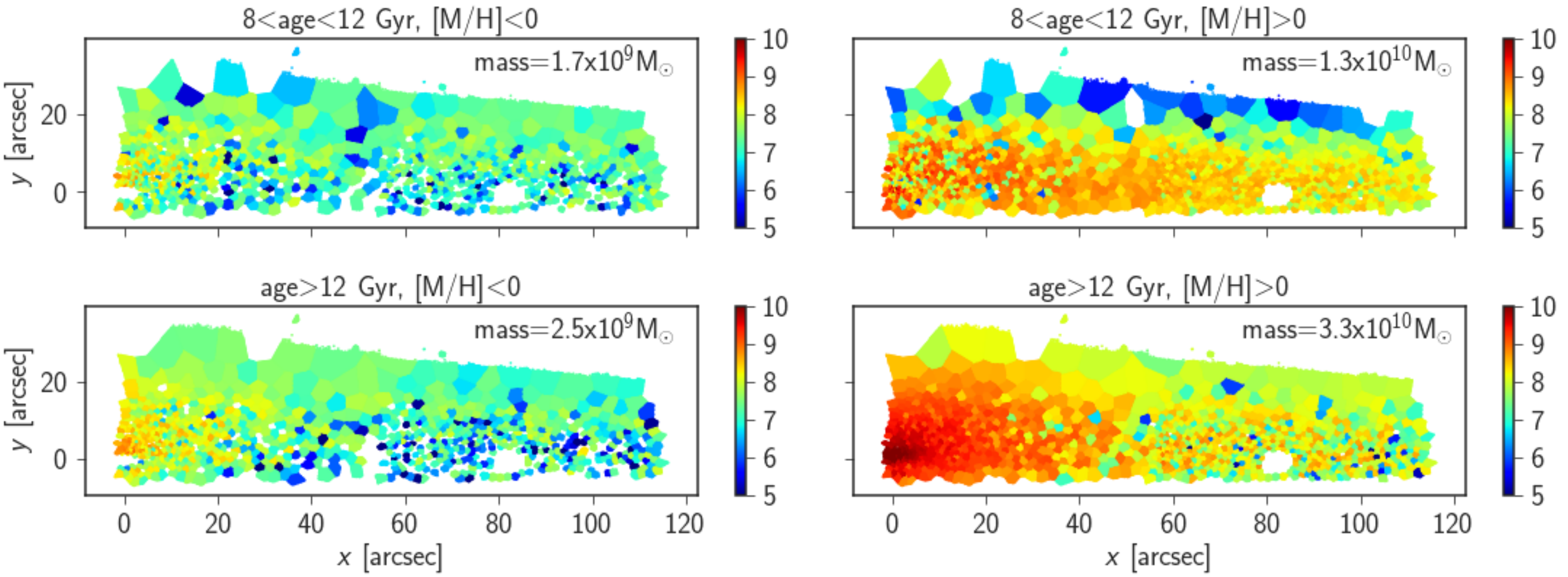}
\caption{Surface density maps of stellar populations binned by age and \Fe. The colorbars correspond to the logarithm of the surface density, in units of \msun kpc$^{-2}$. We interpret the metal-poor population (left column) as mostly made of accreted stars, brought by a $\sim$1:10 merger happening $\sim$ 8 Gyr ago.}
\label{fig:mass_ages_feh}
\end{figure*}

\begin{figure}
\includegraphics[width=\columnwidth]{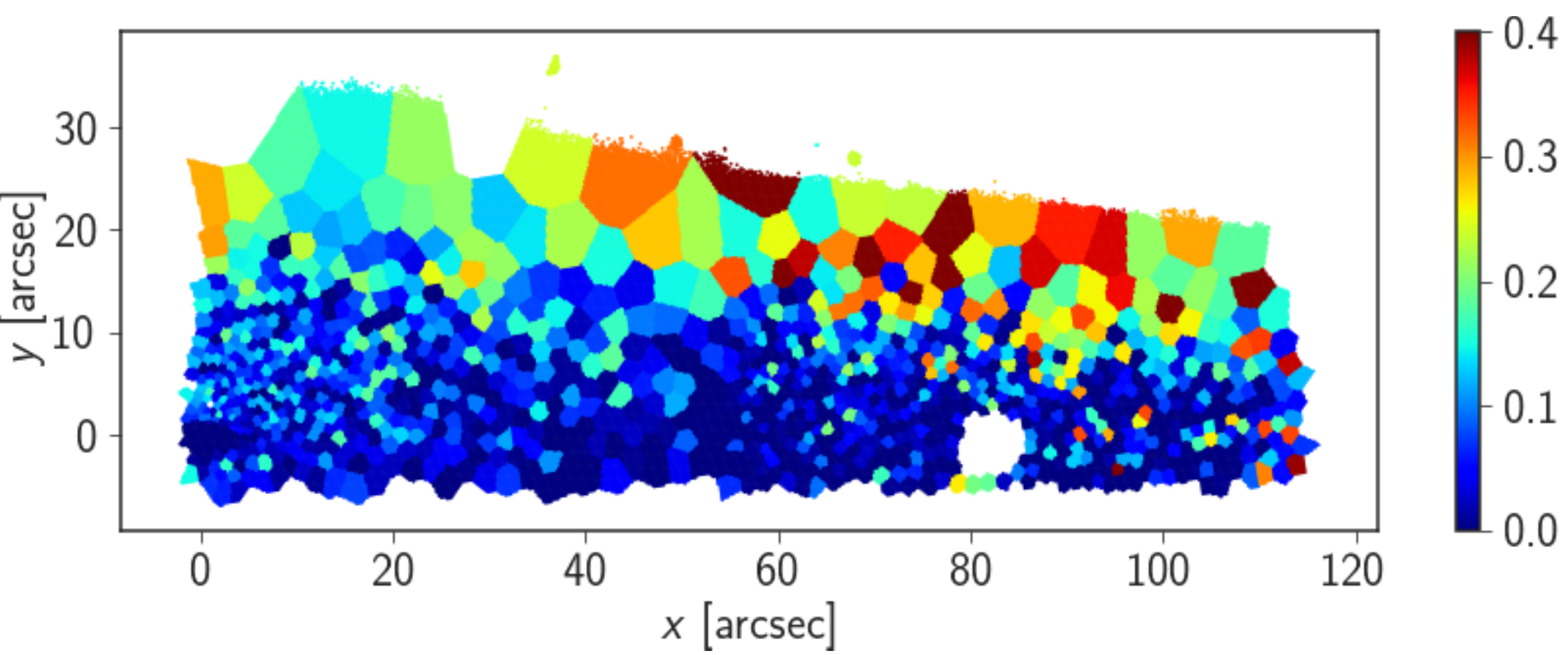}
\caption{Spatial distribution of the fraction of accreted stars (here defined as stars with \Fe$<0$). This fraction increases with height above the midplane, and accreted stars represent 28\% of the thick disc on average. }
\label{fig:frac_accret}
\end{figure}
We argued in Section \ref{sec:thick_disk_pops} that the metal-poor component in the thick disc (also apparent in the B/P bulge), with \Fe between -1 and 0, corresponds to stars that were accreted by NGC 5746. We indeed find that those stars, for a given age, have a chemical composition that is clearly distinct from stellar populations in the rest of NGC 5746, as also discussed for other galaxies by \cite{Pinna2019a,Pinna2019b} and \cite{Davison2021}. If we use the age distribution of the metal-poor stars as an indicator of the time of accretion, this suggests that the merger ended $\sim$8 Gyr ago (since there are no metal-poor stars younger than 8 Gyr old). From the star formation histories of NGC 5746 and its components, we see that this time would coincide with a cessation of star formation in many places in the galaxy, until it picks up again a few Gyr later in the thin disc. The interaction with the satellite galaxy could also be the cause of the enhanced star formation seen in the thin disc $\sim$10 Gyr ago: satellites can trigger starbursts during pericenter passages, not just at the final coalescence \citep{Ruiz2020,DiCintio2021}.

In addition to the bimodal metallicity structure of the thick disc, one argument in favour of an accreted origin for the metal-poor component is its high \al (see Figure \ref{fig:age_fe_components}). Indeed, the only other possible explanation for the presence of stars that are on average younger and more metal-poor than the main thick disc component would be that those stars originate from a metal-poor region of the thin disc and were heated to become part of the thick disc. However, the thin disc does not contain any region where stars are predominantly metal-poor, and the few metal-poor stars present in the thin disc have a low \al value: their chemistry is not consistent with what we observe in the thick disc. As a consequence, the most likely explanation for the presence of younger, metal-poor, $\alpha$-rich stars in the thick disc is that they were accreted from a smaller satellite galaxy (given that lower mass galaxies have a lower metallicity, see e.g. \citealp{Gallazzi2005}). The high \al values could be consistent with an accelerated star formation history truncated at high redshift, before too many generations of Type Ia supernovae could increase the Fe content in the galaxy and decrease its average \al (e.g., \citealp{Matteucci1990}).

While the external origin of the metal-poor stars is quite likely, it is a priori unknown if, as described until now, those stars were all brought in by a satellite galaxy, or if, instead, NGC 5746 could have undergone a merger with a very gas-rich galaxy, bringing a high amount of metal-poor gas. A starburst within the disc of the main galaxy could then have formed a population of metal-poor, $\alpha$-rich stars: this scenario for the formation of thick discs during gas-rich mergers has been proposed by \cite{Brook2004}. It is impossible to infer the precise gas content of the accreted satellite, but we find the scenario of a very gas-rich merger less likely (in the scenario presented by \citealp{Brook2004}, the satellites have a gas fraction of $\sim$80\%). The main reason for this is the chemical bimodality observed at a given age: in a scenario where a large amount of gas is accreted, it is expected that this gas would mix efficiently with gas present in the main galaxy, and that stars formed during the starburst phase would be chemically homogeneous. This was for instance shown in \cite{Brook2005}, where all the stars formed during the merger have a similar chemical composition, and where there is no vertical metallicity gradient in the resulting thick disc. In NGC 5746, we find both several components in the thick disc, and a vertical metallicity gradient (thin disc stars with an age of 8 Gyr are more metal-rich than thick disc stars of the same age --- see Figure \ref{fig:age_fe_components}). We also do not see any dilution of the metallicity in the thin disc around the time of the accretion event. Another argument against a very gas-rich merger is the spatial distribution of metal-poor stars, which form a very extended disc (as we will next show) instead of the centrally-concentrated component expected in a gas-rich merger \citep{Brook2007}. We thus think that it is most likely that a large fraction of the metal-poor stars were accreted during a merger finishing about 8 Gyr ago. The gas fraction in the satellite is of course impossible to determine, and some of the stars now in NGC 5746 probably formed from metal-poor gas from the satellite, either during the interaction, or after the satellite coalesced with the main galaxy.
We will now attempt to better characterize this merger.

Figure \ref{fig:mass_ages_feh} shows the spatial distribution and the total mass contained in four different stellar populations, separated in age (between 8 and 12 Gyr old, and above 12 Gyr old) and \Fe (above and below \Fe = 0). We can first use those masses to estimate the mass ratio of the merger. For this, we assume that the merger took place 8 Gyr ago, that all stars with ages above 8 Gyr and sub-solar metallicities are accreted, and that all stars with ages above 8 Gyr and super-solar metallicities formed in-situ. With those assumptions, the mass of accreted stars is 4.2$\times 10^9$ \msun, and the mass of in-situ stars is \mbox{4.6$\times 10^{10}$ \msun} (within our two MUSE pointings), so that the merger mass ratio would have been around 1:10. It is of course a simplification to assume that all the old  metal-poor stars are accreted: some of them must have formed in situ. If we instead assume that only half of the oldest metal-poor stars are accreted, then the mass of accreted stars is 3$\times 10^9$ \msun, and the merger mass ratio would be $\sim$1:15. The total stellar mass of the satellite galaxy can be estimated using the fact that our pointings cover about half of NGC 5746: this gives a mass of 6--8 $\times 10^9$ \msun for the accreted galaxy (depending on the fraction of metal-poor stars that we consider as being accreted). According to predictions for the mass-metallicity relation at redshift 1--2, this mass is roughly consistent with the relatively high mean metallicity ($\mathrm{[M/H}\simeq-0.5$) of the accreted stars (e.g., \citealp{Fontanot2021}).

In Figure  \ref{fig:mass_ages_feh}, there is a striking difference between the spatial distribution of the metal-rich and metal-poor stars. The metal-poor stars are very uniformly distributed throughout the galaxy, and there is little difference between the two age bins we consider. Those populations are slightly denser in the inner regions of NGC 5746, but overall their density varies little with height and radius. The similarity between the distribution of the metal-poor stars in the two age bins that we consider suggests that they mostly have a similar origin, and that only a small fraction of the oldest metal-poor stars formed in situ (the chemical enrichment in the main progenitor of NGC 5746 was probably very fast at early times). If we identify the metal-poor population as made of accreted stars, this means that the incoming satellite deposited its stars throughout the disc of NGC 5746 (instead of contributing to a compact classical bulge): this suggest the possibility of an orbit with a low inclination, for which the fraction of stars deposited in the disc is typically high (see e.g., \citealp{Penarrubia2006,Read2008,Villalobos2008}). We show in Figure \ref{fig:frac_accret} the fraction of accreted stars in each Voronoi bin. The patterns seen in this Figure, in particular the increase of the fraction of accreted stars with height above the midplane and with galactocentric distance are both consistent with the outcomes of minor merger simulations presented in \cite{Villalobos2008} and with cosmological simulations by \cite{Gomez2017} and \cite{Park2021}. The negligible contribution of accreted stars to the bulge of the galaxy matches what is observed in simulations where the satellite galaxy is less dense than the main disc: in this case, the satellite is efficiently disrupted as it orbits within the main disc, and it deposits most of its stars within the disc instead of the bulge \citep{Villalobos2008}.

The spatial distribution of the metal-rich stars brings another piece of the puzzle (right column of Figure \ref{fig:mass_ages_feh}). The metal-rich stars older than 12 Gyr (that by themselves contribute 60\% of the total mass of the galaxy) are very centrally concentrated, but can still be found throughout the galaxy. They occupy all the regions in the disc, with no strong vertical density gradient from the thin to the thick disc. By contrast, the younger metal-rich stars (ages from 8 to 12 Gyr) are mostly found in the thin disc, and seem to avoid the thick disc. This means that if a 1:10 merger indeed happened $\sim$ 8 Gyr ago, it did not significantly heat vertically this population of 8-12 Gyr old metal-rich stars. By extension, this also suggests that it similarly did not heat the very oldest stars, and that the 1:10 merger is not responsible for the concentrated and thick spatial distribution of the stars older than 12 Gyr, which represent most of the mass in the galaxy. This result if not very surprising given than 1:10 mergers are not expected to contribute much to vertical disc heating, particularly for low-inclination orbits where the vertical structure of the main disc is less disturbed \citep{Villalobos2008, Bien2013}, and particularly if some gas is present in the main galaxy to absorb part of the orbital energy of the satellite \citep{Hopkins2008}. The oldest stars might have been affected by earlier mergers, but we do not have the time resolution to uncover those mergers.

To summarize, our observations have uncovered a $\sim$1:10 to 1:15 merger happening $\sim$ 8 Gyr ago, possibly on a low-inclination orbit. This satellite deposited its stars throughout the whole galaxy, but might not have caused significant disc heating in the vertical direction,  did not contribute to the growth of a classical bulge, and is not responsible for the thick and centrally concentrated structure of the very oldest stars in NGC 5746.

A population of low-metallicity accreted stars was also detected by \cite{Pinna2019a,Pinna2019b} in 3 S0 galaxies, where accreted stars represent 4--5\% of the total stellar mass (this was later confirmed by \citealp{Poci2021}). Given that our data and our analysis based on pPXF are very similar to the ones in \cite{Pinna2019a,Pinna2019b}, it would be fair to wonder if the detection of an accreted component could be an artefact of the methods we both use. However, we have found that a younger, more metal-poor component in the thick disc is also recovered when we analyse our spectra with STECKMAP (see Appendix \ref{appendix:steckmap}), which reinforces our confidences in our results. Moreover, finding an accreted component in the discs of galaxies is not surprising: minor mergers are very frequent in $\Lambda$CDM. For instance,  \cite{Stewart2008} show that 70\% of Milky Way-mass galaxies have experienced a 1:10 merger in the last 10 Gyr, and this fraction increases slightly for more massive galaxies. Accreted stars are thus very frequent in the discs of simulated galaxies (e.g., \citealp{Abadi2003,Penarrubia2006,Read2008,Pillepich2015,Park2021}), and 
\cite{Gomez2017} show that 30\% of their simulated disc galaxies contain at least 5\% of accreted stars in their disc, with some discs containing up to 30\% of accreted stars.
\cite{Gomez2017} also show that accreted discs are built from a small (1 to 3) number of merger events with relatively massive galaxies, which would be consistent with what we observe, although we cannot  disentangle mergers happening close together, or at very early times.
Finally, an interaction with a low inclination is also not unusual: even when satellites start out at the virial radius with a very inclined orbit, the orbit tends to slowly align with the disc, both because the satellite's orbit changes, and because the disc tilts as a response to the interaction \citep{Quinn1986,Abadi2003,Ruchti2014,Gomez2017}. This means that the detection of accreted stars in galactic discs is not a surprising result.

\subsection{Thick disc formation}
The overall properties of the thick disc in NGC 5746 are very similar to thick discs observed in other nearby galaxies, in the sense that it is older, more metal-poor and more $\alpha$-rich than the thin disc (see e.g., \citealp{Mould2005, Rejkuba2009,Yoachim2008b,Comeron2015,Comeron2016, Kasparova2016,Kasparova2020, Pinna2019a,Pinna2019b,Scott2021}). This is also true of the thick disc of the Milky Way in the solar neighbourhood, where we observe vertical gradients in age, \Fe and \al \citep{Gilmore1985,Ivezic2008, Bovy2012b,Schlesinger2012,Casagrande2016}.

However, the radial structure of the thick disc in NGC 5746 appears very unlike the Milky Way. In the Milky Way, the thick disc does not show a radial [M/H] gradient, but strong radial gradients in age and \al \citep{Cheng2012, Nidever2014, Hayden2015,Martig2016}. These gradients are built-up from the flares of mono-age populations, populating the inner thick disc with old stars, and the outer thick disc with younger stars \citep{Minchev2015, Garcia2021}. This scenario does not seem to correspond to what we observe in NGC 5746, where the thick disc does not show any radial trends in its mean age, \Fe and \al. It also does not contain stars younger than 8 Gyr (while those stars are present in the thin disc), which means that flared young populations do not significantly contribute to the populations observed in the thick disc (at least within the radial range we explore here, since we cannot rule out a small amount of flaring for young populations in the outer regions of NGC 5746). Such a uniform thick disc is actually expected for galaxies more massive than $\sim 10^{11}$ \msun, as shown by \cite{Garcia2021} --- this uniformity is often connected to mergers.
As discussed in the previous section, the 1:10 merger we have identified is probably not responsible for the vertical heating of  most of the stars now in the thick disc, although it could have contributed to some enhanced radial mixing, and thus to the uniformity we find in the stellar populations. Most of the thick disc was probably already in place $\sim$12 Gyr ago, but we cannot directly identify the sequence of events that drove its formation.

The 1:10 merger still had an important part to play in the formation of the thick disc in the sense that it brought a significant amount of mass. If we sum the mass in all the bins identified as belonging to the thick disc, we find 1.4$\times 10^9$ \msun (this is of course not the total mass of the thick disc). Of those stars, 1$\times 10^9$ \msun are found at \Fe>0, and 0.4$\times 10^9$ \msun at \Fe<0. If we identify the metal-poor component with the accreted component, then we find that 28\% of the thick disc is accreted (see also Figure \ref{fig:frac_accret}). If we only counted the metal-poor stars with ages between 8 and 12 Gyr (a very conservative approach), this would bring the fraction of accreted stars in the thick disc to 14\%.

The metal-poor, $\alpha$-rich accreted stars thus form a significant fraction of the thick disc, and they are also responsible for most of the vertical gradients in \Fe and \al that we see in the maps presented in Figure \ref{fig:pop_maps}, and for the distinct mean \Fe and \al in the thin and thick disc: given the distributions shown in Figure \ref{fig:age_fe_components}, in the absence of accreted stars, the thin and thick disc would have very similar mean \Fe and \al (possibly a slightly higher mean \al for the thick disc).

The presence of a significant accreted component in thick discs was also discussed by \cite{Pinna2019a,Pinna2019b}, finding that accreted populations represent 17\% of the thick disc mass for FCC 177 and FCC 170 and 23\% for FCC 153. These fractions are also consistent with those found for thick discs in cosmological simulations by \cite{Park2021}. Our analysis shows that using stellar populations to identify this accreted component is a promising direction, given that accreted stars might not be easily identifiable in the global kinematics of the thick discs \citep{Gomez2017,Comeron2019}.

\subsection{Dating bar formation}
We can use the properties of stellar populations in the central regions of NGC 5746 to estimate the time of bar formation.
A first indicator of the age of the bar is the age distribution of stars in the bar and B/P bulge, where we mostly find stars older than 8 Gyr, while the thin disc contains younger stars. If the bar had assembled at late times from material belonging to the thin disc, it should contain a small proportion of younger stars. This thus indicates that the bar formed more than 8 Gyr ago. 

Another piece of evidence comes from the age of stars in the nuclear disc. This technique is being used by the TIMER project  \citep{Gadotti2019} to date bar formation in a number of nearby galaxies, following an idea presented in \cite{Gadotti2015}. Nuclear discs are indeed expected to form from bar-driven gas inflows towards the central regions of galaxies, settling around the bar's inner Lindblad resonance \citep{Combes1985,Athanassoula1992,Cole2014,Emsellem2015,Sormani2015}. There is a slight hint in Figure \ref{fig:cumul_mass} that the growth of the nuclear disc is delayed compared to the growth of the bar, which is consistent with the idea of the nuclear disc being built from the inflow of gas along the bar.

The oldest stars in the nuclear disc trace the onset of star formation in that region, and place a lower limit on the time of bar formation \citep{Gadotti2015,Baba2020}. In the case of NGC 5746, the nuclear disc only contains very old stars, from 8 to 14 Gyr old: this suggests that the bar might have already been in place well before $z=2$.  While the case of NGC 5746 is probably extreme, it is not unique. For instance, from the age of stars in the nuclear disc of NGC 4371 (a barred galaxy with a  stellar mass of $10^{10.8}$ \msun), \cite{Gadotti2015} estimate that its bar formed 10 Gyr ago. 
Bars have also been observed in galaxies up to $z\sim2$ \citep{Simmons2014}, and possibly even at $z\sim3$ \citep{Hodge2019} in massive galaxies: massive galaxies form their bars earlier because their discs form (and become dynamically cold) earlier \citep{Sheth2008,Sheth2012,Rosas2020}. Our observations support this picture, and also confirm that  bars can be very long-lived features (see also \citealp{Debattista2006,Curir2008,Kraljic2012,Gadotti2015,Perez2017}).

What is maybe more unexpected is the absence of younger stars in the nuclear disc: NGC 5746 still hosts star formation in its main disc. This probably means that gas inflows towards the bar and the nuclear disc have been strongly suppressed in the last 8 Gyr. This would be consistent with the central hole observed in the HI disc \citep{Rand2008}. One additional explanation could be the effect of the B/P bulge: \cite{Fragkoudi2016} have indeed shown in simulations that when a bar buckles and a peanut is formed, the gravitational potential in the midplane of the bar is reduced, and gas inflows are suppressed. If true, this could place the formation of the B/P bulge about 8 Gyr ago.

To summarize, we find that the bar and nuclear disc in NGC 5746 have been in place for the last $\sim$ 12 Gyr, and must then have formed very soon after the main disc formed. Such an accelerated evolution is probably only possible in the very most massive disc galaxies. 
If this is true, it also means that the 1:10 merger that happened $\sim$8 Gyr ago not only did not heat the pre-existing disc very much (at least not in the vertical direction), and did not create a classical bulge, it also did not destroy the bar and the nuclear disc. Simulations by \cite{Sarzi2015} indeed suggest that nuclear discs are not disrupted during a 1:10 merger; NGC 5746  provides support to this idea.

\section{Conclusion}
Massive disc galaxies that lack a classical bulge have often been presented as a challenge to $\Lambda$CDM given the expected high frequency of mergers, particularly in massive galaxies \citep[e.g.,][]{Kautsch2006,Shen2010,Kormendy2010b}. While mergers do not always result in classical bulges \citep{Gargiulo2019}, they still have on average a strong impact on galaxy morphology. It is thus important to obtain observational constrains on the formation histories of massive disc galaxies.

NGC 5746 is one such massive ($\sim 10^{11}$ \msun), nearly edge-on disc galaxy with no signs of a classical bulge \citep{Barentine2012}: in its central regions, it contains a nuclear disc, a bar, and a B/P bulge. The disc itself contains both a thin and a thick component. We obtained MUSE observations of the central region and the inner disc of NGC 5746 and fit the spectra with pPXF to study the kinematics and stellar populations of this galaxy, and to uncover its formation history. 

The kinematic maps highlight the different components of NGC 5746: we have identified kinematic signatures of the nuclear disc, bar and B/P bulge that had already been observed by \cite{Falcon2003}, \cite{Chung2004}, and \cite{Molaeinezhad2016}. We also extend previous kinematic analyses into the disc region, where we find a velocity dispersion decreasing with radius and increasing with height above the midplane, and the corresponding asymmetric drift in the rotation curve.

We find that the thin disc is overall younger, more metal-rich, and more $\alpha$-poor than the thick disc. The bar and B/P bulge are old, metal-rich and slightly more $\alpha$-enhanced than the thin disc. Finally, the nuclear disc is old, metal-rich and $\alpha$-poor. There are no very obvious radial gradients in the properties of each component. 

When we look in more detail into the populations in each component, we find that the thin disc has an extended star formation history. Most of the stars are older than 8 Gyr (with possibly a peak of star formation around 10 Gyr), but the thin disc also shows more recent star formation, from about 4-5 Gyr ago down to the current time.
The thick disc hosts complex stellar populations: a main component that is metal-rich and very old (typically above 12 Gyr), and a second component that is more metal-poor (\Fe between -1 and 0 dex), more $\alpha$-rich, and with an age distribution  roughly flat from 8 to 14 Gyr.
This  is not consistent with the global chemical evolution of the disc, and probably corresponds to stars accreted from a smaller satellite galaxy.
Finally, the B/P bulge is showing stellar populations that are very similar to the thick disc, and the nuclear disc is showing some of the simplest stellar populations in the galaxy: nearly all of the stars are older than 10 Gyr.

Piecing together all the available evidence, we find that a massive and extended disc formed very early on (80\% of the stellar mass formed before $z=2$). The main component of the thick disc might have formed already thick, for instance following an intense phase of gas-rich mergers \citep{Brook2004}, but this is not something that we can test. This massive disc developed a bar very fast, which drove gas towards the center of the galaxy and triggered the formation of the nuclear disc. At some point after the bar formed, a B/P bulge formed, either following the buckling of the bar, or because of vertical resonances. Around $\sim$ 8 Gyr ago,  a $\sim$1:10 merger happened, possibly on a low-inclination orbit. The satellite deposited its stars throughout the whole galaxy, but might not have caused significant disc heating  in the vertical direction and did not contribute to the growth of a classical bulge. It also did not destroy the bar and the nuclear disc: NGC 5746 is an example of a galaxy with a very long-lived bar. The merger still had an important part to play in the formation of the thick disc in the sense that it brought a significant amount of mass: $\sim 30$\% of the thick disc is made of accreted stars. The merger also possibly had an effect on the radial re-distribution of stars within the disc, which would explain the absence of radial metallicity gradients in the thin and thick discs. After this, we do not detect signs of significant other mergers, and star formation continues within the thin disc down to the present time.

It thus seems that NGC 5746 did not completely escape mergers, but that the only relatively recent significant merger did not create much damage to the galaxy (while still contributing $~\sim 30$\% of accreted stars to the thick disc). This is still a relatively quiescent merger history, with no merger detected in the past $\sim 8$ Gyr. \cite{Jackson2020} argue that this might be the case for $~\sim 30$\% of massive disc galaxies: an unusually quiet merger history allows them to form a disc early and maintain it down to $z=0$. Further studies of nearby disc galaxies will be needed to establish this from an observational point of view, and to establish how unusual or common NGC 5746's formation history might be.

\section*{Acknowledgements}
We thank Ryan Leaman for useful discussions, and the referee for suggestions that improved the paper.
FP and JFB acknowledge support from grant
PID2019-107427GB-C32 from the Spanish Ministry of Science and Innovation. JFB  also acknowledges support through the IAC project TRACES which is partially supported through the state budget and the regional budget of the Consejer\'ia de Econom\'ia, Industria, Comercio y Conocimiento of the Canary Islands Autonomous Community. The Science, Technology and Facilities Council is acknowledged by JN for support through the Consolidated Grant Cosmology and Astrophysics at Portsmouth, ST/S000550/1. TRL acknowledges support from a Spinoza grant (NWO) awarded to A. Helmi. 
Based on observations made with ESO Telescopes at the La Silla Paranal Observatory under programme ID 095.B-0760(A)

\section*{Data Availability}

The MUSE data are accessible via the ESO archive.



\bibliographystyle{mnras}
\bibliography{library_5746}

\begin{thebibliography}{}
\makeatletter
\relax
\def\mn@urlcharsother{\let\do\@makeother \do\$\do\&\do\#\do\^\do\_\do\%\do\~}
\def\mn@doi{\begingroup\mn@urlcharsother \@ifnextchar [ {\mn@doi@}
  {\mn@doi@[]}}
\def\mn@doi@[#1]#2{\def\@tempa{#1}\ifx\@tempa\@empty \href
  {http://dx.doi.org/#2} {doi:#2}\else \href {http://dx.doi.org/#2} {#1}\fi
  \endgroup}
\def\mn@eprint#1#2{\mn@eprint@#1:#2::\@nil}
\def\mn@eprint@arXiv#1{\href {http://arxiv.org/abs/#1} {{\tt arXiv:#1}}}
\def\mn@eprint@dblp#1{\href {http://dblp.uni-trier.de/rec/bibtex/#1.xml}
  {dblp:#1}}
\def\mn@eprint@#1:#2:#3:#4\@nil{\def\@tempa {#1}\def\@tempb {#2}\def\@tempc
  {#3}\ifx \@tempc \@empty \let \@tempc \@tempb \let \@tempb \@tempa \fi \ifx
  \@tempb \@empty \def\@tempb {arXiv}\fi \@ifundefined
  {mn@eprint@\@tempb}{\@tempb:\@tempc}{\expandafter \expandafter \csname
  mn@eprint@\@tempb\endcsname \expandafter{\@tempc}}}

\bibitem[\protect\citeauthoryear{{Abadi}, {Navarro}, {Steinmetz}  \&
  {Eke}}{{Abadi} et~al.}{2003}]{Abadi2003}
{Abadi} M.~G.,  {Navarro} J.~F.,  {Steinmetz} M.,   {Eke} V.~R.,  2003, \mn@doi
  [\apj] {10.1086/378316}, \href
  {https://ui.adsabs.harvard.edu/abs/2003ApJ...597...21A} {597, 21}

\bibitem[\protect\citeauthoryear{{Aguerri}, {Balcells}  \&
  {Peletier}}{{Aguerri} et~al.}{2001}]{Aguerri2001}
{Aguerri} J.~A.~L.,  {Balcells} M.,   {Peletier} R.~F.,  2001, \mn@doi [\aap]
  {10.1051/0004-6361:20000441}, \href
  {https://ui.adsabs.harvard.edu/abs/2001A&A...367..428A} {367, 428}

\bibitem[\protect\citeauthoryear{{Alabi}, {Forbes}, {Romanowsky}  \&
  {Brodie}}{{Alabi} et~al.}{2020}]{Alabi2020}
{Alabi} A.~B.,  {Forbes} D.~A.,  {Romanowsky} A.~J.,   {Brodie} J.~P.,  2020,
  \mn@doi [\mnras] {10.1093/mnras/stz3382}, \href
  {https://ui.adsabs.harvard.edu/abs/2020MNRAS.491.5693A} {491, 5693}

\bibitem[\protect\citeauthoryear{{Athanassoula}}{{Athanassoula}}{1992}]{Athanassoula1992}
{Athanassoula} E.,  1992, \mn@doi [\mnras] {10.1093/mnras/259.2.328}, \href
  {https://ui.adsabs.harvard.edu/abs/1992MNRAS.259..328A} {259, 328}

\bibitem[\protect\citeauthoryear{{Athanassoula}}{{Athanassoula}}{2005}]{Athanassoula2005}
{Athanassoula} E.,  2005, \mn@doi [\mnras] {10.1111/j.1365-2966.2005.08872.x},
  \href {https://ui.adsabs.harvard.edu/abs/2005MNRAS.358.1477A} {358, 1477}

\bibitem[\protect\citeauthoryear{{Baba} \& {Kawata}}{{Baba} \&
  {Kawata}}{2020}]{Baba2020}
{Baba} J.,  {Kawata} D.,  2020, \mn@doi [\mnras] {10.1093/mnras/staa140}, \href
  {https://ui.adsabs.harvard.edu/abs/2020MNRAS.492.4500B} {492, 4500}

\bibitem[\protect\citeauthoryear{{Bacon} et~al.,}{{Bacon}
  et~al.}{2010}]{Bacon2010}
{Bacon} R.,  et~al., 2010, in {McLean} I.~S.,  {Ramsay} S.~K.,   {Takami} H.,
  eds,  Society of Photo-Optical Instrumentation Engineers (SPIE) Conference
  Series Vol. 7735, Ground-based and Airborne Instrumentation for Astronomy
  III. p. 773508, \mn@doi{10.1117/12.856027}

\bibitem[\protect\citeauthoryear{{Bailin}, {Bell}, {Chappell}, {Radburn-Smith}
  \& {de Jong}}{{Bailin} et~al.}{2011}]{Bailin2011}
{Bailin} J.,  {Bell} E.~F.,  {Chappell} S.~N.,  {Radburn-Smith} D.~J.,   {de
  Jong} R.~S.,  2011, \mn@doi [\apj] {10.1088/0004-637X/736/1/24}, \href
  {https://ui.adsabs.harvard.edu/abs/2011ApJ...736...24B} {736, 24}

\bibitem[\protect\citeauthoryear{{Balcells}, {Graham}  \&
  {Peletier}}{{Balcells} et~al.}{2007}]{Balcells2007}
{Balcells} M.,  {Graham} A.~W.,   {Peletier} R.~F.,  2007, \mn@doi [\apj]
  {10.1086/519752}, \href
  {https://ui.adsabs.harvard.edu/abs/2007ApJ...665.1084B} {665, 1084}

\bibitem[\protect\citeauthoryear{{Barentine} \& {Kormendy}}{{Barentine} \&
  {Kormendy}}{2012}]{Barentine2012}
{Barentine} J.~C.,  {Kormendy} J.,  2012, \mn@doi [\apj]
  {10.1088/0004-637X/754/2/140}, \href
  {https://ui.adsabs.harvard.edu/abs/2012ApJ...754..140B} {754, 140}

\bibitem[\protect\citeauthoryear{{Beasley}, {Trujillo}, {Leaman}  \&
  {Montes}}{{Beasley} et~al.}{2018}]{Beasley2018}
{Beasley} M.~A.,  {Trujillo} I.,  {Leaman} R.,   {Montes} M.,  2018, \mn@doi
  [\nat] {10.1038/nature25756}, \href
  {https://ui.adsabs.harvard.edu/abs/2018Natur.555..483B} {555, 483}

\bibitem[\protect\citeauthoryear{{Bell}, {Monachesi}, {Harmsen}, {de Jong},
  {Bailin}, {Radburn-Smith}, {D'Souza}  \& {Holwerda}}{{Bell}
  et~al.}{2017}]{Bell2017}
{Bell} E.~F.,  {Monachesi} A.,  {Harmsen} B.,  {de Jong} R.~S.,  {Bailin} J.,
  {Radburn-Smith} D.~J.,  {D'Souza} R.,   {Holwerda} B.~W.,  2017, \mn@doi
  [\apjl] {10.3847/2041-8213/aa6158}, \href
  {https://ui.adsabs.harvard.edu/abs/2017ApJ...837L...8B} {837, L8}

\bibitem[\protect\citeauthoryear{{Bender}, {Saglia}  \& {Gerhard}}{{Bender}
  et~al.}{1994}]{Bender1994}
{Bender} R.,  {Saglia} R.~P.,   {Gerhard} O.~E.,  1994, \mn@doi [\mnras]
  {10.1093/mnras/269.3.785}, \href
  {https://ui.adsabs.harvard.edu/abs/1994MNRAS.269..785B} {269, 785}

\bibitem[\protect\citeauthoryear{{Bianchi}}{{Bianchi}}{2007}]{Bianchi2007}
{Bianchi} S.,  2007, \mn@doi [\aap] {10.1051/0004-6361:20077649}, \href
  {https://ui.adsabs.harvard.edu/abs/2007A&A...471..765B} {471, 765}

\bibitem[\protect\citeauthoryear{{Bien}, {Brandt}  \& {Just}}{{Bien}
  et~al.}{2013}]{Bien2013}
{Bien} R.,  {Brandt} T.,   {Just} A.,  2013, \mn@doi [\mnras]
  {10.1093/mnras/sts141}, \href
  {https://ui.adsabs.harvard.edu/abs/2013MNRAS.428.1631B} {428, 1631}

\bibitem[\protect\citeauthoryear{{Blom}, {Forbes}, {Brodie}, {Foster},
  {Romanowsky}, {Spitler}  \& {Strader}}{{Blom} et~al.}{2012}]{Blom2012}
{Blom} C.,  {Forbes} D.~A.,  {Brodie} J.~P.,  {Foster} C.,  {Romanowsky} A.~J.,
   {Spitler} L.~R.,   {Strader} J.,  2012, \mn@doi [\mnras]
  {10.1111/j.1365-2966.2012.21795.x}, \href
  {https://ui.adsabs.harvard.edu/abs/2012MNRAS.426.1959B} {426, 1959}

\bibitem[\protect\citeauthoryear{{Boecker}, {Leaman}, {van de Ven}, {Norris},
  {Mackereth}  \& {Crain}}{{Boecker} et~al.}{2020}]{Boecker2020}
{Boecker} A.,  {Leaman} R.,  {van de Ven} G.,  {Norris} M.~A.,  {Mackereth}
  J.~T.,   {Crain} R.~A.,  2020, \mn@doi [\mnras] {10.1093/mnras/stz3077},
  \href {https://ui.adsabs.harvard.edu/abs/2020MNRAS.491..823B} {491, 823}

\bibitem[\protect\citeauthoryear{{Bournaud}, {Elmegreen}  \&
  {Elmegreen}}{{Bournaud} et~al.}{2007}]{Bournaud2007}
{Bournaud} F.,  {Elmegreen} B.~G.,   {Elmegreen} D.~M.,  2007, \mn@doi [\apj]
  {10.1086/522077}, \href
  {https://ui.adsabs.harvard.edu/abs/2007ApJ...670..237B} {670, 237}

\bibitem[\protect\citeauthoryear{{Bournaud}, {Elmegreen}  \&
  {Martig}}{{Bournaud} et~al.}{2009}]{Bournaud2009}
{Bournaud} F.,  {Elmegreen} B.~G.,   {Martig} M.,  2009, \mn@doi [\apjl]
  {10.1088/0004-637X/707/1/L1}, \href
  {https://ui.adsabs.harvard.edu/abs/2009ApJ...707L...1B} {707, L1}

\bibitem[\protect\citeauthoryear{{Bovy}, {Rix}  \& {Hogg}}{{Bovy}
  et~al.}{2012a}]{Bovy2012a}
{Bovy} J.,  {Rix} H.-W.,   {Hogg} D.~W.,  2012a, \mn@doi [\apj]
  {10.1088/0004-637X/751/2/131}, \href
  {https://ui.adsabs.harvard.edu/abs/2012ApJ...751..131B} {751, 131}

\bibitem[\protect\citeauthoryear{{Bovy}, {Rix}, {Liu}, {Hogg}, {Beers}  \&
  {Lee}}{{Bovy} et~al.}{2012b}]{Bovy2012b}
{Bovy} J.,  {Rix} H.-W.,  {Liu} C.,  {Hogg} D.~W.,  {Beers} T.~C.,   {Lee}
  Y.~S.,  2012b, \mn@doi [\apj] {10.1088/0004-637X/753/2/148}, \href
  {https://ui.adsabs.harvard.edu/abs/2012ApJ...753..148B} {753, 148}

\bibitem[\protect\citeauthoryear{{Brook}, {Kawata}, {Gibson}  \&
  {Freeman}}{{Brook} et~al.}{2004}]{Brook2004}
{Brook} C.~B.,  {Kawata} D.,  {Gibson} B.~K.,   {Freeman} K.~C.,  2004, \mn@doi
  [\apj] {10.1086/422709}, \href
  {https://ui.adsabs.harvard.edu/abs/2004ApJ...612..894B} {612, 894}

\bibitem[\protect\citeauthoryear{{Brook}, {Gibson}, {Martel}  \&
  {Kawata}}{{Brook} et~al.}{2005}]{Brook2005}
{Brook} C.~B.,  {Gibson} B.~K.,  {Martel} H.,   {Kawata} D.,  2005, \mn@doi
  [\apj] {10.1086/431924}, \href
  {https://ui.adsabs.harvard.edu/abs/2005ApJ...630..298B} {630, 298}

\bibitem[\protect\citeauthoryear{{Brook}, {Richard}, {Kawata}, {Martel}  \&
  {Gibson}}{{Brook} et~al.}{2007}]{Brook2007}
{Brook} C.,  {Richard} S.,  {Kawata} D.,  {Martel} H.,   {Gibson} B.~K.,  2007,
  \mn@doi [\apj] {10.1086/511056}, \href
  {https://ui.adsabs.harvard.edu/abs/2007ApJ...658...60B} {658, 60}

\bibitem[\protect\citeauthoryear{{Bureau} \& {Athanassoula}}{{Bureau} \&
  {Athanassoula}}{2005}]{Bureau2005}
{Bureau} M.,  {Athanassoula} E.,  2005, \mn@doi [\apj] {10.1086/430056}, \href
  {https://ui.adsabs.harvard.edu/abs/2005ApJ...626..159B} {626, 159}

\bibitem[\protect\citeauthoryear{{Bureau} \& {Freeman}}{{Bureau} \&
  {Freeman}}{1999}]{Bureau1999}
{Bureau} M.,  {Freeman} K.~C.,  1999, \mn@doi [\aj] {10.1086/300922}, \href
  {https://ui.adsabs.harvard.edu/abs/1999AJ....118..126B} {118, 126}

\bibitem[\protect\citeauthoryear{{Buta} et~al.,}{{Buta}
  et~al.}{2015}]{Buta2015}
{Buta} R.~J.,  et~al., 2015, \mn@doi [\apjs] {10.1088/0067-0049/217/2/32},
  \href {https://ui.adsabs.harvard.edu/abs/2015ApJS..217...32B} {217, 32}

\bibitem[\protect\citeauthoryear{{Cappellari}}{{Cappellari}}{2017}]{Cappellari2017}
{Cappellari} M.,  2017, \mn@doi [\mnras] {10.1093/mnras/stw3020}, \href
  {https://ui.adsabs.harvard.edu/abs/2017MNRAS.466..798C} {466, 798}

\bibitem[\protect\citeauthoryear{{Cappellari} \& {Copin}}{{Cappellari} \&
  {Copin}}{2003}]{Cappellari2003}
{Cappellari} M.,  {Copin} Y.,  2003, \mn@doi [\mnras]
  {10.1046/j.1365-8711.2003.06541.x}, \href
  {https://ui.adsabs.harvard.edu/abs/2003MNRAS.342..345C} {342, 345}

\bibitem[\protect\citeauthoryear{{Cappellari} \& {Emsellem}}{{Cappellari} \&
  {Emsellem}}{2004}]{Cappellari2004}
{Cappellari} M.,  {Emsellem} E.,  2004, \mn@doi [\pasp] {10.1086/381875}, \href
  {https://ui.adsabs.harvard.edu/abs/2004PASP..116..138C} {116, 138}

\bibitem[\protect\citeauthoryear{{Casagrande} et~al.,}{{Casagrande}
  et~al.}{2016}]{Casagrande2016}
{Casagrande} L.,  et~al., 2016, \mn@doi [\mnras] {10.1093/mnras/stv2320}, \href
  {https://ui.adsabs.harvard.edu/abs/2016MNRAS.455..987C} {455, 987}

\bibitem[\protect\citeauthoryear{{Ceverino}, {Dekel}, {Tweed}  \&
  {Primack}}{{Ceverino} et~al.}{2015}]{Ceverino2015}
{Ceverino} D.,  {Dekel} A.,  {Tweed} D.,   {Primack} J.,  2015, \mn@doi
  [\mnras] {10.1093/mnras/stu2694}, \href
  {https://ui.adsabs.harvard.edu/abs/2015MNRAS.447.3291C} {447, 3291}

\bibitem[\protect\citeauthoryear{{Cheng} et~al.,}{{Cheng}
  et~al.}{2012}]{Cheng2012}
{Cheng} J.~Y.,  et~al., 2012, \mn@doi [\apj] {10.1088/0004-637X/746/2/149},
  \href {https://ui.adsabs.harvard.edu/abs/2012ApJ...746..149C} {746, 149}

\bibitem[\protect\citeauthoryear{{Chung} \& {Bureau}}{{Chung} \&
  {Bureau}}{2004}]{Chung2004}
{Chung} A.,  {Bureau} M.,  2004, \mn@doi [\aj] {10.1086/420988}, \href
  {https://ui.adsabs.harvard.edu/abs/2004AJ....127.3192C} {127, 3192}

\bibitem[\protect\citeauthoryear{{Clark} et~al.,}{{Clark}
  et~al.}{2018}]{Clark2018}
{Clark} C.~J.~R.,  et~al., 2018, \mn@doi [\aap] {10.1051/0004-6361/201731419},
  \href {https://ui.adsabs.harvard.edu/abs/2018A&A...609A..37C} {609, A37}

\bibitem[\protect\citeauthoryear{{Coccato} et~al.,}{{Coccato}
  et~al.}{2015}]{Coccato2015}
{Coccato} L.,  et~al., 2015, \mn@doi [\aap] {10.1051/0004-6361/201526560},
  \href {https://ui.adsabs.harvard.edu/abs/2015A&A...581A..65C} {581, A65}

\bibitem[\protect\citeauthoryear{{Cole}, {Debattista}, {Erwin}, {Earp}  \&
  {Ro{\v{s}}kar}}{{Cole} et~al.}{2014}]{Cole2014}
{Cole} D.~R.,  {Debattista} V.~P.,  {Erwin} P.,  {Earp} S. W.~F.,
  {Ro{\v{s}}kar} R.,  2014, \mn@doi [\mnras] {10.1093/mnras/stu1985}, \href
  {https://ui.adsabs.harvard.edu/abs/2014MNRAS.445.3352C} {445, 3352}

\bibitem[\protect\citeauthoryear{{Combes} \& {Gerin}}{{Combes} \&
  {Gerin}}{1985}]{Combes1985}
{Combes} F.,  {Gerin} M.,  1985, \aap, \href
  {https://ui.adsabs.harvard.edu/abs/1985A&A...150..327C} {150, 327}

\bibitem[\protect\citeauthoryear{{Combes}, {Debbasch}, {Friedli}  \&
  {Pfenniger}}{{Combes} et~al.}{1990}]{Combes1990}
{Combes} F.,  {Debbasch} F.,  {Friedli} D.,   {Pfenniger} D.,  1990, \aap,
  \href {https://ui.adsabs.harvard.edu/abs/1990A&A...233...82C} {233, 82}

\bibitem[\protect\citeauthoryear{{Comer{\'o}n} et~al.,}{{Comer{\'o}n}
  et~al.}{2014}]{Comeron2014}
{Comer{\'o}n} S.,  et~al., 2014, \mn@doi [\aap] {10.1051/0004-6361/201321633},
  \href {https://ui.adsabs.harvard.edu/abs/2014A&A...562A.121C} {562, A121}

\bibitem[\protect\citeauthoryear{{Comer{\'o}n}, {Salo}, {Janz}, {Laurikainen}
  \& {Yoachim}}{{Comer{\'o}n} et~al.}{2015}]{Comeron2015}
{Comer{\'o}n} S.,  {Salo} H.,  {Janz} J.,  {Laurikainen} E.,   {Yoachim} P.,
  2015, \mn@doi [\aap] {10.1051/0004-6361/201526815}, \href
  {https://ui.adsabs.harvard.edu/abs/2015A&A...584A..34C} {584, A34}

\bibitem[\protect\citeauthoryear{{Comer{\'o}n}, {Salo}, {Peletier}  \&
  {Mentz}}{{Comer{\'o}n} et~al.}{2016}]{Comeron2016}
{Comer{\'o}n} S.,  {Salo} H.,  {Peletier} R.~F.,   {Mentz} J.,  2016, \mn@doi
  [\aap] {10.1051/0004-6361/201629292}, \href
  {https://ui.adsabs.harvard.edu/abs/2016A&A...593L...6C} {593, L6}

\bibitem[\protect\citeauthoryear{{Comer{\'o}n}, {Salo}  \&
  {Knapen}}{{Comer{\'o}n} et~al.}{2018}]{Comeron2018}
{Comer{\'o}n} S.,  {Salo} H.,   {Knapen} J.~H.,  2018, \mn@doi [\aap]
  {10.1051/0004-6361/201731415}, \href
  {https://ui.adsabs.harvard.edu/abs/2018A&A...610A...5C} {610, A5}

\bibitem[\protect\citeauthoryear{{Comer{\'o}n}, {Salo}, {Knapen}  \&
  {Peletier}}{{Comer{\'o}n} et~al.}{2019}]{Comeron2019}
{Comer{\'o}n} S.,  {Salo} H.,  {Knapen} J.~H.,   {Peletier} R.~F.,  2019,
  \mn@doi [\aap] {10.1051/0004-6361/201833653}, \href
  {https://ui.adsabs.harvard.edu/abs/2019A&A...623A..89C} {623, A89}

\bibitem[\protect\citeauthoryear{{Crocker}, {Jeong}, {Komugi}, {Combes},
  {Bureau}, {Young}  \& {Yi}}{{Crocker} et~al.}{2009}]{Crocker2009}
{Crocker} A.~F.,  {Jeong} H.,  {Komugi} S.,  {Combes} F.,  {Bureau} M.,
  {Young} L.~M.,   {Yi} S.,  2009, \mn@doi [\mnras]
  {10.1111/j.1365-2966.2008.14295.x}, \href
  {https://ui.adsabs.harvard.edu/abs/2009MNRAS.393.1255C} {393, 1255}

\bibitem[\protect\citeauthoryear{{Curir}, {Mazzei}  \& {Murante}}{{Curir}
  et~al.}{2008}]{Curir2008}
{Curir} A.,  {Mazzei} P.,   {Murante} G.,  2008, \mn@doi [\aap]
  {10.1051/0004-6361:20078285}, \href
  {https://ui.adsabs.harvard.edu/abs/2008A&A...481..651C} {481, 651}

\bibitem[\protect\citeauthoryear{{D'Souza} \& {Bell}}{{D'Souza} \&
  {Bell}}{2018}]{DSouza2018}
{D'Souza} R.,  {Bell} E.~F.,  2018, \mn@doi [\mnras] {10.1093/mnras/stx3081},
  \href {https://ui.adsabs.harvard.edu/abs/2018MNRAS.474.5300D} {474, 5300}

\bibitem[\protect\citeauthoryear{{Dalcanton} \& {Bernstein}}{{Dalcanton} \&
  {Bernstein}}{2002}]{Dalcanton2002}
{Dalcanton} J.~J.,  {Bernstein} R.~A.,  2002, \mn@doi [\aj] {10.1086/342286},
  \href {https://ui.adsabs.harvard.edu/abs/2002AJ....124.1328D} {124, 1328}

\bibitem[\protect\citeauthoryear{{Davison}, {Norris}, {Pfeffer}, {Davies}  \&
  {Crain}}{{Davison} et~al.}{2020}]{Davison2020}
{Davison} T.~A.,  {Norris} M.~A.,  {Pfeffer} J.~L.,  {Davies} J.~J.,   {Crain}
  R.~A.,  2020, \mn@doi [\mnras] {10.1093/mnras/staa1816}, \href
  {https://ui.adsabs.harvard.edu/abs/2020MNRAS.497...81D} {497, 81}

\bibitem[\protect\citeauthoryear{{Davison} et~al.,}{{Davison}
  et~al.}{2021}]{Davison2021}
{Davison} T.~A.,  et~al., 2021, \mn@doi [\mnras] {10.1093/mnras/stab162}, \href
  {https://ui.adsabs.harvard.edu/abs/2021MNRAS.502.2296D} {502, 2296}

\bibitem[\protect\citeauthoryear{{De Vis} et~al.,}{{De Vis}
  et~al.}{2019}]{DeVis2019}
{De Vis} P.,  et~al., 2019, \mn@doi [\aap] {10.1051/0004-6361/201834444}, \href
  {https://ui.adsabs.harvard.edu/abs/2019A&A...623A...5D} {623, A5}

\bibitem[\protect\citeauthoryear{{Debattista}, {Carollo}, {Mayer}  \&
  {Moore}}{{Debattista} et~al.}{2004}]{Debattista2004}
{Debattista} V.~P.,  {Carollo} C.~M.,  {Mayer} L.,   {Moore} B.,  2004, \mn@doi
  [\apjl] {10.1086/386332}, \href
  {https://ui.adsabs.harvard.edu/abs/2004ApJ...604L..93D} {604, L93}

\bibitem[\protect\citeauthoryear{{Debattista}, {Mayer}, {Carollo}, {Moore},
  {Wadsley}  \& {Quinn}}{{Debattista} et~al.}{2006}]{Debattista2006}
{Debattista} V.~P.,  {Mayer} L.,  {Carollo} C.~M.,  {Moore} B.,  {Wadsley} J.,
   {Quinn} T.,  2006, \mn@doi [\apj] {10.1086/504147}, \href
  {https://ui.adsabs.harvard.edu/abs/2006ApJ...645..209D} {645, 209}

\bibitem[\protect\citeauthoryear{{Di Cintio}, {Mostoghiu}, {Knebe}  \&
  {Navarro}}{{Di Cintio} et~al.}{2021}]{DiCintio2021}
{Di Cintio} A.,  {Mostoghiu} R.,  {Knebe} A.,   {Navarro} J.~F.,  2021, \mn@doi
  [\mnras] {10.1093/mnras/stab1682}, \href
  {https://ui.adsabs.harvard.edu/abs/2021MNRAS.506..531D} {506, 531}

\bibitem[\protect\citeauthoryear{{Donohoe-Keyes}, {Martig}, {James}  \&
  {Kraljic}}{{Donohoe-Keyes} et~al.}{2019}]{Donohoe2019}
{Donohoe-Keyes} C.~E.,  {Martig} M.,  {James} P.~A.,   {Kraljic} K.,  2019,
  \mn@doi [\mnras] {10.1093/mnras/stz2474}, \href
  {https://ui.adsabs.harvard.edu/abs/2019MNRAS.489.4992D} {489, 4992}

\bibitem[\protect\citeauthoryear{{Duc} et~al.,}{{Duc} et~al.}{2015}]{Duc2015}
{Duc} P.-A.,  et~al., 2015, \mn@doi [\mnras] {10.1093/mnras/stu2019}, \href
  {https://ui.adsabs.harvard.edu/abs/2015MNRAS.446..120D} {446, 120}

\bibitem[\protect\citeauthoryear{{Elias}, {Sales}, {Creasey}, {Cooper},
  {Bullock}, {Rich}  \& {Hernquist}}{{Elias} et~al.}{2018}]{Elias2018}
{Elias} L.~M.,  {Sales} L.~V.,  {Creasey} P.,  {Cooper} M.~C.,  {Bullock}
  J.~S.,  {Rich} R.~M.,   {Hernquist} L.,  2018, \mn@doi [\mnras]
  {10.1093/mnras/sty1718}, \href
  {https://ui.adsabs.harvard.edu/abs/2018MNRAS.479.4004E} {479, 4004}

\bibitem[\protect\citeauthoryear{{Emsellem}, {Renaud}, {Bournaud}, {Elmegreen},
  {Combes}  \& {Gabor}}{{Emsellem} et~al.}{2015}]{Emsellem2015}
{Emsellem} E.,  {Renaud} F.,  {Bournaud} F.,  {Elmegreen} B.,  {Combes} F.,
  {Gabor} J.~M.,  2015, \mn@doi [\mnras] {10.1093/mnras/stu2209}, \href
  {https://ui.adsabs.harvard.edu/abs/2015MNRAS.446.2468E} {446, 2468}

\bibitem[\protect\citeauthoryear{{Erwin}}{{Erwin}}{2015}]{Erwin2015}
{Erwin} P.,  2015, \mn@doi [\apj] {10.1088/0004-637X/799/2/226}, \href
  {https://ui.adsabs.harvard.edu/abs/2015ApJ...799..226E} {799, 226}

\bibitem[\protect\citeauthoryear{{Erwin} et~al.,}{{Erwin}
  et~al.}{2015}]{Erwin2015b}
{Erwin} P.,  et~al., 2015, \mn@doi [\mnras] {10.1093/mnras/stu2376}, \href
  {https://ui.adsabs.harvard.edu/abs/2015MNRAS.446.4039E} {446, 4039}

\bibitem[\protect\citeauthoryear{{Falc{\'o}n-Barroso}, {Balcells}, {Peletier}
  \& {Vazdekis}}{{Falc{\'o}n-Barroso} et~al.}{2003}]{Falcon2003}
{Falc{\'o}n-Barroso} J.,  {Balcells} M.,  {Peletier} R.~F.,   {Vazdekis} A.,
  2003, \mn@doi [\aap] {10.1051/0004-6361:20030470}, \href
  {https://ui.adsabs.harvard.edu/abs/2003A&A...405..455F} {405, 455}

\bibitem[\protect\citeauthoryear{{Falc{\'o}n-Barroso}
  et~al.,}{{Falc{\'o}n-Barroso} et~al.}{2006}]{Falcon2006}
{Falc{\'o}n-Barroso} J.,  et~al., 2006, \mn@doi [\mnras]
  {10.1111/j.1365-2966.2006.10261.x}, \href
  {https://ui.adsabs.harvard.edu/abs/2006MNRAS.369..529F} {369, 529}

\bibitem[\protect\citeauthoryear{{Falc{\'o}n-Barroso},
  {S{\'a}nchez-Bl{\'a}zquez}, {Vazdekis}, {Ricciardelli}, {Cardiel}, {Cenarro},
  {Gorgas}  \& {Peletier}}{{Falc{\'o}n-Barroso} et~al.}{2011}]{Falcon2011}
{Falc{\'o}n-Barroso} J.,  {S{\'a}nchez-Bl{\'a}zquez} P.,  {Vazdekis} A.,
  {Ricciardelli} E.,  {Cardiel} N.,  {Cenarro} A.~J.,  {Gorgas} J.,
  {Peletier} R.~F.,  2011, \mn@doi [\aap] {10.1051/0004-6361/201116842}, \href
  {https://ui.adsabs.harvard.edu/abs/2011A&A...532A..95F} {532, A95}

\bibitem[\protect\citeauthoryear{{Fensch} et~al.,}{{Fensch}
  et~al.}{2020}]{Fensch2020}
{Fensch} J.,  et~al., 2020, \mn@doi [\aap] {10.1051/0004-6361/202038550}, \href
  {https://ui.adsabs.harvard.edu/abs/2020A&A...644A.164F} {644, A164}

\bibitem[\protect\citeauthoryear{{Fisher} \& {Drory}}{{Fisher} \&
  {Drory}}{2011}]{Fisher2011}
{Fisher} D.~B.,  {Drory} N.,  2011, \mn@doi [\apjl]
  {10.1088/2041-8205/733/2/L47}, \href
  {https://ui.adsabs.harvard.edu/abs/2011ApJ...733L..47F} {733, L47}

\bibitem[\protect\citeauthoryear{{Fisher} \& {Drory}}{{Fisher} \&
  {Drory}}{2016}]{Fisher2016}
{Fisher} D.~B.,  {Drory} N.,  2016, {An Observational Guide to Identifying
  Pseudobulges and Classical Bulges in Disc Galaxies}.
p.~41, \mn@doi{10.1007/978-3-319-19378-6\_3}

\bibitem[\protect\citeauthoryear{{Font}, {McCarthy}, {Le Brun}, {Crain}  \&
  {Kelvin}}{{Font} et~al.}{2017}]{Font2017}
{Font} A.~S.,  {McCarthy} I.~G.,  {Le Brun} A. M.~C.,  {Crain} R.~A.,
  {Kelvin} L.~S.,  2017, \mn@doi [\pasa] {10.1017/pasa.2017.50}, \href
  {https://ui.adsabs.harvard.edu/abs/2017PASA...34...50F} {34, e050}

\bibitem[\protect\citeauthoryear{{Fontanot} et~al.,}{{Fontanot}
  et~al.}{2021}]{Fontanot2021}
{Fontanot} F.,  et~al., 2021, \mn@doi [\mnras] {10.1093/mnras/stab1213}, \href
  {https://ui.adsabs.harvard.edu/abs/2021MNRAS.504.4481F} {504, 4481}

\bibitem[\protect\citeauthoryear{{Foreman-Mackey}, {Hogg}, {Lang}  \&
  {Goodman}}{{Foreman-Mackey} et~al.}{2013}]{Foreman2013}
{Foreman-Mackey} D.,  {Hogg} D.~W.,  {Lang} D.,   {Goodman} J.,  2013, \mn@doi
  [\pasp] {10.1086/670067}, \href
  {https://ui.adsabs.harvard.edu/abs/2013PASP..125..306F} {125, 306}

\bibitem[\protect\citeauthoryear{{Fragkoudi}, {Athanassoula}  \&
  {Bosma}}{{Fragkoudi} et~al.}{2016}]{Fragkoudi2016}
{Fragkoudi} F.,  {Athanassoula} E.,   {Bosma} A.,  2016, \mn@doi [\mnras]
  {10.1093/mnrasl/slw120}, \href
  {https://ui.adsabs.harvard.edu/abs/2016MNRAS.462L..41F} {462, L41}

\bibitem[\protect\citeauthoryear{{Gadotti}}{{Gadotti}}{2009}]{Gadotti2009}
{Gadotti} D.~A.,  2009, \mn@doi [\mnras] {10.1111/j.1365-2966.2008.14257.x},
  \href {https://ui.adsabs.harvard.edu/abs/2009MNRAS.393.1531G} {393, 1531}

\bibitem[\protect\citeauthoryear{{Gadotti}, {Seidel},
  {S{\'a}nchez-Bl{\'a}zquez}, {Falc{\'o}n-Barroso}, {Husemann}, {Coelho}  \&
  {P{\'e}rez}}{{Gadotti} et~al.}{2015}]{Gadotti2015}
{Gadotti} D.~A.,  {Seidel} M.~K.,  {S{\'a}nchez-Bl{\'a}zquez} P.,
  {Falc{\'o}n-Barroso} J.,  {Husemann} B.,  {Coelho} P.,   {P{\'e}rez} I.,
  2015, \mn@doi [\aap] {10.1051/0004-6361/201526677}, \href
  {https://ui.adsabs.harvard.edu/abs/2015A&A...584A..90G} {584, A90}

\bibitem[\protect\citeauthoryear{{Gadotti} et~al.,}{{Gadotti}
  et~al.}{2019}]{Gadotti2019}
{Gadotti} D.~A.,  et~al., 2019, \mn@doi [\mnras] {10.1093/mnras/sty2666}, \href
  {https://ui.adsabs.harvard.edu/abs/2019MNRAS.482..506G} {482, 506}

\bibitem[\protect\citeauthoryear{{Gadotti} et~al.,}{{Gadotti}
  et~al.}{2020}]{Gadotti2020}
{Gadotti} D.~A.,  et~al., 2020, \mn@doi [\aap] {10.1051/0004-6361/202038448},
  \href {https://ui.adsabs.harvard.edu/abs/2020A&A...643A..14G} {643, A14}

\bibitem[\protect\citeauthoryear{{Gallazzi}, {Charlot}, {Brinchmann}, {White}
  \& {Tremonti}}{{Gallazzi} et~al.}{2005}]{Gallazzi2005}
{Gallazzi} A.,  {Charlot} S.,  {Brinchmann} J.,  {White} S. D.~M.,   {Tremonti}
  C.~A.,  2005, \mn@doi [\mnras] {10.1111/j.1365-2966.2005.09321.x}, \href
  {https://ui.adsabs.harvard.edu/abs/2005MNRAS.362...41G} {362, 41}

\bibitem[\protect\citeauthoryear{{Garc{\'\i}a-Benito}
  et~al.,}{{Garc{\'\i}a-Benito} et~al.}{2015}]{Garcia2015}
{Garc{\'\i}a-Benito} R.,  et~al., 2015, \mn@doi [\aap]
  {10.1051/0004-6361/201425080}, \href
  {https://ui.adsabs.harvard.edu/abs/2015A&A...576A.135G} {576, A135}

\bibitem[\protect\citeauthoryear{{Garc{\'\i}a de la Cruz}, {Martig}, {Minchev}
  \& {James}}{{Garc{\'\i}a de la Cruz} et~al.}{2021}]{Garcia2021}
{Garc{\'\i}a de la Cruz} J.,  {Martig} M.,  {Minchev} I.,   {James} P.,  2021,
  \mn@doi [\mnras] {10.1093/mnras/staa3906}, \href
  {https://ui.adsabs.harvard.edu/abs/2021MNRAS.501.5105G} {501, 5105}

\bibitem[\protect\citeauthoryear{{Gargiulo} et~al.,}{{Gargiulo}
  et~al.}{2019}]{Gargiulo2019}
{Gargiulo} I.~D.,  et~al., 2019, \mn@doi [\mnras] {10.1093/mnras/stz2536},
  \href {https://ui.adsabs.harvard.edu/abs/2019MNRAS.489.5742G} {489, 5742}

\bibitem[\protect\citeauthoryear{{Gilmore} \& {Reid}}{{Gilmore} \&
  {Reid}}{1983}]{Gilmore1983}
{Gilmore} G.,  {Reid} N.,  1983, \mn@doi [\mnras] {10.1093/mnras/202.4.1025},
  \href {https://ui.adsabs.harvard.edu/abs/1983MNRAS.202.1025G} {202, 1025}

\bibitem[\protect\citeauthoryear{{Gilmore} \& {Wyse}}{{Gilmore} \&
  {Wyse}}{1985}]{Gilmore1985}
{Gilmore} G.,  {Wyse} R.~F.~G.,  1985, \mn@doi [\aj] {10.1086/113907}, \href
  {https://ui.adsabs.harvard.edu/abs/1985AJ.....90.2015G} {90, 2015}

\bibitem[\protect\citeauthoryear{{Girardi}, {Bressan}, {Bertelli}  \&
  {Chiosi}}{{Girardi} et~al.}{2000}]{Girardi2000}
{Girardi} L.,  {Bressan} A.,  {Bertelli} G.,   {Chiosi} C.,  2000, \mn@doi
  [\aaps] {10.1051/aas:2000126}, \href
  {https://ui.adsabs.harvard.edu/abs/2000A&AS..141..371G} {141, 371}

\bibitem[\protect\citeauthoryear{{G{\'o}mez} et~al.,}{{G{\'o}mez}
  et~al.}{2017}]{Gomez2017}
{G{\'o}mez} F.~A.,  et~al., 2017, \mn@doi [\mnras] {10.1093/mnras/stx2149},
  \href {https://ui.adsabs.harvard.edu/abs/2017MNRAS.472.3722G} {472, 3722}

\bibitem[\protect\citeauthoryear{{Gonz{\'a}lez Delgado} et~al.,}{{Gonz{\'a}lez
  Delgado} et~al.}{2014}]{Delgado2014}
{Gonz{\'a}lez Delgado} R.~M.,  et~al., 2014, \mn@doi [\apjl]
  {10.1088/2041-8205/791/1/L16}, \href
  {https://ui.adsabs.harvard.edu/abs/2014ApJ...791L..16G} {791, L16}

\bibitem[\protect\citeauthoryear{{Harmsen}, {Monachesi}, {Bell}, {de Jong},
  {Bailin}, {Radburn-Smith}  \& {Holwerda}}{{Harmsen}
  et~al.}{2017}]{Harmsen2017}
{Harmsen} B.,  {Monachesi} A.,  {Bell} E.~F.,  {de Jong} R.~S.,  {Bailin} J.,
  {Radburn-Smith} D.~J.,   {Holwerda} B.~W.,  2017, \mn@doi [\mnras]
  {10.1093/mnras/stw2992}, \href
  {https://ui.adsabs.harvard.edu/abs/2017MNRAS.466.1491H} {466, 1491}

\bibitem[\protect\citeauthoryear{{Hayden} et~al.,}{{Hayden}
  et~al.}{2015}]{Hayden2015}
{Hayden} M.~R.,  et~al., 2015, \mn@doi [\apj] {10.1088/0004-637X/808/2/132},
  \href {https://ui.adsabs.harvard.edu/abs/2015ApJ...808..132H} {808, 132}

\bibitem[\protect\citeauthoryear{{Hodge} et~al.,}{{Hodge}
  et~al.}{2019}]{Hodge2019}
{Hodge} J.~A.,  et~al., 2019, \mn@doi [\apj] {10.3847/1538-4357/ab1846}, \href
  {https://ui.adsabs.harvard.edu/abs/2019ApJ...876..130H} {876, 130}

\bibitem[\protect\citeauthoryear{{Hood}, {Kannappan}, {Stark}, {Dell'Antonio},
  {Moffett}, {Eckert}, {Norris}  \& {Hendel}}{{Hood} et~al.}{2018}]{Hood2018}
{Hood} C.~E.,  {Kannappan} S.~J.,  {Stark} D.~V.,  {Dell'Antonio} I.~P.,
  {Moffett} A.~J.,  {Eckert} K.~D.,  {Norris} M.~A.,   {Hendel} D.,  2018,
  \mn@doi [\apj] {10.3847/1538-4357/aab719}, \href
  {https://ui.adsabs.harvard.edu/abs/2018ApJ...857..144H} {857, 144}

\bibitem[\protect\citeauthoryear{{Hopkins}, {Hernquist}, {Cox}, {Younger}  \&
  {Besla}}{{Hopkins} et~al.}{2008}]{Hopkins2008}
{Hopkins} P.~F.,  {Hernquist} L.,  {Cox} T.~J.,  {Younger} J.~D.,   {Besla} G.,
   2008, \mn@doi [\apj] {10.1086/592087}, \href
  {https://ui.adsabs.harvard.edu/abs/2008ApJ...688..757H} {688, 757}

\bibitem[\protect\citeauthoryear{{Hughes}, {Pfeffer}, {Martig}, {Bastian},
  {Crain}, {Kruijssen}  \& {Reina-Campos}}{{Hughes} et~al.}{2019}]{Hughes2019}
{Hughes} M.~E.,  {Pfeffer} J.,  {Martig} M.,  {Bastian} N.,  {Crain} R.~A.,
  {Kruijssen} J.~M.~D.,   {Reina-Campos} M.,  2019, \mn@doi [\mnras]
  {10.1093/mnras/sty2889}, \href
  {https://ui.adsabs.harvard.edu/abs/2019MNRAS.482.2795H} {482, 2795}

\bibitem[\protect\citeauthoryear{{Iannuzzi} \& {Athanassoula}}{{Iannuzzi} \&
  {Athanassoula}}{2015}]{Iannuzzi2015}
{Iannuzzi} F.,  {Athanassoula} E.,  2015, \mn@doi [\mnras]
  {10.1093/mnras/stv764}, \href
  {https://ui.adsabs.harvard.edu/abs/2015MNRAS.450.2514I} {450, 2514}

\bibitem[\protect\citeauthoryear{{Ivezi{\'c}} et~al.,}{{Ivezi{\'c}}
  et~al.}{2008}]{Ivezic2008}
{Ivezi{\'c}} {\v{Z}}.,  et~al., 2008, \mn@doi [\apj] {10.1086/589678}, \href
  {https://ui.adsabs.harvard.edu/abs/2008ApJ...684..287I} {684, 287}

\bibitem[\protect\citeauthoryear{{Jackson}, {Martin}, {Kaviraj}, {Laigle},
  {Devriendt}, {Dubois}  \& {Pichon}}{{Jackson} et~al.}{2020}]{Jackson2020}
{Jackson} R.~A.,  {Martin} G.,  {Kaviraj} S.,  {Laigle} C.,  {Devriendt}
  J.~E.~G.,  {Dubois} Y.,   {Pichon} C.,  2020, \mn@doi [\mnras]
  {10.1093/mnras/staa970}, \href
  {https://ui.adsabs.harvard.edu/abs/2020MNRAS.494.5568J} {494, 5568}

\bibitem[\protect\citeauthoryear{{James} \& {Percival}}{{James} \&
  {Percival}}{2018}]{James2018}
{James} P.~A.,  {Percival} S.~M.,  2018, \mn@doi [\mnras]
  {10.1093/mnras/stx2990}, \href
  {https://ui.adsabs.harvard.edu/abs/2018MNRAS.474.3101J} {474, 3101}

\bibitem[\protect\citeauthoryear{{Jiang}, {Li}, {Fang}  \& {Wang}}{{Jiang}
  et~al.}{2019}]{Jiang2019}
{Jiang} X.,  {Li} J.,  {Fang} T.,   {Wang} Q.~D.,  2019, \mn@doi [\apj]
  {10.3847/1538-4357/ab44b4}, \href
  {https://ui.adsabs.harvard.edu/abs/2019ApJ...885...38J} {885, 38}

\bibitem[\protect\citeauthoryear{{Kacharov}, {Neumayer}, {Seth}, {Cappellari},
  {McDermid}, {Walcher}  \& {B{\"o}ker}}{{Kacharov}
  et~al.}{2018}]{Kacharov2018}
{Kacharov} N.,  {Neumayer} N.,  {Seth} A.~C.,  {Cappellari} M.,  {McDermid} R.,
   {Walcher} C.~J.,   {B{\"o}ker} T.,  2018, \mn@doi [\mnras]
  {10.1093/mnras/sty1985}, \href
  {https://ui.adsabs.harvard.edu/abs/2018MNRAS.480.1973K} {480, 1973}

\bibitem[\protect\citeauthoryear{{Kasparova}, {Katkov}, {Chilingarian},
  {Silchenko}, {Moiseev}  \& {Borisov}}{{Kasparova}
  et~al.}{2016}]{Kasparova2016}
{Kasparova} A.~V.,  {Katkov} I.~Y.,  {Chilingarian} I.~V.,  {Silchenko} O.~K.,
  {Moiseev} A.~V.,   {Borisov} S.~B.,  2016, \mn@doi [\mnras]
  {10.1093/mnrasl/slw083}, \href
  {https://ui.adsabs.harvard.edu/abs/2016MNRAS.460L..89K} {460, L89}

\bibitem[\protect\citeauthoryear{{Kasparova}, {Katkov}  \&
  {Chilingarian}}{{Kasparova} et~al.}{2020}]{Kasparova2020}
{Kasparova} A.~V.,  {Katkov} I.~Y.,   {Chilingarian} I.~V.,  2020, \mn@doi
  [\mnras] {10.1093/mnras/staa611}, \href
  {https://ui.adsabs.harvard.edu/abs/2020MNRAS.493.5464K} {493, 5464}

\bibitem[\protect\citeauthoryear{{Kautsch}, {Grebel}, {Barazza}  \&
  {Gallagher}}{{Kautsch} et~al.}{2006}]{Kautsch2006}
{Kautsch} S.~J.,  {Grebel} E.~K.,  {Barazza} F.~D.,   {Gallagher} J.~S. I.,
  2006, \mn@doi [\aap] {10.1051/0004-6361:20053981}, \href
  {https://ui.adsabs.harvard.edu/abs/2006A&A...445..765K} {445, 765}

\bibitem[\protect\citeauthoryear{{Kirby}, {Cohen}, {Guhathakurta}, {Cheng},
  {Bullock}  \& {Gallazzi}}{{Kirby} et~al.}{2013}]{Kirby2013}
{Kirby} E.~N.,  {Cohen} J.~G.,  {Guhathakurta} P.,  {Cheng} L.,  {Bullock}
  J.~S.,   {Gallazzi} A.,  2013, \mn@doi [\apj] {10.1088/0004-637X/779/2/102},
  \href {https://ui.adsabs.harvard.edu/abs/2013ApJ...779..102K} {779, 102}

\bibitem[\protect\citeauthoryear{{Kormendy}}{{Kormendy}}{2013}]{Kormendy2013}
{Kormendy} J.,  2013, {Secular Evolution in Disk Galaxies}.
p.~1

\bibitem[\protect\citeauthoryear{{Kormendy} \& {Barentine}}{{Kormendy} \&
  {Barentine}}{2010}]{Kormendy2010a}
{Kormendy} J.,  {Barentine} J.~C.,  2010, \mn@doi [\apjl]
  {10.1088/2041-8205/715/2/L176}, \href
  {https://ui.adsabs.harvard.edu/abs/2010ApJ...715L.176K} {715, L176}

\bibitem[\protect\citeauthoryear{{Kormendy} \& {Bender}}{{Kormendy} \&
  {Bender}}{2019}]{Kormendy2019}
{Kormendy} J.,  {Bender} R.,  2019, \mn@doi [\apj] {10.3847/1538-4357/aafdff},
  \href {https://ui.adsabs.harvard.edu/abs/2019ApJ...872..106K} {872, 106}

\bibitem[\protect\citeauthoryear{{Kormendy} \& {Kennicutt}}{{Kormendy} \&
  {Kennicutt}}{2004}]{Kormendy2004}
{Kormendy} J.,  {Kennicutt} Robert~C. J.,  2004, \mn@doi [\araa]
  {10.1146/annurev.astro.42.053102.134024}, \href
  {https://ui.adsabs.harvard.edu/abs/2004ARA&A..42..603K} {42, 603}

\bibitem[\protect\citeauthoryear{{Kormendy}, {Drory}, {Bender}  \&
  {Cornell}}{{Kormendy} et~al.}{2010}]{Kormendy2010b}
{Kormendy} J.,  {Drory} N.,  {Bender} R.,   {Cornell} M.~E.,  2010, \mn@doi
  [\apj] {10.1088/0004-637X/723/1/54}, \href
  {https://ui.adsabs.harvard.edu/abs/2010ApJ...723...54K} {723, 54}

\bibitem[\protect\citeauthoryear{{Kraljic}, {Bournaud}  \& {Martig}}{{Kraljic}
  et~al.}{2012}]{Kraljic2012}
{Kraljic} K.,  {Bournaud} F.,   {Martig} M.,  2012, \mn@doi [\apj]
  {10.1088/0004-637X/757/1/60}, \href
  {https://ui.adsabs.harvard.edu/abs/2012ApJ...757...60K} {757, 60}

\bibitem[\protect\citeauthoryear{{Kroupa}}{{Kroupa}}{2001}]{Kroupa2001}
{Kroupa} P.,  2001, \mn@doi [\mnras] {10.1046/j.1365-8711.2001.04022.x}, \href
  {https://ui.adsabs.harvard.edu/abs/2001MNRAS.322..231K} {322, 231}

\bibitem[\protect\citeauthoryear{{Kuijken} \& {Merrifield}}{{Kuijken} \&
  {Merrifield}}{1995}]{Kuijken1995}
{Kuijken} K.,  {Merrifield} M.~R.,  1995, \mn@doi [\apjl] {10.1086/187824},
  \href {https://ui.adsabs.harvard.edu/abs/1995ApJ...443L..13K} {443, L13}

\bibitem[\protect\citeauthoryear{{Kuntschner} et~al.,}{{Kuntschner}
  et~al.}{2010}]{Kuntschner2010}
{Kuntschner} H.,  et~al., 2010, \mn@doi [\mnras]
  {10.1111/j.1365-2966.2010.17161.x}, \href
  {https://ui.adsabs.harvard.edu/abs/2010MNRAS.408...97K} {408, 97}

\bibitem[\protect\citeauthoryear{{Lackner}, {Cen}, {Ostriker}  \&
  {Joung}}{{Lackner} et~al.}{2012}]{Lackner2012}
{Lackner} C.~N.,  {Cen} R.,  {Ostriker} J.~P.,   {Joung} M.~R.,  2012, \mn@doi
  [\mnras] {10.1111/j.1365-2966.2012.21525.x}, \href
  {https://ui.adsabs.harvard.edu/abs/2012MNRAS.425..641L} {425, 641}

\bibitem[\protect\citeauthoryear{{Lee} \& {Yi}}{{Lee} \& {Yi}}{2013}]{Lee2013}
{Lee} J.,  {Yi} S.~K.,  2013, \mn@doi [\apj] {10.1088/0004-637X/766/1/38},
  \href {https://ui.adsabs.harvard.edu/abs/2013ApJ...766...38L} {766, 38}

\bibitem[\protect\citeauthoryear{{Li}, {Shen}, {Bureau}, {Zhou}, {Du}  \&
  {Debattista}}{{Li} et~al.}{2018}]{Li2018}
{Li} Z.-Y.,  {Shen} J.,  {Bureau} M.,  {Zhou} Y.,  {Du} M.,   {Debattista}
  V.~P.,  2018, \mn@doi [\apj] {10.3847/1538-4357/aaa771}, \href
  {https://ui.adsabs.harvard.edu/abs/2018ApJ...854...65L} {854, 65}

\bibitem[\protect\citeauthoryear{{L{\"u}tticke}, {Dettmar}  \&
  {Pohlen}}{{L{\"u}tticke} et~al.}{2000}]{Lutticke2000}
{L{\"u}tticke} R.,  {Dettmar} R.~J.,   {Pohlen} M.,  2000, \mn@doi [\aaps]
  {10.1051/aas:2000354}, \href
  {https://ui.adsabs.harvard.edu/abs/2000A&AS..145..405L} {145, 405}

\bibitem[\protect\citeauthoryear{{Mackereth} et~al.,}{{Mackereth}
  et~al.}{2017}]{Mackereth2017}
{Mackereth} J.~T.,  et~al., 2017, \mn@doi [\mnras] {10.1093/mnras/stx1774},
  \href {https://ui.adsabs.harvard.edu/abs/2017MNRAS.471.3057M} {471, 3057}

\bibitem[\protect\citeauthoryear{{Mackey} et~al.,}{{Mackey}
  et~al.}{2019}]{Mackey2019}
{Mackey} D.,  et~al., 2019, \mn@doi [\nat] {10.1038/s41586-019-1597-1}, \href
  {https://ui.adsabs.harvard.edu/abs/2019Natur.574...69M} {574, 69}

\bibitem[\protect\citeauthoryear{{Mancillas}, {Duc}, {Combes}, {Bournaud},
  {Emsellem}, {Martig}  \& {Michel-Dansac}}{{Mancillas}
  et~al.}{2019}]{Mancillas2019}
{Mancillas} B.,  {Duc} P.-A.,  {Combes} F.,  {Bournaud} F.,  {Emsellem} E.,
  {Martig} M.,   {Michel-Dansac} L.,  2019, \mn@doi [\aap]
  {10.1051/0004-6361/201936320}, \href
  {https://ui.adsabs.harvard.edu/abs/2019A&A...632A.122M} {632, A122}

\bibitem[\protect\citeauthoryear{{Martig}, {Minchev}, {Ness}, {Fouesneau}  \&
  {Rix}}{{Martig} et~al.}{2016}]{Martig2016}
{Martig} M.,  {Minchev} I.,  {Ness} M.,  {Fouesneau} M.,   {Rix} H.-W.,  2016,
  \mn@doi [\apj] {10.3847/0004-637X/831/2/139}, \href
  {https://ui.adsabs.harvard.edu/abs/2016ApJ...831..139M} {831, 139}

\bibitem[\protect\citeauthoryear{{Martin}, {Kaviraj}, {Devriendt}, {Dubois}  \&
  {Pichon}}{{Martin} et~al.}{2018}]{Martin2018}
{Martin} G.,  {Kaviraj} S.,  {Devriendt} J.~E.~G.,  {Dubois} Y.,   {Pichon} C.,
   2018, \mn@doi [\mnras] {10.1093/mnras/sty1936}, \href
  {https://ui.adsabs.harvard.edu/abs/2018MNRAS.480.2266M} {480, 2266}

\bibitem[\protect\citeauthoryear{{Mart{\'\i}nez-Delgado}
  et~al.,}{{Mart{\'\i}nez-Delgado} et~al.}{2010}]{Martinez2010}
{Mart{\'\i}nez-Delgado} D.,  et~al., 2010, \mn@doi [\aj]
  {10.1088/0004-6256/140/4/962}, \href
  {https://ui.adsabs.harvard.edu/abs/2010AJ....140..962M} {140, 962}

\bibitem[\protect\citeauthoryear{{Martinez-Valpuesta}, {Shlosman}  \&
  {Heller}}{{Martinez-Valpuesta} et~al.}{2006}]{Martinez2006}
{Martinez-Valpuesta} I.,  {Shlosman} I.,   {Heller} C.,  2006, \mn@doi [\apj]
  {10.1086/498338}, \href
  {https://ui.adsabs.harvard.edu/abs/2006ApJ...637..214M} {637, 214}

\bibitem[\protect\citeauthoryear{{Matteucci} \& {Brocato}}{{Matteucci} \&
  {Brocato}}{1990}]{Matteucci1990}
{Matteucci} F.,  {Brocato} E.,  1990, \mn@doi [\apj] {10.1086/169508}, \href
  {https://ui.adsabs.harvard.edu/abs/1990ApJ...365..539M} {365, 539}

\bibitem[\protect\citeauthoryear{{McConnachie} et~al.,}{{McConnachie}
  et~al.}{2009}]{McConnachie2009}
{McConnachie} A.~W.,  et~al., 2009, \mn@doi [\nat] {10.1038/nature08327}, \href
  {https://ui.adsabs.harvard.edu/abs/2009Natur.461...66M} {461, 66}

\bibitem[\protect\citeauthoryear{{M{\'e}ndez-Abreu}, {Debattista}, {Corsini}
  \& {Aguerri}}{{M{\'e}ndez-Abreu} et~al.}{2014}]{Mendez2014}
{M{\'e}ndez-Abreu} J.,  {Debattista} V.~P.,  {Corsini} E.~M.,   {Aguerri}
  J.~A.~L.,  2014, \mn@doi [\aap] {10.1051/0004-6361/201423955}, \href
  {https://ui.adsabs.harvard.edu/abs/2014A&A...572A..25M} {572, A25}

\bibitem[\protect\citeauthoryear{{Merritt}, {van Dokkum}, {Abraham}  \&
  {Zhang}}{{Merritt} et~al.}{2016}]{Merritt2016}
{Merritt} A.,  {van Dokkum} P.,  {Abraham} R.,   {Zhang} J.,  2016, \mn@doi
  [\apj] {10.3847/0004-637X/830/2/62}, \href
  {https://ui.adsabs.harvard.edu/abs/2016ApJ...830...62M} {830, 62}

\bibitem[\protect\citeauthoryear{{Minchev}, {Martig}, {Streich}, {Scannapieco},
  {de Jong}  \& {Steinmetz}}{{Minchev} et~al.}{2015}]{Minchev2015}
{Minchev} I.,  {Martig} M.,  {Streich} D.,  {Scannapieco} C.,  {de Jong} R.~S.,
    {Steinmetz} M.,  2015, \mn@doi [\apjl] {10.1088/2041-8205/804/1/L9}, \href
  {https://ui.adsabs.harvard.edu/abs/2015ApJ...804L...9M} {804, L9}

\bibitem[\protect\citeauthoryear{{Molaeinezhad}, {Falc{\'o}n-Barroso},
  {Mart{\'\i}nez-Valpuesta}, {Khosroshahi}, {Balcells}  \&
  {Peletier}}{{Molaeinezhad} et~al.}{2016}]{Molaeinezhad2016}
{Molaeinezhad} A.,  {Falc{\'o}n-Barroso} J.,  {Mart{\'\i}nez-Valpuesta} I.,
  {Khosroshahi} H.~G.,  {Balcells} M.,   {Peletier} R.~F.,  2016, \mn@doi
  [\mnras] {10.1093/mnras/stv2697}, \href
  {https://ui.adsabs.harvard.edu/abs/2016MNRAS.456..692M} {456, 692}

\bibitem[\protect\citeauthoryear{{Molaeinezhad}, {Falc{\'o}n-Barroso},
  {Mart{\'\i}nez-Valpuesta}, {Khosroshahi}, {Vazdekis}, {La Barbera},
  {Peletier}  \& {Balcells}}{{Molaeinezhad} et~al.}{2017}]{Molaeinezhad2017}
{Molaeinezhad} A.,  {Falc{\'o}n-Barroso} J.,  {Mart{\'\i}nez-Valpuesta} I.,
  {Khosroshahi} H.~G.,  {Vazdekis} A.,  {La Barbera} F.,  {Peletier} R.~F.,
  {Balcells} M.,  2017, \mn@doi [\mnras] {10.1093/mnras/stx051}, \href
  {https://ui.adsabs.harvard.edu/abs/2017MNRAS.467..353M} {467, 353}

\bibitem[\protect\citeauthoryear{{Monachesi}, {Bell}, {Radburn-Smith},
  {Bailin}, {de Jong}, {Holwerda}, {Streich}  \& {Silverstein}}{{Monachesi}
  et~al.}{2016}]{Monachesi2016}
{Monachesi} A.,  {Bell} E.~F.,  {Radburn-Smith} D.~J.,  {Bailin} J.,  {de Jong}
  R.~S.,  {Holwerda} B.,  {Streich} D.,   {Silverstein} G.,  2016, \mn@doi
  [\mnras] {10.1093/mnras/stv2987}, \href
  {https://ui.adsabs.harvard.edu/abs/2016MNRAS.457.1419M} {457, 1419}

\bibitem[\protect\citeauthoryear{{Monachesi} et~al.,}{{Monachesi}
  et~al.}{2019}]{Monachesi2019}
{Monachesi} A.,  et~al., 2019, \mn@doi [\mnras] {10.1093/mnras/stz538}, \href
  {https://ui.adsabs.harvard.edu/abs/2019MNRAS.485.2589M} {485, 2589}

\bibitem[\protect\citeauthoryear{{Morales}, {Mart{\'\i}nez-Delgado}, {Grebel},
  {Cooper}, {Javanmardi}  \& {Miskolczi}}{{Morales} et~al.}{2018}]{Morales2018}
{Morales} G.,  {Mart{\'\i}nez-Delgado} D.,  {Grebel} E.~K.,  {Cooper} A.~P.,
  {Javanmardi} B.,   {Miskolczi} A.,  2018, \mn@doi [\aap]
  {10.1051/0004-6361/201732271}, \href
  {https://ui.adsabs.harvard.edu/abs/2018A&A...614A.143M} {614, A143}

\bibitem[\protect\citeauthoryear{{Mould}}{{Mould}}{2005}]{Mould2005}
{Mould} J.,  2005, \mn@doi [\aj] {10.1086/427248}, \href
  {https://ui.adsabs.harvard.edu/abs/2005AJ....129..698M} {129, 698}

\bibitem[\protect\citeauthoryear{{Nersesian} et~al.,}{{Nersesian}
  et~al.}{2019}]{Nersesian2019}
{Nersesian} A.,  et~al., 2019, \mn@doi [\aap] {10.1051/0004-6361/201935118},
  \href {https://ui.adsabs.harvard.edu/abs/2019A&A...624A..80N} {624, A80}

\bibitem[\protect\citeauthoryear{{Nidever} et~al.,}{{Nidever}
  et~al.}{2014}]{Nidever2014}
{Nidever} D.~L.,  et~al., 2014, \mn@doi [\apj] {10.1088/0004-637X/796/1/38},
  \href {https://ui.adsabs.harvard.edu/abs/2014ApJ...796...38N} {796, 38}

\bibitem[\protect\citeauthoryear{{Ocvirk}, {Pichon}, {Lan{\c{c}}on}  \&
  {Thi{\'e}baut}}{{Ocvirk} et~al.}{2006}]{Ocvirk2006}
{Ocvirk} P.,  {Pichon} C.,  {Lan{\c{c}}on} A.,   {Thi{\'e}baut} E.,  2006,
  \mn@doi [\mnras] {10.1111/j.1365-2966.2005.09323.x}, \href
  {https://ui.adsabs.harvard.edu/abs/2006MNRAS.365...74O} {365, 74}

\bibitem[\protect\citeauthoryear{{Park} et~al.,}{{Park}
  et~al.}{2021}]{Park2021}
{Park} M.~J.,  et~al., 2021, \mn@doi [\apjs] {10.3847/1538-4365/abe937}, \href
  {https://ui.adsabs.harvard.edu/abs/2021ApJS..254....2P} {254, 2}

\bibitem[\protect\citeauthoryear{{Pe{\~n}arrubia}, {McConnachie}  \&
  {Babul}}{{Pe{\~n}arrubia} et~al.}{2006}]{Penarrubia2006}
{Pe{\~n}arrubia} J.,  {McConnachie} A.,   {Babul} A.,  2006, \mn@doi [\apjl]
  {10.1086/508656}, \href
  {https://ui.adsabs.harvard.edu/abs/2006ApJ...650L..33P} {650, L33}

\bibitem[\protect\citeauthoryear{{Peebles}}{{Peebles}}{2020}]{Peebles2020}
{Peebles} P.~J.~E.,  2020, \mn@doi [\mnras] {10.1093/mnras/staa2649}, \href
  {https://ui.adsabs.harvard.edu/abs/2020MNRAS.498.4386P} {498, 4386}

\bibitem[\protect\citeauthoryear{{Peletier}, {Balcells}, {Davies},
  {Andredakis}, {Vazdekis}, {Burkert}  \& {Prada}}{{Peletier}
  et~al.}{1999}]{Peletier1999}
{Peletier} R.~F.,  {Balcells} M.,  {Davies} R.~L.,  {Andredakis} Y.,
  {Vazdekis} A.,  {Burkert} A.,   {Prada} F.,  1999, \mn@doi [\mnras]
  {10.1046/j.1365-8711.1999.02980.x}, \href
  {https://ui.adsabs.harvard.edu/abs/1999MNRAS.310..703P} {310, 703}

\bibitem[\protect\citeauthoryear{{P{\'e}rez} et~al.,}{{P{\'e}rez}
  et~al.}{2017}]{Perez2017}
{P{\'e}rez} I.,  et~al., 2017, \mn@doi [\mnras] {10.1093/mnrasl/slx087}, \href
  {https://ui.adsabs.harvard.edu/abs/2017MNRAS.470L.122P} {470, L122}

\bibitem[\protect\citeauthoryear{{Peters} \& {Kuzio de Naray}}{{Peters} \&
  {Kuzio de Naray}}{2017}]{Peters2017}
{Peters} W.,  {Kuzio de Naray} R.,  2017, \mn@doi [\mnras]
  {10.1093/mnras/stx821}, \href
  {https://ui.adsabs.harvard.edu/abs/2017MNRAS.469.3541P} {469, 3541}

\bibitem[\protect\citeauthoryear{{Pietrinferni}, {Cassisi}, {Salaris}  \&
  {Castelli}}{{Pietrinferni} et~al.}{2004}]{Pietrinferni2004}
{Pietrinferni} A.,  {Cassisi} S.,  {Salaris} M.,   {Castelli} F.,  2004,
  \mn@doi [\apj] {10.1086/422498}, \href
  {https://ui.adsabs.harvard.edu/abs/2004ApJ...612..168P} {612, 168}

\bibitem[\protect\citeauthoryear{{Pillepich}, {Madau}  \& {Mayer}}{{Pillepich}
  et~al.}{2015}]{Pillepich2015}
{Pillepich} A.,  {Madau} P.,   {Mayer} L.,  2015, \mn@doi [\apj]
  {10.1088/0004-637X/799/2/184}, \href
  {https://ui.adsabs.harvard.edu/abs/2015ApJ...799..184P} {799, 184}

\bibitem[\protect\citeauthoryear{{Pinna} et~al.,}{{Pinna}
  et~al.}{2019a}]{Pinna2019a}
{Pinna} F.,  et~al., 2019a, \mn@doi [\aap] {10.1051/0004-6361/201833193}, \href
  {https://ui.adsabs.harvard.edu/abs/2019A&A...623A..19P} {623, A19}

\bibitem[\protect\citeauthoryear{{Pinna} et~al.,}{{Pinna}
  et~al.}{2019b}]{Pinna2019b}
{Pinna} F.,  et~al., 2019b, \mn@doi [\aap] {10.1051/0004-6361/201935154}, \href
  {https://ui.adsabs.harvard.edu/abs/2019A&A...625A..95P} {625, A95}

\bibitem[\protect\citeauthoryear{{Poci} et~al.,}{{Poci}
  et~al.}{2021}]{Poci2021}
{Poci} A.,  et~al., 2021, \mn@doi [\aap] {10.1051/0004-6361/202039644}, \href
  {https://ui.adsabs.harvard.edu/abs/2021A&A...647A.145P} {647, A145}

\bibitem[\protect\citeauthoryear{{Quinn} \& {Goodman}}{{Quinn} \&
  {Goodman}}{1986}]{Quinn1986}
{Quinn} P.~J.,  {Goodman} J.,  1986, \mn@doi [\apj] {10.1086/164619}, \href
  {https://ui.adsabs.harvard.edu/abs/1986ApJ...309..472Q} {309, 472}

\bibitem[\protect\citeauthoryear{{Quinn}, {Hernquist}  \& {Fullagar}}{{Quinn}
  et~al.}{1993}]{Quinn1993}
{Quinn} P.~J.,  {Hernquist} L.,   {Fullagar} D.~P.,  1993, \mn@doi [\apj]
  {10.1086/172184}, \href
  {https://ui.adsabs.harvard.edu/abs/1993ApJ...403...74Q} {403, 74}

\bibitem[\protect\citeauthoryear{{Rand} \& {Benjamin}}{{Rand} \&
  {Benjamin}}{2008}]{Rand2008}
{Rand} R.~J.,  {Benjamin} R.~A.,  2008, \mn@doi [\apj] {10.1086/528952}, \href
  {https://ui.adsabs.harvard.edu/abs/2008ApJ...676..991R} {676, 991}

\bibitem[\protect\citeauthoryear{{Read}, {Lake}, {Agertz}  \&
  {Debattista}}{{Read} et~al.}{2008}]{Read2008}
{Read} J.~I.,  {Lake} G.,  {Agertz} O.,   {Debattista} V.~P.,  2008, \mn@doi
  [\mnras] {10.1111/j.1365-2966.2008.13643.x}, \href
  {https://ui.adsabs.harvard.edu/abs/2008MNRAS.389.1041R} {389, 1041}

\bibitem[\protect\citeauthoryear{{Rejkuba}, {Mouhcine}  \& {Ibata}}{{Rejkuba}
  et~al.}{2009}]{Rejkuba2009}
{Rejkuba} M.,  {Mouhcine} M.,   {Ibata} R.,  2009, \mn@doi [\mnras]
  {10.1111/j.1365-2966.2009.14821.x}, \href
  {https://ui.adsabs.harvard.edu/abs/2009MNRAS.396.1231R} {396, 1231}

\bibitem[\protect\citeauthoryear{{Remus} \& {Forbes}}{{Remus} \&
  {Forbes}}{2021}]{Remus2021}
{Remus} R.-S.,  {Forbes} D.~A.,  2021, arXiv e-prints, \href
  {https://ui.adsabs.harvard.edu/abs/2021arXiv210112216R} {p. arXiv:2101.12216}

\bibitem[\protect\citeauthoryear{{Rich} et~al.,}{{Rich}
  et~al.}{2019}]{Rich2019}
{Rich} R.~M.,  et~al., 2019, \mn@doi [\mnras] {10.1093/mnras/stz2106}, \href
  {https://ui.adsabs.harvard.edu/abs/2019MNRAS.490.1539R} {490, 1539}

\bibitem[\protect\citeauthoryear{{Rix}, {Franx}, {Fisher}  \&
  {Illingworth}}{{Rix} et~al.}{1992}]{Rix1992}
{Rix} H.-W.,  {Franx} M.,  {Fisher} D.,   {Illingworth} G.,  1992, \mn@doi
  [\apjl] {10.1086/186635}, \href
  {https://ui.adsabs.harvard.edu/abs/1992ApJ...400L...5R} {400, L5}

\bibitem[\protect\citeauthoryear{{Rodriguez-Gomez} et~al.,}{{Rodriguez-Gomez}
  et~al.}{2016}]{Rodriguez2016}
{Rodriguez-Gomez} V.,  et~al., 2016, \mn@doi [\mnras] {10.1093/mnras/stw456},
  \href {https://ui.adsabs.harvard.edu/abs/2016MNRAS.458.2371R} {458, 2371}

\bibitem[\protect\citeauthoryear{{Rosas-Guevara} et~al.,}{{Rosas-Guevara}
  et~al.}{2020}]{Rosas2020}
{Rosas-Guevara} Y.,  et~al., 2020, \mn@doi [\mnras] {10.1093/mnras/stz3180},
  \href {https://ui.adsabs.harvard.edu/abs/2020MNRAS.491.2547R} {491, 2547}

\bibitem[\protect\citeauthoryear{{Rubin}, {Graham}  \& {Kenney}}{{Rubin}
  et~al.}{1992}]{Rubin1992}
{Rubin} V.~C.,  {Graham} J.~A.,   {Kenney} J. D.~P.,  1992, \mn@doi [\apjl]
  {10.1086/186460}, \href
  {https://ui.adsabs.harvard.edu/abs/1992ApJ...394L...9R} {394, L9}

\bibitem[\protect\citeauthoryear{{Ruchti}, {Read}, {Feltzing}, {Pipino}  \&
  {Bensby}}{{Ruchti} et~al.}{2014}]{Ruchti2014}
{Ruchti} G.~R.,  {Read} J.~I.,  {Feltzing} S.,  {Pipino} A.,   {Bensby} T.,
  2014, \mn@doi [\mnras] {10.1093/mnras/stu1435}, \href
  {https://ui.adsabs.harvard.edu/abs/2014MNRAS.444..515R} {444, 515}

\bibitem[\protect\citeauthoryear{{Ruiz-Lara} et~al.,}{{Ruiz-Lara}
  et~al.}{2015}]{Ruiz2015}
{Ruiz-Lara} T.,  et~al., 2015, \mn@doi [\aap] {10.1051/0004-6361/201526752},
  \href {https://ui.adsabs.harvard.edu/abs/2015A&A...583A..60R} {583, A60}

\bibitem[\protect\citeauthoryear{{Ruiz-Lara} et~al.,}{{Ruiz-Lara}
  et~al.}{2017}]{Ruiz2017}
{Ruiz-Lara} T.,  et~al., 2017, \mn@doi [\aap] {10.1051/0004-6361/201730705},
  \href {https://ui.adsabs.harvard.edu/abs/2017A&A...604A...4R} {604, A4}

\bibitem[\protect\citeauthoryear{{Ruiz-Lara} et~al.,}{{Ruiz-Lara}
  et~al.}{2018}]{Ruiz2018}
{Ruiz-Lara} T.,  et~al., 2018, \mn@doi [\aap] {10.1051/0004-6361/201732398},
  \href {https://ui.adsabs.harvard.edu/abs/2018A&A...617A..18R} {617, A18}

\bibitem[\protect\citeauthoryear{{Ruiz-Lara}, {Gallart}, {Bernard}  \&
  {Cassisi}}{{Ruiz-Lara} et~al.}{2020}]{Ruiz2020}
{Ruiz-Lara} T.,  {Gallart} C.,  {Bernard} E.~J.,   {Cassisi} S.,  2020, \mn@doi
  [Nature Astronomy] {10.1038/s41550-020-1097-0}, \href
  {https://ui.adsabs.harvard.edu/abs/2020NatAs...4..965R} {4, 965}

\bibitem[\protect\citeauthoryear{{S{\'a}nchez-Bl{\'a}zquez}
  et~al.,}{{S{\'a}nchez-Bl{\'a}zquez} et~al.}{2014}]{Sanchez2014}
{S{\'a}nchez-Bl{\'a}zquez} P.,  et~al., 2014, \mn@doi [\aap]
  {10.1051/0004-6361/201423635}, \href
  {https://ui.adsabs.harvard.edu/abs/2014A&A...570A...6S} {570, A6}

\bibitem[\protect\citeauthoryear{{Sarzi} et~al.,}{{Sarzi}
  et~al.}{2006}]{Sarzi2006}
{Sarzi} M.,  et~al., 2006, \mn@doi [\mnras] {10.1111/j.1365-2966.2005.09839.x},
  \href {https://ui.adsabs.harvard.edu/abs/2006MNRAS.366.1151S} {366, 1151}

\bibitem[\protect\citeauthoryear{{Sarzi}, {Ledo}  \& {Dotti}}{{Sarzi}
  et~al.}{2015}]{Sarzi2015}
{Sarzi} M.,  {Ledo} H.~R.,   {Dotti} M.,  2015, \mn@doi [\mnras]
  {10.1093/mnras/stv1497}, \href
  {https://ui.adsabs.harvard.edu/abs/2015MNRAS.453.1070S} {453, 1070}

\bibitem[\protect\citeauthoryear{{Sarzi} et~al.,}{{Sarzi}
  et~al.}{2018}]{Sarzi2018}
{Sarzi} M.,  et~al., 2018, \mn@doi [\aap] {10.1051/0004-6361/201833137}, \href
  {https://ui.adsabs.harvard.edu/abs/2018A&A...616A.121S} {616, A121}

\bibitem[\protect\citeauthoryear{{Schlesinger} et~al.,}{{Schlesinger}
  et~al.}{2012}]{Schlesinger2012}
{Schlesinger} K.~J.,  et~al., 2012, \mn@doi [\apj]
  {10.1088/0004-637X/761/2/160}, \href
  {https://ui.adsabs.harvard.edu/abs/2012ApJ...761..160S} {761, 160}

\bibitem[\protect\citeauthoryear{{Scott}, {van de Sande}, {Sharma},
  {Bland-Hawthorn}, {Freeman}, {Gerhard}, {Hayden}  \& {McDermid}}{{Scott}
  et~al.}{2021}]{Scott2021}
{Scott} N.,  {van de Sande} J.,  {Sharma} S.,  {Bland-Hawthorn} J.,  {Freeman}
  K.,  {Gerhard} O.,  {Hayden} M.~R.,   {McDermid} R.,  2021, \mn@doi [\apjl]
  {10.3847/2041-8213/abfc57}, \href
  {https://ui.adsabs.harvard.edu/abs/2021ApJ...913L..11S} {913, L11}

\bibitem[\protect\citeauthoryear{{Sellwood}}{{Sellwood}}{2014}]{Sellwood2014}
{Sellwood} J.~A.,  2014, \mn@doi [Reviews of Modern Physics]
  {10.1103/RevModPhys.86.1}, \href
  {https://ui.adsabs.harvard.edu/abs/2014RvMP...86....1S} {86, 1}

\bibitem[\protect\citeauthoryear{{Sellwood} \& {Gerhard}}{{Sellwood} \&
  {Gerhard}}{2020}]{Sellwood2020}
{Sellwood} J.~A.,  {Gerhard} O.,  2020, \mn@doi [\mnras]
  {10.1093/mnras/staa1336}, \href
  {https://ui.adsabs.harvard.edu/abs/2020MNRAS.495.3175S} {495, 3175}

\bibitem[\protect\citeauthoryear{{Serra} \& {Trager}}{{Serra} \&
  {Trager}}{2007}]{Serra2007}
{Serra} P.,  {Trager} S.~C.,  2007, \mn@doi [\mnras]
  {10.1111/j.1365-2966.2006.11188.x}, \href
  {https://ui.adsabs.harvard.edu/abs/2007MNRAS.374..769S} {374, 769}

\bibitem[\protect\citeauthoryear{{Shen}, {Rich}, {Kormendy}, {Howard}, {De
  Propris}  \& {Kunder}}{{Shen} et~al.}{2010}]{Shen2010}
{Shen} J.,  {Rich} R.~M.,  {Kormendy} J.,  {Howard} C.~D.,  {De Propris} R.,
  {Kunder} A.,  2010, \mn@doi [\apjl] {10.1088/2041-8205/720/1/L72}, \href
  {https://ui.adsabs.harvard.edu/abs/2010ApJ...720L..72S} {720, L72}

\bibitem[\protect\citeauthoryear{{Sheth} et~al.,}{{Sheth}
  et~al.}{2008}]{Sheth2008}
{Sheth} K.,  et~al., 2008, \mn@doi [\apj] {10.1086/524980}, \href
  {https://ui.adsabs.harvard.edu/abs/2008ApJ...675.1141S} {675, 1141}

\bibitem[\protect\citeauthoryear{{Sheth} et~al.,}{{Sheth}
  et~al.}{2010}]{Sheth2010}
{Sheth} K.,  et~al., 2010, \mn@doi [\pasp] {10.1086/657638}, \href
  {https://ui.adsabs.harvard.edu/abs/2010PASP..122.1397S} {122, 1397}

\bibitem[\protect\citeauthoryear{{Sheth}, {Melbourne}, {Elmegreen},
  {Elmegreen}, {Athanassoula}, {Abraham}  \& {Weiner}}{{Sheth}
  et~al.}{2012}]{Sheth2012}
{Sheth} K.,  {Melbourne} J.,  {Elmegreen} D.~M.,  {Elmegreen} B.~G.,
  {Athanassoula} E.,  {Abraham} R.~G.,   {Weiner} B.~J.,  2012, \mn@doi [\apj]
  {10.1088/0004-637X/758/2/136}, \href
  {https://ui.adsabs.harvard.edu/abs/2012ApJ...758..136S} {758, 136}

\bibitem[\protect\citeauthoryear{{Simmons} et~al.,}{{Simmons}
  et~al.}{2014}]{Simmons2014}
{Simmons} B.~D.,  et~al., 2014, \mn@doi [\mnras] {10.1093/mnras/stu1817}, \href
  {https://ui.adsabs.harvard.edu/abs/2014MNRAS.445.3466S} {445, 3466}

\bibitem[\protect\citeauthoryear{{Sormani}, {Binney}  \& {Magorrian}}{{Sormani}
  et~al.}{2015}]{Sormani2015}
{Sormani} M.~C.,  {Binney} J.,   {Magorrian} J.,  2015, \mn@doi [\mnras]
  {10.1093/mnras/stv441}, \href
  {https://ui.adsabs.harvard.edu/abs/2015MNRAS.449.2421S} {449, 2421}

\bibitem[\protect\citeauthoryear{{Springel} \& {Hernquist}}{{Springel} \&
  {Hernquist}}{2005}]{Springel2005}
{Springel} V.,  {Hernquist} L.,  2005, \mn@doi [\apjl] {10.1086/429486}, \href
  {https://ui.adsabs.harvard.edu/abs/2005ApJ...622L...9S} {622, L9}

\bibitem[\protect\citeauthoryear{{Springob}, {Masters}, {Haynes}, {Giovanelli}
  \& {Marinoni}}{{Springob} et~al.}{2007}]{Springob2007}
{Springob} C.~M.,  {Masters} K.~L.,  {Haynes} M.~P.,  {Giovanelli} R.,
  {Marinoni} C.,  2007, \mn@doi [\apjs] {10.1086/519527}, \href
  {https://ui.adsabs.harvard.edu/abs/2007ApJS..172..599S} {172, 599}

\bibitem[\protect\citeauthoryear{{Stewart}, {Bullock}, {Wechsler}, {Maller}  \&
  {Zentner}}{{Stewart} et~al.}{2008}]{Stewart2008}
{Stewart} K.~R.,  {Bullock} J.~S.,  {Wechsler} R.~H.,  {Maller} A.~H.,
  {Zentner} A.~R.,  2008, \mn@doi [\apj] {10.1086/588579}, \href
  {https://ui.adsabs.harvard.edu/abs/2008ApJ...683..597S} {683, 597}

\bibitem[\protect\citeauthoryear{{Str{\"o}mberg}}{{Str{\"o}mberg}}{1946}]{Stroemberg1946}
{Str{\"o}mberg} G.,  1946, \mn@doi [\apj] {10.1086/144830}, \href
  {https://ui.adsabs.harvard.edu/abs/1946ApJ...104...12S} {104, 12}

\bibitem[\protect\citeauthoryear{{Trager} \& {Somerville}}{{Trager} \&
  {Somerville}}{2009}]{Trager2009}
{Trager} S.~C.,  {Somerville} R.~S.,  2009, \mn@doi [\mnras]
  {10.1111/j.1365-2966.2009.14571.x}, \href
  {https://ui.adsabs.harvard.edu/abs/2009MNRAS.395..608T} {395, 608}

\bibitem[\protect\citeauthoryear{{Trager}, {Faber}, {Worthey}  \&
  {Gonz{\'a}lez}}{{Trager} et~al.}{2000}]{Trager2000}
{Trager} S.~C.,  {Faber} S.~M.,  {Worthey} G.,   {Gonz{\'a}lez} J.~J.,  2000,
  \mn@doi [\aj] {10.1086/301299}, \href
  {https://ui.adsabs.harvard.edu/abs/2000AJ....119.1645T} {119, 1645}

\bibitem[\protect\citeauthoryear{{Tully}, {Courtois}  \& {Sorce}}{{Tully}
  et~al.}{2016}]{Tully2016}
{Tully} R.~B.,  {Courtois} H.~M.,   {Sorce} J.~G.,  2016, \mn@doi [\aj]
  {10.3847/0004-6256/152/2/50}, \href
  {https://ui.adsabs.harvard.edu/abs/2016AJ....152...50T} {152, 50}

\bibitem[\protect\citeauthoryear{{Vazdekis}, {S{\'a}nchez-Bl{\'a}zquez},
  {Falc{\'o}n-Barroso}, {Cenarro}, {Beasley}, {Cardiel}, {Gorgas}  \&
  {Peletier}}{{Vazdekis} et~al.}{2010}]{Vazdekis2010}
{Vazdekis} A.,  {S{\'a}nchez-Bl{\'a}zquez} P.,  {Falc{\'o}n-Barroso} J.,
  {Cenarro} A.~J.,  {Beasley} M.~A.,  {Cardiel} N.,  {Gorgas} J.,   {Peletier}
  R.~F.,  2010, \mn@doi [\mnras] {10.1111/j.1365-2966.2010.16407.x}, \href
  {https://ui.adsabs.harvard.edu/abs/2010MNRAS.404.1639V} {404, 1639}

\bibitem[\protect\citeauthoryear{{Vazdekis} et~al.,}{{Vazdekis}
  et~al.}{2015}]{Vazdekis2015}
{Vazdekis} A.,  et~al., 2015, \mn@doi [\mnras] {10.1093/mnras/stv151}, \href
  {https://ui.adsabs.harvard.edu/abs/2015MNRAS.449.1177V} {449, 1177}

\bibitem[\protect\citeauthoryear{{Vera-Casanova} et~al.,}{{Vera-Casanova}
  et~al.}{2021}]{Vera2021}
{Vera-Casanova} A.,  et~al., 2021, arXiv e-prints, \href
  {https://ui.adsabs.harvard.edu/abs/2021arXiv210506467V} {p. arXiv:2105.06467}

\bibitem[\protect\citeauthoryear{{Villalobos} \& {Helmi}}{{Villalobos} \&
  {Helmi}}{2008}]{Villalobos2008}
{Villalobos} {\'A}.,  {Helmi} A.,  2008, \mn@doi [\mnras]
  {10.1111/j.1365-2966.2008.13979.x}, \href
  {https://ui.adsabs.harvard.edu/abs/2008MNRAS.391.1806V} {391, 1806}

\bibitem[\protect\citeauthoryear{{Walker}, {Mihos}  \& {Hernquist}}{{Walker}
  et~al.}{1996}]{Walker1996}
{Walker} I.~R.,  {Mihos} J.~C.,   {Hernquist} L.,  1996, \mn@doi [\apj]
  {10.1086/176956}, \href
  {https://ui.adsabs.harvard.edu/abs/1996ApJ...460..121W} {460, 121}

\bibitem[\protect\citeauthoryear{{Weilbacher}, {Streicher}, {Urrutia},
  {P{\'e}contal-Rousset}, {Jarno}  \& {Bacon}}{{Weilbacher}
  et~al.}{2014}]{Weilbacher2014}
{Weilbacher} P.~M.,  {Streicher} O.,  {Urrutia} T.,  {P{\'e}contal-Rousset} A.,
   {Jarno} A.,   {Bacon} R.,  2014, in {Manset} N.,  {Forshay} P.,  eds,
  Astronomical Society of the Pacific Conference Series Vol. 485, Astronomical
  Data Analysis Software and Systems XXIII. p.~451 (\mn@eprint {arXiv}
  {1507.00034})

\bibitem[\protect\citeauthoryear{{Weilbacher} et~al.,}{{Weilbacher}
  et~al.}{2020}]{Weilbacher2020}
{Weilbacher} P.~M.,  et~al., 2020, \mn@doi [\aap]
  {10.1051/0004-6361/202037855}, \href
  {https://ui.adsabs.harvard.edu/abs/2020A&A...641A..28W} {641, A28}

\bibitem[\protect\citeauthoryear{{Williams}, {Zamojski}, {Bureau},
  {Kuntschner}, {Merrifield}, {de Zeeuw}  \& {Kuijken}}{{Williams}
  et~al.}{2011}]{Williams2011}
{Williams} M.~J.,  {Zamojski} M.~A.,  {Bureau} M.,  {Kuntschner} H.,
  {Merrifield} M.~R.,  {de Zeeuw} P.~T.,   {Kuijken} K.,  2011, \mn@doi
  [\mnras] {10.1111/j.1365-2966.2011.18535.x}, \href
  {https://ui.adsabs.harvard.edu/abs/2011MNRAS.414.2163W} {414, 2163}

\bibitem[\protect\citeauthoryear{{Yoachim} \& {Dalcanton}}{{Yoachim} \&
  {Dalcanton}}{2006}]{Yoachim2006}
{Yoachim} P.,  {Dalcanton} J.~J.,  2006, \mn@doi [\aj] {10.1086/497970}, \href
  {https://ui.adsabs.harvard.edu/abs/2006AJ....131..226Y} {131, 226}

\bibitem[\protect\citeauthoryear{{Yoachim} \& {Dalcanton}}{{Yoachim} \&
  {Dalcanton}}{2008a}]{Yoachim2008a}
{Yoachim} P.,  {Dalcanton} J.~J.,  2008a, \mn@doi [\apj] {10.1086/589553},
  \href {https://ui.adsabs.harvard.edu/abs/2008ApJ...682.1004Y} {682, 1004}

\bibitem[\protect\citeauthoryear{{Yoachim} \& {Dalcanton}}{{Yoachim} \&
  {Dalcanton}}{2008b}]{Yoachim2008b}
{Yoachim} P.,  {Dalcanton} J.~J.,  2008b, \mn@doi [\apj] {10.1086/590246},
  \href {https://ui.adsabs.harvard.edu/abs/2008ApJ...683..707Y} {683, 707}

\bibitem[\protect\citeauthoryear{{de Vaucouleurs}, {de Vaucouleurs}, {Corwin},
  {Buta}, {Paturel}  \& {Fouque}}{{de Vaucouleurs} et~al.}{1991}]{RC3}
{de Vaucouleurs} G.,  {de Vaucouleurs} A.,  {Corwin} Herold~G. J.,  {Buta}
  R.~J.,  {Paturel} G.,   {Fouque} P.,  1991, {Third Reference Catalogue of
  Bright Galaxies}

\makeatother
\end{thebibliography}

\appendix
\section{Emission lines analysis}\label{appendix:emission}
\begin{figure}
\includegraphics[width=\columnwidth]{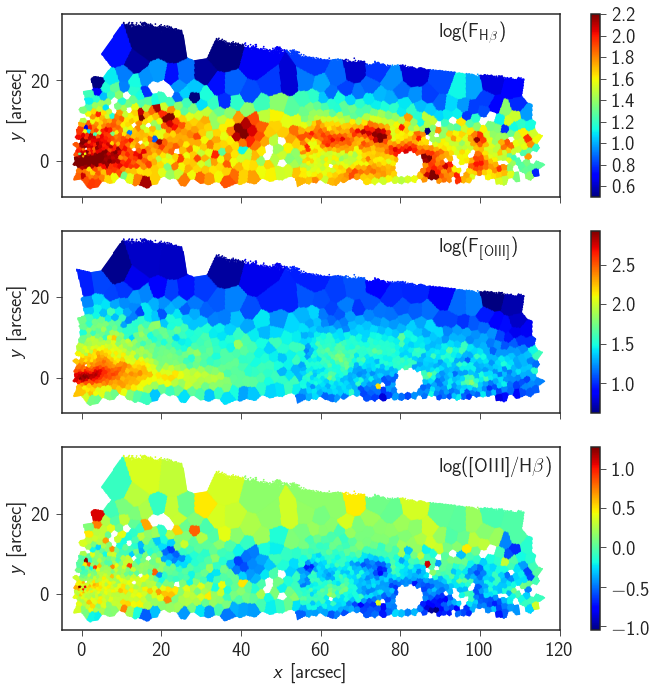}
\caption{Spatial distribution of the ionized gas (the fluxes are shown using arbitrary units). The top panel shows the distribution of H$\beta$ emission, the middle panel, [OIII], while the bottom panel shows the ratio [OIII]/H$\beta$.  }
\label{fig:emission}
\end{figure}

In this Appendix, we present the results from our GandALF analysis of emission lines: Figure \ref{fig:emission} shows the spatial distribution of the  H$\beta$ and [OIII] fluxes, as well as the ratio between the two. We note that a similar analysis (using the same MUSE data) was performed by \cite{DeVis2019}, showing more emission lines, and deriving oxygen abundances: we do not aim to redo this analysis here.

Figure \ref{fig:emission} shows that ionized gas is present in the disc of NGC 5746 (with negligible emission at large height above the midplane, as expected). A series of star forming regions (characterized by a low [OIII]/H$\beta$ ratio) is found in a ring configuration. This ring corresponds to the ring seen in PAH emission (Figure \ref{fig:pointings} and \citealp{Barentine2012}), which is another tracer of star forming regions. There are also signs of young populations in this ring in our stellar populations analysis (particularly in the distribution of young stars in Figure \ref{fig:mass_maps} and the SSP-equivalent age map in Figure \ref{fig:comp_indices_maps}).

\section{Stellar populations recovery using line-strength indices}\label{appendix:indices}
\begin{figure}
\includegraphics[width=\columnwidth]{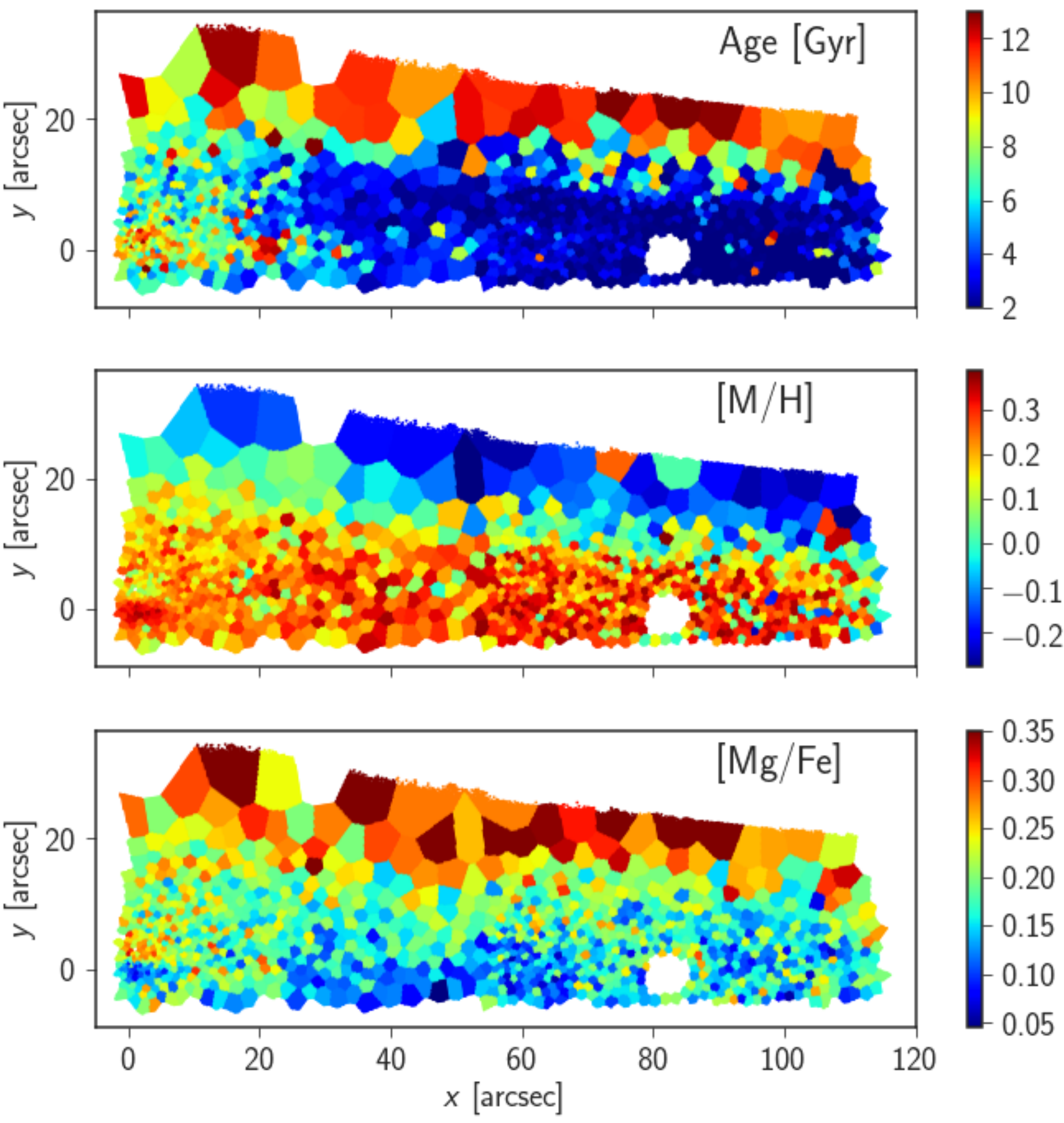}
\caption{Spatial distribution of SSP-equivalent age, [M/H] and [Mg/Fe] obtained from our line-strength analysis. Those maps are qualitatively similar to the ones obtained from pPXF (Figure \ref{fig:pop_maps}). }
\label{fig:comp_indices_maps}
\end{figure}

\begin{figure}
\includegraphics[width=\columnwidth]{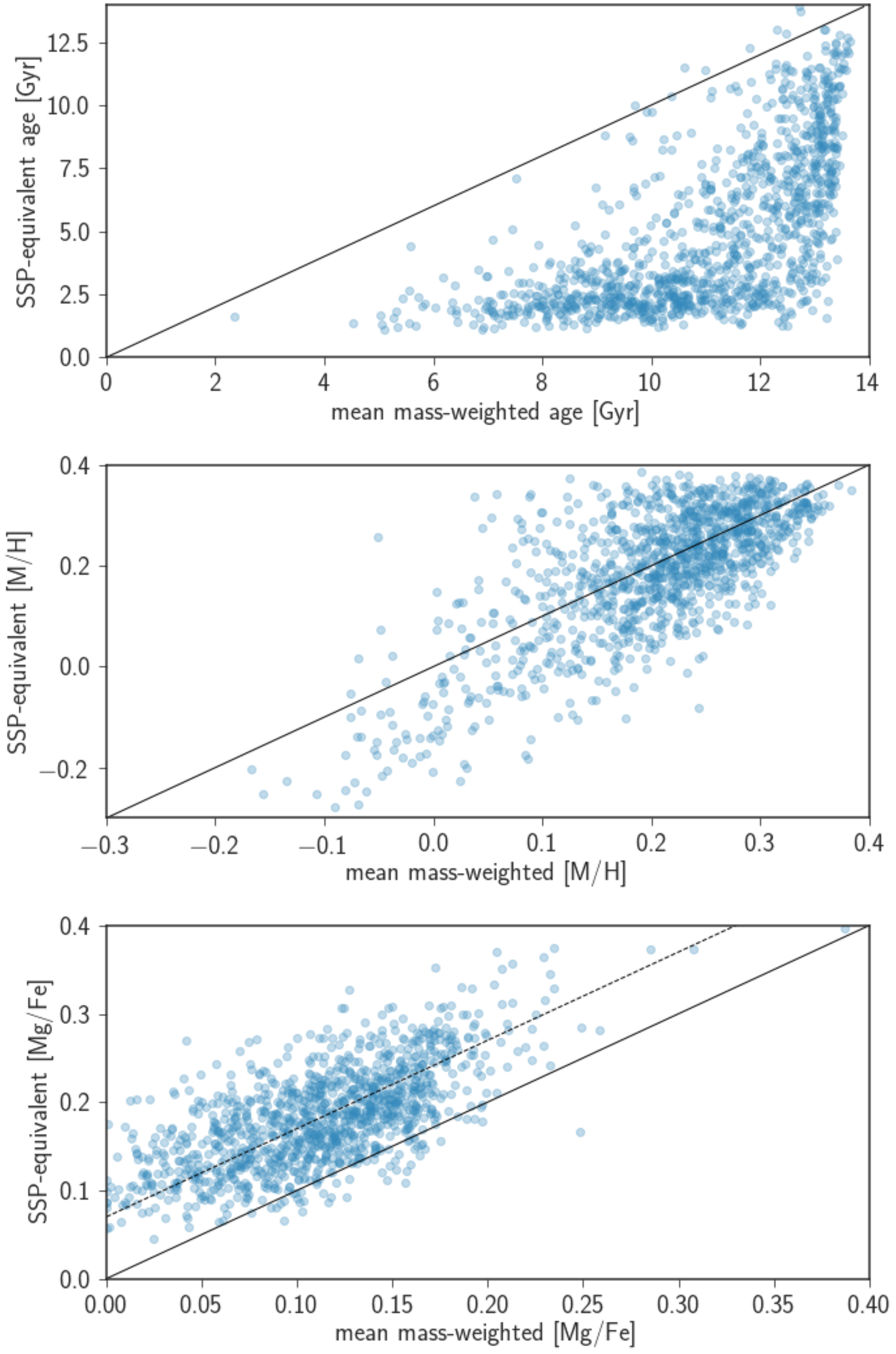}
\caption{Comparison of the SSP-equivalent age (top panel), [M/H] (middle panel) and [Mg/Fe] (bottom panel) obtained from our line-strength analysis to the mass-weighted values obtained from full spectrum fitting with pPXF. The solid line is the 1-to-1 line, while in the bottom panel, the dashed line shows a constant offset of 0.07 to guide the eye. }
\label{fig:comp_ppxf_indices}
\end{figure}

We compute line-strength indices in the LIS system \citep{Vazdekis2010}, using a constant resolution of 8.4 \AA\ as a function of wavelength. We use H$\beta_\mathrm{o}$ as an age indicator and Fe5015, Fe5270, Fe5335,  and Mg$b$ to trace metallicity and [Mg/Fe]. Our analysis is based on the same MILES stellar population models as the ones used with pPXF for full spectrum fitting. 
We derive SSP-equivalent age, metallicity and [Mg/Fe] using a Markov Chain Monte Carlo analysis, using flat priors for all variables and a Gaussian likelihood. The analysis is run using the emcee code \citep{Foreman2013}, with 100 walkers and 1000 iterations. 

For each spaxel, we derive the SSP-equivalent age, metallicity and [Mg/Fe] from the median of the posterior distribution of each parameter. We show in Figure \ref{fig:comp_indices_maps} the spatial distribution of those three quantities, and in Figure \ref{fig:comp_ppxf_indices} a comparison between the SSP-equivalent values and the average mass-weighted values from our standard full spectrum fitting analysis with pPXF.

Figure \ref{fig:comp_indices_maps} shows that the main trends we find with full spectrum fitting are also recovered with the line-strength analysis. From the thin to the thick disc, we see an increase in age, a decrease in metallicity and an increase in [Mg/Fe], similarly to what we found with pPXF. We also find that the nuclear disc is old, metal-rich and $\alpha$-poor. The main difference with our pPXF analysis is that the thin disc shows very low SSP-equivalent ages, from 1 to 4 Gyr over most of the thin disc.

This is also apparent in Figure \ref{fig:comp_ppxf_indices}, where in the top panel we compare the SSP equivalent ages to the mass-weighted mean ages obtained with pPXF. While the old SSP-equivalent ages correspond quite well to old mass-weighted mean ages, the SSP-equivalent ages are driven to low values as soon as the spectra contain a contribution from younger stars. This has already been shown, for instance by \cite{Serra2007} and \cite{Trager2009}: the Balmer line indices are dominated by young stars and the SSP-equivalent ages reflect the fraction of stars formed within the past Gyr.

By contrast, hot young stars do not contribute much to metal lines, so that the SSP-equivalent [M/H] and [Mg/Fe] are much closer to the mass-weighted averages from pPXF (see middle and bottom panels in Figure \ref{fig:comp_ppxf_indices}). There is a slight offset in [Mg/Fe], with SSP-equivalent values larger by $\sim 0.07$, but there are no additional biases. This also confirms that the pPXF analysis interpolating between only two values of [Mg/Fe] recovers reasonable values for the mass-weighted [Mg/Fe] of a stellar population.
 
\section{Stellar populations recovery using STECKMAP}\label{appendix:steckmap}
\begin{figure}
\includegraphics[width=0.9\columnwidth]{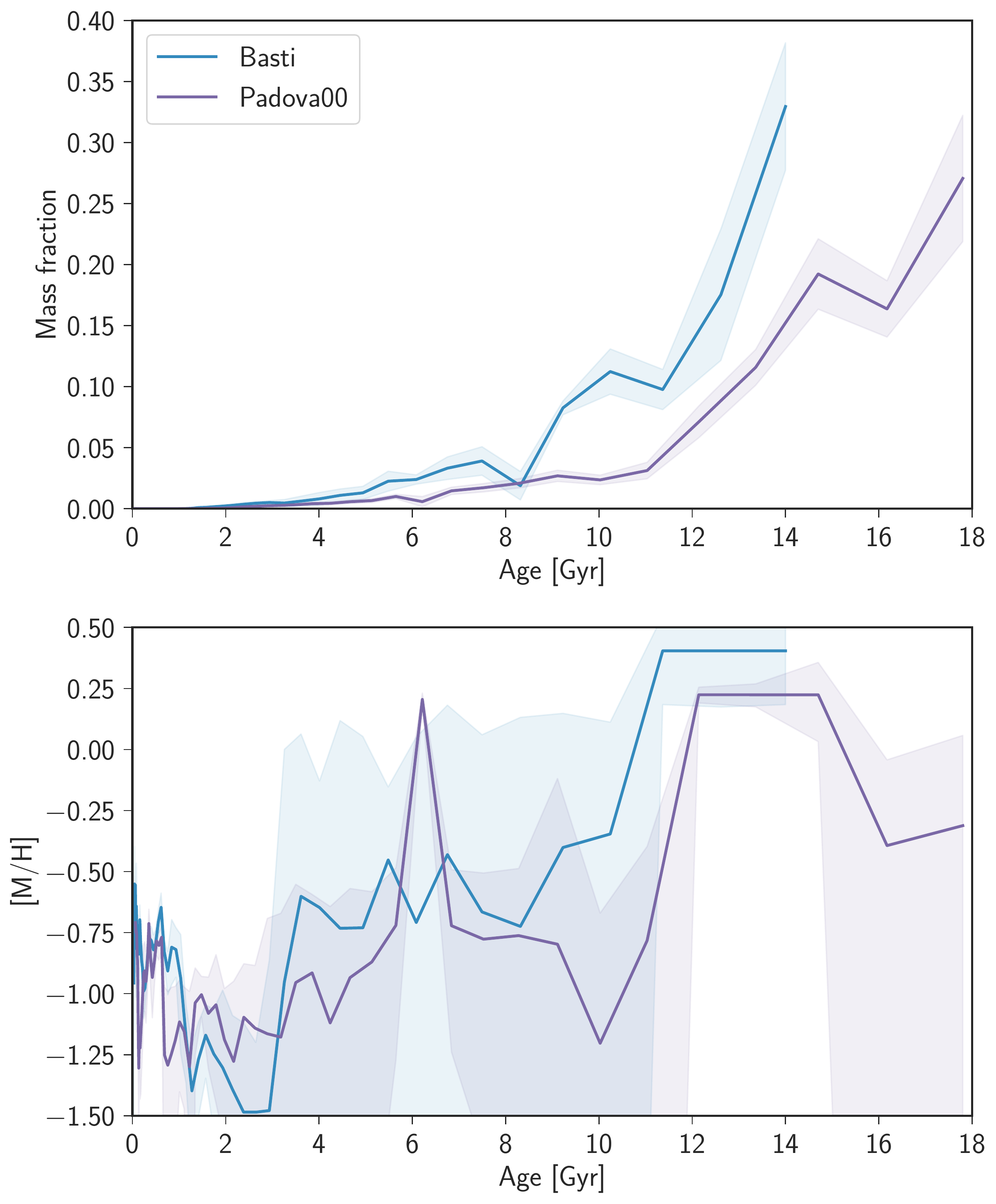}
\caption{Results of our STECKMAP analysis of stellar populations in the thick disc. We show the mass fraction as a function of age (top panel), and the mean [M/H] as a function of age (bottom panel) for two different sets of isochrones, the  BaSTI and Padova00 isochrones (the shaded regions represent the standard deviation obtained from our Monte Carlo analysis). Those results are consistent with our  pPXF analysis, with the detection of both an old metal-rich and a younger more metal-poor population in the thick disc of NGC 5746. }
\label{fig:comp_steckmap}
\end{figure}

As a sanity check for the reported findings on the stellar populations inhabiting the thick disc of NGC 5746, we performed a parallel (and blind) test using another widely used and well tested code, STECKMAP (``STEllar Content and Kinematics via Maximum A Posteriori likelihood'', \citealp{Ocvirk2006}). STECKMAP uses a Bayesian minimization method based on a penalized $\chi^2$ to reproduce the observed spectrum via a maximum {\it a posteriori} algorithm. Like pPXF, the non-parametric nature of STECKMAP allows us to characterise the stellar populations shaping observed spectra with no assumptions on the shape of the solution (i.e. the star formation histories). 

We combined the emission-cleaned spectra of all spaxels belonging to the thick disc into a single spectrum, which we fed to STECKMAP. For a fair comparison with the pPXF results, we use the same wavelength range (4750 to 5500 \AA) and the same stellar evolution models: MILES models \citep[][]{Vazdekis2015} using the  BaSTI stellar isochrones \citep{Pietrinferni2004}. The fact that STECKMAP cannot fit simultaneously age, metallicity, and [Mg/Fe] forces us to use in this case the MILES models in their {\it baseFe} version. We also use a value of 1 and 0.01 for the age and metallicity distribution smoothing parameters ($\mu_x$ and $\mu_Z$; regularization is implemented in STECKMAP using these parameters). 

Errors in the mass fractions and metallicities of the different subpopulations with STECKMAP are computed through a series of 25 Monte Carlo simulations as described in \citet[][]{Ruiz2017}: in each of the Monte Carlo simulation we add noise consistent with the quality of the original data to the best fit of the observed spectrum and run STECKMAP  again on this noisy, best fit spectrum. The standard deviation of the 25 mass fractions or metallicities from each realisation is considered the error in each magnitude.

The results of this test can be seen in Figure \ref{fig:comp_steckmap} (blue lines).  Consistently with the analysis using  pPXF, we also find that, while the thick disc mostly contains old and metal-rich stars, a younger ($\sim$ 10 Gyr old) and more metal-poor ($\mathrm{[M/H]}\simeq-0.6$) population is also present.

We tested as well the effect of a different choice of stellar isochrones, the Padova00 isochrones \citep[][]{Girardi2000}: the results of this analysis are shown as purple lines in Figure \ref{fig:comp_steckmap}. While the different age range covered by the Padova00 isochrones does not allow a direct comparison with the results obtained using the  BaSTI isochrones, we still find the same overall trend. In particular, we find the same drop in mean [M/H] from the oldest stars to the younger stars. 

To conclude, we still find the presence of younger metal-poor stars in the thick disc when we perform our analysis with a totally different code (STECKMAP), using a different version of the BaSTI MILES models (the {\it baseFe} version), and a different set of stellar isochrones (the Padova00 isochrones). This strongly reinforces our confidence in the results discussed in this paper.

\section{Bootstrap}\label{appendix:bootstrap}
\begin{figure*}
\includegraphics[width=2\columnwidth]{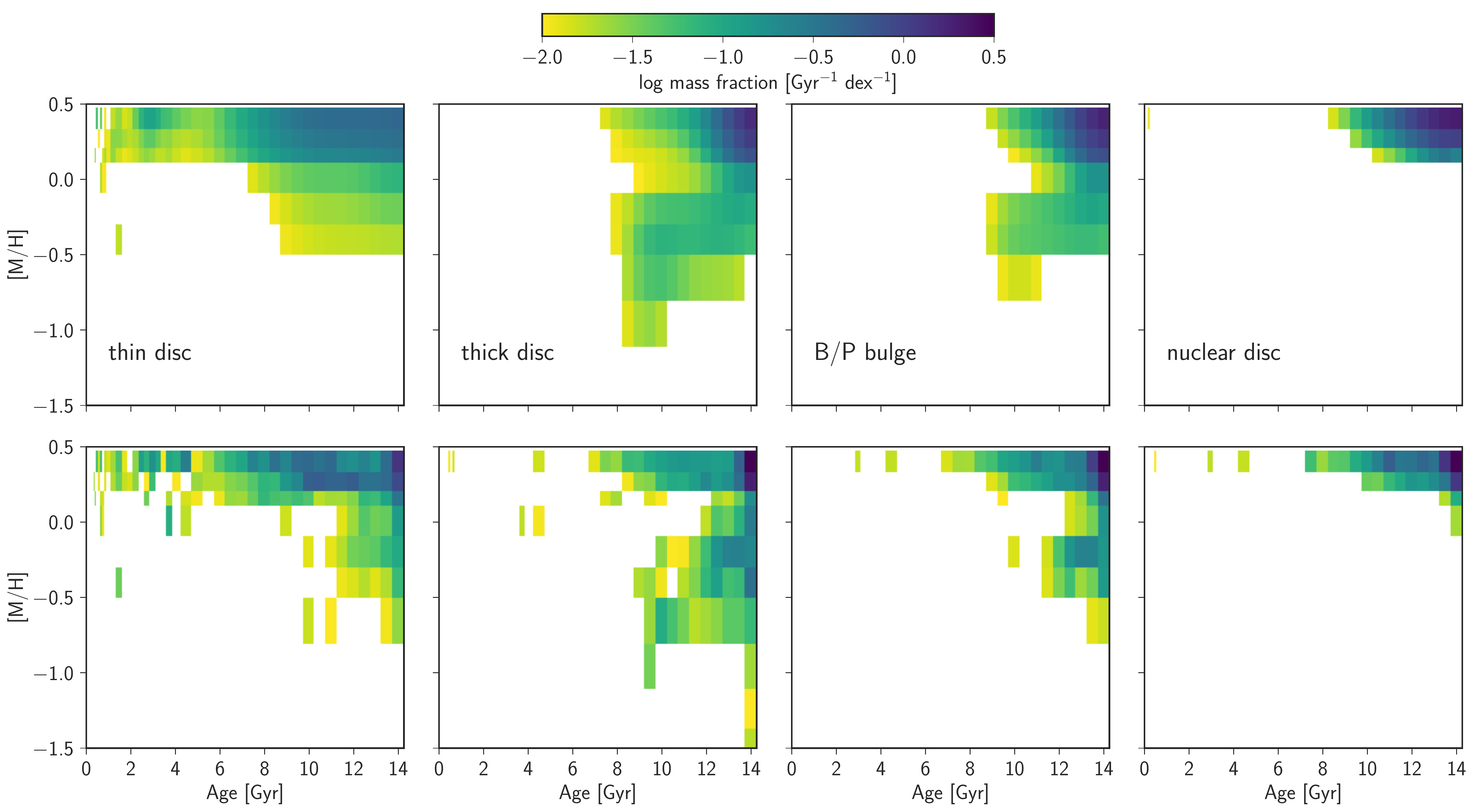}
\caption{Comparison of the age and \Fe distribution for the regularized pPXF solution (top row) and for the mean of 100 bootstrap samples (bottom row) for the thin disc, thick disc, B/P bulge and nuclear disc (from left to right)}
\label{fig:comp_bootstrap}
\end{figure*}
 Figure \ref{fig:comp_bootstrap} compares the age and \Fe distributions obtained from the pPXF best fit with regularization to the distribution obtained from our bootstrap analysis, for the thin and thick disc, the B/P bulge and the nuclear disc. We find very comparable results for the two analyses, as discussed in Section \ref{sec:pops_components}.


\bsp	
\label{lastpage}
\end{document}